\documentclass[12pt,superscriptaddress,letter]{revtex4}%
\usepackage{amssymb}
\usepackage{color}
\usepackage{graphicx}
\usepackage{dcolumn}
\usepackage{bm}
\usepackage[header,title,page,titletoc]{appendix}
\usepackage{amsmath}
\usepackage{subfigure}
\usepackage{amsfonts}%
\providecommand{\U}[1]{\protect \rule{.1in}{.1in}}
\begin{document}
\title{Knot Physics on Entangled Vortex-Membranes: Classification, Dynamics and
Effective Theory}
\author{Su-Peng Kou}
\thanks{Corresponding author}
\email{spkou@bnu.edu.cn}
\affiliation{Center for Advanced Quantum Studies, Department of Physics, Beijing Normal
University, Beijing, 100875, P. R. China}

\begin{abstract}
In this paper, knot physics on entangled vortex-membranes are studied
including classification, knot dynamics and effective theory. The physics
objects in this paper are entangled vortex-membranes that are called composite
knot-crystals. Under projection, a composite knot-crystal is reduced into
coupled zero-lattices. In the continuum limit, the effective theories of
coupled zero-lattices become quantum field theories. After considering the
topological interplay between knots and different types of zero-lattices,
gauge interactions emerge. Based on a particular composite knot-crystal with
($\mathcal{N}=4$, $\mathcal{M}=3$) (we call it standard knot-crystal), the
derived effective model becomes the (one-flavor) Standard model. As a result,
the knot physics may provide an alternative interpretation on quantum field theory.

\end{abstract}
\maketitle

\section{Introduction}

A vortex (point-vortex, vortex-line, vortex-membrane) is among the most
important and most studied objects of fluid mechanics that consists of the
rotating motion of fluid around a common center (a point, a line, or a
membrane). In three dimensional (3D) superfluid (SF), it has been known that a
vortex-line is subject to wavy distortions called Kelvin
waves\cite{Thomson1880a,Donnelly1991a}. Because Kelvin waves are relevant to
Kolmogorov-like turbulence\cite{S95,V00}, a variety of approaches have been
used to study this phenomenon. For two entangled vortex-rings, there may exist
the leapfrogging motion in classical fluids according to the works of
Helmholtz and Kelvin\cite{dys93, hic22, bor13,wac14, cap14,1}. Kelvin came to
the idea that classical atoms were knots of swirling vortices in the
luminiferous aether. Chemical elements correspond to knots and links. The
study of knotted vortex-lines and their dynamics has attracted scientists from
diverse settings, including classical fluid dynamics and superfluid
dynamics\cite{Kleckner2013,Hall2016}.

In the paper\cite{kou}, the Kelvin wave and knot dynamics on three dimensional
vortex-membranes in five dimensional fluid were studied. A new theory -
\emph{knot physics} is developed to characterize the entanglement evolution of
3D leapfrogging vortex-membranes. Owning to the the conservation conditions of
the volume of the knot in 5D space, the shape and the volume of knot is never
changed and the knot can only split and knot-pieces evolutes following the
equation of motion of Schr\"{o}dinger equation. Three dimensional quantum
Dirac model is derived to describe the entanglement evolution of the entangled
vortex-membranes: The elementary excitations are knots with a projected zero;
The physics quality to describe local deformation is knot density or the
density of zeros between two projected vortex-membranes; The Biot-Savart
equation for Kelvin waves becomes Schr\"{o}dinger equation for probability
waves of knots; The angular frequency for leapfrogging motion turns into the
mass of knots; etc.\emph{ }The knot physics may give an alternative
interpretation on quantum mechanics.

In this paper, we will study the Kelvin wave and knot dynamics on complex
entangled vortex-membranes -- a composite knot-crystal. By projecting
entangled vortex-membranes into several coupled zero-lattices (T-zero-lattices
and W-zero-lattices), the information of the system becomes the coupled
zero-lattices with internal degrees of freedom. After considering the
topological interplay between knots and zero-lattices, different kinds of
gauge interactions emerge. In particular, it is the 3D quantum gauge field
theories that characterize the knot dynamics of the composite knot-crystal.
The knot physics may give a complete interpretation on quantum chromodynamics
(QCD) and quantum electrodynamics (QED): The collective motions of composite
knot-crystal are described by quantum fluctuations of zero-lattices: fermionic
elementary particles are knots that are topological defects of zero-lattices
with internal-twistings; \textrm{U(1)} gauge field (Maxwell field) are phase
fluctuations of internal twistings of internal T-type zero-lattice;
\textrm{SU(N)} Yang-Mills field are number fluctuations of internal twistings
of T-type zero-lattice, ... Two composite knots interact by exchanging
fluctuations of the internal-twistings. Based on a particular composite
knot-crystal with ($\mathcal{N}=4$, $\mathcal{M}=3$) (we call it standard
knot-crystal), the derived effective model is just the (one-flavor) Standard model.

The paper is organized as below. In Sec. II, we review Biot-Savart mechanics.
In Sec. III, we give the mathematical definition of composite knot-crystal and
show the classification of composite knot-crystals. In Sec. IV, we introduce
the zero-lattices by projecting knot-crystals and show a topological
constraint -- twist-writhe locking condition for a composite knot-crystal. In
Sec. V, we obtain the effective theory for knot on a 1-level knot-crystal with
($\mathcal{N}=1$, $\mathcal{M}=1$) and the knots are described by Weyl
equation. In Sec. VI, we show the dynamics of 1-level knot-crystal with
($\mathcal{N}=2$, $\mathcal{M}=1$) and the knots are described by Dirac
equation. In VII, we obtain the effective theory of knots on a 2-level
knot-crystal with ($\mathcal{N}=2$, $\mathcal{M}=2$) and the collective
motions of composite knot-crystal are described by a chiral \textrm{SU}%
$_{\mathrm{weak}}$\textrm{(2)} gauge theory with Weyl fermions and Dirac
fermions. In VIII, we obtain the effect theory of knots on a 2-level
knot-crystal with ($\mathcal{N}=4$, $\mathcal{M}=2$) and the collective
motions of composite knot-crystal are described by \textrm{SU}%
$_{\mathrm{strong}}$\textrm{(n)}$\times$\textrm{U}$_{\mathrm{em}}$\textrm{(1)}
gauge field theory. In this section, the \textrm{SU}$_{\mathrm{strong}}%
$\textrm{(n)}$\times$\textrm{U}$_{\mathrm{em}}$\textrm{(1)} gauge fields can
be viewed as fluctuations of the internal twistings. The knots correspond to
quarks and electrons. In Sec. IX, based on a 3-level composite knot-crystal
with ($\mathcal{N}=4$, $\mathcal{M}=3$) (we call it standard knot-crystal),
the derived effective model becomes the (one-flavor) Standard model with
\textrm{SU}$_{\mathrm{strong}}$\textrm{(n)}$\times$\textrm{U}$_{\mathrm{em}}%
$\textrm{(1)}$\times$\textrm{SU}$_{\mathrm{weak}}$\textrm{(2) }gauge fields.
The knots correspond to quarks, electrons, and neutrinos. Finally, the
conclusions are drawn in Sec. X.

\section{Review on Biot-Savart mechanics}

In the paper\cite{kou}, to characterize the entanglement evolution of
vortex-membranes, a new theory -- knot physics was developed from Biot-Savart
mechanics. In this paper, firstly we review the key points of the Biot-Savart mechanics.

The 3D vortex-membrane is defined by a given singular vorticity
$\mathbf{\Omega}=\kappa \delta_{P}$ in the 5D inviscid incompressible fluid
($\nabla \cdot \mathbf{v}\equiv0$), the singular $\delta$-type vorticity denotes
the sub-manifold $P$ in 5D space, and $\kappa$ is the constant circulation
strength. For 5D case, we have 3D vortex-membranes with Marsden-Weinstein (MW)
symplectic structure\cite{leap}.

The generalized Biot-Savart equation for a 3D vortex-piece under local
induction approximation (LIA) can be described by Hamiltonian formula
\begin{equation}
\dot{\xi}=\frac{\partial \mathrm{H}_{\text{\textrm{volume}}}(P)}{\partial \eta
},\text{ }\dot{\eta}=-\frac{\partial \mathrm{H}_{\text{\textrm{volume}}}%
(P)}{\partial \xi}%
\end{equation}
where the Hamiltonian on the vortex-membranes is just $3$-volume
\begin{equation}
\mathrm{H}_{\text{\textrm{volume}}}(P)=(\kappa \alpha \ln \epsilon)\cdot
\mathrm{volume}(P)
\end{equation}
with $\mathrm{volume}(P)=\int_{P}dV_{P}$ and $\alpha=\frac{\Gamma(\frac{5}%
{2})}{6\pi^{\frac{5}{2}}}$. Here $\epsilon$ is defined by $\epsilon
={\frac{\ell}{a_{0}}}$ where $\ell$ is the length of the order of the
curvature radius (or inter-vortex distance when the considered infinitesimal
vortex is a part of a vortex tangle) and $a_{0}$ denotes the infinitesimal
vortex radius which is much smaller than any other characteristic size in the
system. In complex description, $\mathrm{z}=\xi+i\eta,$ above equation can
also be written into\cite{leap} $i\frac{d\mathrm{z}}{dt}=\frac{\delta
\mathrm{H}_{\text{\textrm{volume}}}(P)}{\delta \mathrm{z}^{\ast}}.$

For Kelvin waves on a 3D helical vortex-membrane, the plane-wave is described
by a complex field, $\mathrm{z}(\vec{x},t)=r_{0}e^{\pm i\vec{k}\cdot \vec
{x}-i\omega t+i\phi_{0}}$ where $\vec{k}$ is the winding wave vector along a
direction on 3D vortex-membrane with $\left \vert \vec{k}\right \vert =\frac
{\pi}{a}$ and $a$ is a fixed length that denotes the half pitch of the
windings. The (Lamb impulse) momentum and the (Lamb impulse) angular momentum
along $\mathbf{\vec{e}}$-direction on vortex-membrane with a plane Kelvin wave
are given by $\vec{p}_{\mathrm{Lamb}}=\mathbf{P}_{\mathrm{Lamb}}%
\cdot \mathbf{\vec{e}}=\pm \frac{1}{2}\rho \kappa V_{P}a^{2}\vec{k}$ and
$\left \vert J_{\mathrm{Lamb}}\right \vert =\left \vert \mathbf{J}_{\mathrm{Lamb}%
}\cdot \mathbf{\vec{e}}\right \vert =\frac{1}{2}\rho \kappa V_{P}a^{2},$
respectively. $V_{p}$ is the total volume length of the vortex-membrane.
Because the projected (Lamb impulse) angular momentum is a constant on the
vortex-membrane, the effective Planck constant $\hbar_{\mathrm{knot}}$ is
derived as angular momentum in extra space that is proportional to the volume
of the knot in 5D space, i.e., $\hbar_{\mathrm{eff}}=J_{\mathrm{Lamb}}%
=\frac{1}{2}\rho_{0}\kappa V_{P}r_{0}^{2}$ where $V_{p}$ is the total volume
of the vortex-membrane and $\rho_{0}$ is the superfluid mass density.

For two entangled vortex-membranes, the nonlocal interaction leads to
leapfrogging motion\cite{1}. For leapfrogging motion, the entangled
vortex-membranes exchange energy in a periodic fashion. The winding radii of
two vortex-membranes oscillate with a fixed leapfrogging angular frequency
$\omega^{\ast}=(\alpha \kappa \ln \epsilon)\frac{2}{r_{0}^{2}}$ where $r_{0}$ is
the distance between two vortex-membranes.

A knot is an elementary entanglement between two vortex-membranes with fixed
volume. On the one hand, a knot is $\pi$-phase changing -- a \emph{sharp,
time-independent, topological} phase changing, on the other hand, a knot has a
phase angle. Quantum mechanics describes the dynamics of \emph{smooth, slow,
non-topological} phase changing of knots. From point view of information, the
elementary volume-changing with a zero is a knot. Under fixed-volume
condition, the knot becomes fragmentized and obeys quantum mechanics rather
than pseudo-quantum mechanics. This is the fundamental principle of quantum
mechanics. The effective Planck constant $\hbar_{\mathrm{knot}}$ is derived as
angular momentum in extra space (the volume of the knot in 5D space)
\begin{equation}
\hbar_{\mathrm{knot}}=\frac{1}{2}\rho_{0}\kappa V_{P}r_{0}^{2}%
\end{equation}
where $V_{p}$ is the total volume of the vortex-membrane and $\rho_{0}$ is the
superfluid mass density.

We pointed out that the function of a Kelvin wave with an extra fragmentized
knot describes the distribution of the knot-pieces and $\mathrm{z}(\vec{x},t)$
plays the role of the wave-function in quantum mechanics as
\begin{equation}
\frac{1}{\sqrt{V_{P}}}\frac{\mathrm{z}(\vec{x},t)}{r_{0}}=\sqrt{\rho
_{\mathrm{knot}}(x,t)}e^{i\Delta \phi(\vec{x},t)}\Longleftrightarrow \psi
(\vec{x},t).
\end{equation}
The angle $\Delta \phi(\vec{x},t)$ becomes the quantum phase angle of
wave-function, the knot density $\rho_{\mathrm{knot}}=\left \langle
\frac{\Delta \hat{K}}{\Delta V_{P}}\right \rangle $ becomes the probability
density of knot-pieces $n_{\mathrm{knot}}(\vec{x})$. For a plane wave,
$\psi(\vec{x},t)=\frac{1}{\sqrt{V_{P}}}e^{-i\omega \cdot t+i\vec{k}\cdot \vec
{x}}$, the projected (Lamb impulse) energy of a knot is
\begin{equation}
E_{\mathrm{knot}}=\hbar_{\mathrm{knot}}\omega
\end{equation}
and the projected (Lamb impulse) momentum of a knot is
\begin{equation}
\vec{p}_{\mathrm{knot}}=\hbar_{\mathrm{knot}}\vec{k}%
\end{equation}
where the effective Planck constant $\hbar_{\mathrm{knot}}$ is obtained as
projected (Lamb impulse) angular momentum of a knot (the elementary
volume-changing of two entangled vortex-membranes)
\[
\hbar_{\mathrm{knot}}=J_{\mathrm{knot}}=\frac{1}{2}\rho_{0}\kappa V_{P}%
r_{0}^{2}.
\]
In general, the energy and momentum for a knot are described by operators%
\begin{equation}
E_{\mathrm{knot}}\rightarrow \hat{E}_{\mathrm{knot}}=i\hbar_{\mathrm{eff}}%
\frac{d}{dt}\text{, }\vec{p}_{\mathrm{knot}}\rightarrow \hat{p}_{\mathrm{knot}%
}=-i\hbar_{\mathrm{eff}}\frac{d}{d\vec{x}}.
\end{equation}
\  \ The energy-momentum relationship $E=H(\vec{p})$ becomes the equation of
motion for wave-function,
\begin{equation}
i\hbar_{\mathrm{eff}}\frac{d}{dt}\psi(\vec{x})=\hat{H}_{\mathrm{knot}}(\hat
{p})\psi(\vec{x}).
\end{equation}

We also use path-integral formulation to describe quantum processes in knot
physics. For a multi-knot system, the probability amplitude is defined by
\begin{align}
&  \langle t_{f},\vec{x}_{M}^{\prime},...,\vec{x}_{2}^{\prime},\vec{x}%
_{1}^{\prime}\left \vert t_{i},\vec{x}_{M},...,\vec{x}_{2},\vec{x}%
_{1}\right \rangle \nonumber \\
&  =\int \mathcal{D}\psi^{\dagger}(\vec{x},t)\mathcal{D}\psi(\vec
{x},t)e^{i\mathcal{S}_{\mathrm{knot}}/\hbar_{\mathrm{eff}}}%
\end{align}
where $\mathcal{S}_{\mathrm{knot}}=%
%TCIMACRO{\dsum \limits_{\omega,\vec{p}}}%
%BeginExpansion
{\displaystyle \sum \limits_{\omega,\vec{p}}}
%EndExpansion
S_{\omega,\vec{p}}=\int \mathcal{L}_{\mathrm{knot}}dtd^{3}x$ with
$S_{\omega,\vec{p}}=\psi_{\vec{p}}^{\dagger}(i\hbar_{\mathrm{eff}}\omega
(\vec{p})-H_{\mathrm{knot}}(\vec{p}))\psi_{\vec{p}}$ and $\mathcal{L}%
_{\mathrm{knot}}=i\psi^{\dagger}\partial_{t}\psi-\mathcal{\hat{H}%
}_{\mathrm{knot}}.$

However, there is an unsolved problem in Ref.\cite{kou}: because free Dirac
model is non-interacting, we don't know how to do a physical dynamic
projection. In this paper, we develop an effective quantum gauge field theory
and solve above problem.

\section{Composite knot-crystal: definition, classification and generalized
translation symmetry}

In solid state physics, a basic theory is about atom-crystal and its lattices.
The atom-crystal has its inherent symmetry, by which we may classify different
types of symmetries (translational symmetry, rotation symmetry, mirror
symmetry). For example, there are $230$ distinct space groups in 3D space. In
addition to simple crystal with monatomic lattices, there exist composite
crystals with polyatomic lattices that have more than one type of atoms in a
unit cell.

In the paper \cite{kou}, a periodic entanglement-pattern between two
vortex-membranes is called \emph{knot-crystal}. The definition of a
"knot-crystal" is based on periodic structures of knots that is similar to
atom-crystal where the atoms form a periodic arrangement. However,
knot-crystal is different from traditional atom-crystal. Fig.1(a) shows a 1D
knot-crystal in 3D space. Due to the existence of rotation symmetry and
generalized translation symmetry, the properties of knot-crystals are much
different from that of atom-crystals. In this paper, under local induction
approximation (LIA), we discuss the properties of composite knot-crystals, a
periodic entanglement-pattern between multi-vortex-membrane. The name of
composite knot-crystal comes from the similarity to the polyatomic
atom-crystal with a composite lattice.

\subsection{Definition}

Firstly we consider an arbitrary $d$ dimensional composite knot-crystal that
comes from a periodic entanglement pattern by entangled vortex-membranes in
$d+2$ dimensional space $(x_{1},x_{2},...,x_{d},x_{d+1},x_{d+2})$. To
characterize a composite knot-crystal, we define the function by an
$\mathcal{N}\times \mathcal{M}$ matrix
\begin{equation}
\mathbf{Z(}\vec{x},t\mathbf{)=}\left(
\begin{array}
[c]{c}%
\mathbf{z}_{1}\mathbf{(}\vec{x},t\mathbf{)}\\
\mathbf{z}_{2}\mathbf{(}\vec{x},t\mathbf{)}\\
\mathbf{...}\\
\mathbf{z}_{\mathcal{N}}\mathbf{(}\vec{x},t\mathbf{)}%
\end{array}
\right)  =\left(
\begin{array}
[c]{cccc}%
z_{1,1}(\vec{x},t) & z_{1,2}(\vec{x},t) & ... & z_{1,\mathcal{M}}(\vec{x},t)\\
z_{2,1}(\vec{x},t) & z_{2,2}(\vec{x},t) & ... & ...\\
... & ... & ... & ...\\
z_{\mathcal{N},1}(\vec{x},t) & z_{\mathcal{N},2}(\vec{x},t) & ... &
z_{\mathcal{N},\mathcal{M}}(\vec{x},t)
\end{array}
\right)
\end{equation}
where $\vec{x}=(x_{1},x_{2},...,x_{d})$. $\mathcal{N}$ and $\mathcal{M}$
denote the vortex-index and the level-index, respectively. We point out that
$\mathbf{Z(}\vec{x},t\mathbf{)}$ is just matrix representation to show the
properties of knot-crystal and there exist $\mathcal{N}\times \mathcal{M}$
independent functions. $\mathbf{z}_{j}\mathbf{(}\vec{x},t\mathbf{)}$ denotes
the function of j-th vortex-membrane. A generalized definition of a
knot-crystal is given by $\mathcal{N}\times \mathcal{M}$ independent functions
of j-th level i-th vortex-membrane $z_{i,j}(\vec{x},t)$ that is an element of
the matrix\ as
\begin{equation}
z_{i,j}(\vec{x},t)=%
%TCIMACRO{\dprod \limits_{I}}%
%BeginExpansion
{\displaystyle \prod \limits_{I}}
%EndExpansion
r_{i,j}(\alpha_{i,j}^{I}e^{i\phi_{i,j}^{I}}+\beta_{i,j}^{I}e^{-i\phi_{i,j}%
^{I}})e^{i\omega_{i,j}t+i(\phi_{i,j}^{I})_{0}}%
\end{equation}
where $\left \vert \alpha_{i,j}^{I}\right \vert ^{2}+\left \vert \beta_{i,j}%
^{I}\right \vert ^{2}=1$ and $I=x_{1},x_{2},...,x_{d}$. $\phi_{i,j}^{I}$ and
$(\phi_{i,j}^{I})_{0}$ denote the winding phase angle and the constant phase
angle along given $x^{I}$-direction j-th level i-th vortex-membrane,
respectively. $\omega_{i,j}$ is rotating velocity and $r_{i,j}$ is the radius
of j-th level winding of i-th vortex-membrane, respectively. In general, to
guarantee the stability of the composite knot-crystal, we consider LIA,
\begin{equation}
r_{i,j-1}\gg r_{i,j}.
\end{equation}
In particular, j-th level i-th vortex-membrane (that is a Kelvin wave) becomes
center membrane of (j--1)th level i-th vortex-membrane.

\begin{figure}[ptb]
\includegraphics[clip,width=0.53\textwidth]{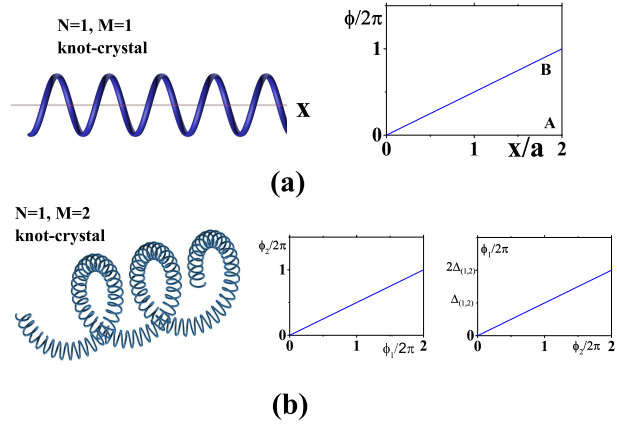}\caption{(a) An
illustration of a knot-crystal with ($\mathcal{N}=1$ and $\mathcal{M}=1$). The
linear relationship between winding angle $\phi$ and $x$ indicates the
generalized spatial translation symmetry; (b) An illustration of a composite
knot-crystal with ($\mathcal{N}=1$ and $\mathcal{M}=1$). There exists linear
relationship between winding angle of level-1 Kelvin wave $\phi_{1}$ and $x$
and linear relationship between winding angle of level-2 Kelvin wave $\phi
_{2}$ and winding angle of level-1 Kelvin wave $\phi_{1}$. In particular,
level-2 vortex-membrane (that is a Kelvin wave) becomes center membrane of
level-1 vortex-membrane.}%
\end{figure}

To determine a\ composite knot-crystal, there is a \emph{hierarchy recurrence
relationship} between two nearest-neighbor levels along $x^{I}$-direction
\begin{align}
\phi_{i,\mathcal{M}}^{I}  &  =k_{i,0}^{I}\cdot x^{I},\nonumber \\
\phi_{i,\mathcal{M}-1}^{I}  &  =\Delta_{i,(\mathcal{M}-1,\mathcal{M})}^{I}%
\phi_{i,\mathcal{M}}^{I},\nonumber \\
&  ...\nonumber \\
\phi_{i,j-1}^{I}  &  =\Delta_{i,(j-1,j)}^{I}\phi_{i,j}^{I},\nonumber \\
&  ...\\
\phi_{i,1}^{I}  &  =\Delta_{i,(1,2)}^{I}\phi_{i,2}^{I},\nonumber
\end{align}
where $k_{i,0}^{I}=\frac{\pi}{a_{i,\mathcal{M}}^{I}}$ denotes winding vector
along the winding direction, and $a_{i,\mathcal{M}}^{I}$ is the length that
denotes the half pitch of largest windings of i-th vortex-membrane. Each
number of the vector $\Delta_{i,(j-1,j)}^{I}$ is the positive winding number
of (j-1)-th level windings in a j-th level winding for i-th vortex-membrane
along $x^{I}$-direction. Thus, according to the hierarchy recurrence
relationship, we have
\begin{equation}
z_{i,j}(\vec{x},t)=%
%TCIMACRO{\dprod \limits_{I}}%
%BeginExpansion
{\displaystyle \prod \limits_{I}}
%EndExpansion
r_{i,j}(\alpha_{i,j}^{I}e^{i\phi_{i,j}^{I}}+\beta_{i,j}^{I}e^{-i\phi_{i,j}%
^{I}})e^{i\omega_{i,j}t+i(\phi_{i,j}^{I})_{0}}%
\end{equation}
where $\phi_{i,j}^{I}=[%
%TCIMACRO{\dprod \nolimits_{j}^{\mathcal{M}}}%
%BeginExpansion
{\displaystyle \prod \nolimits_{j}^{\mathcal{M}}}
%EndExpansion
\Delta_{i,(j-1,j)}^{I}]\cdot \phi_{i,\mathcal{M}}^{I}.$ As a result, the
function for a knot-crystal with $\mathcal{M}$-level $\mathcal{N}$
vortex-membranes is described by the $\mathcal{N}\times \mathcal{M}$ winding
angle $\phi_{ij}^{I}$,
\begin{equation}
\mathbf{Z(}x^{I},t\mathbf{)}=\left(
\begin{array}
[c]{cccc}%
%TCIMACRO{\dprod \limits_{I}}%
%BeginExpansion
{\displaystyle \prod \limits_{I}}
%EndExpansion
z_{1,1}(\phi_{1,1}^{I},t) &
%TCIMACRO{\dprod \limits_{I}}%
%BeginExpansion
{\displaystyle \prod \limits_{I}}
%EndExpansion
z_{1,2}(\phi_{1,2}^{I},t) & ... &
%TCIMACRO{\dprod \limits_{I}}%
%BeginExpansion
{\displaystyle \prod \limits_{I}}
%EndExpansion
z_{1,\mathcal{M}}(\phi_{1,\mathcal{M}}^{I},t)\\%
%TCIMACRO{\dprod \limits_{I}}%
%BeginExpansion
{\displaystyle \prod \limits_{I}}
%EndExpansion
z_{2,1}(\phi_{2,1}^{I},t) &
%TCIMACRO{\dprod \limits_{I}}%
%BeginExpansion
{\displaystyle \prod \limits_{I}}
%EndExpansion
z_{2,2}(\phi_{2,2}^{I},t) & ... & ...\\
... & ... & ... & ...\\%
%TCIMACRO{\dprod \limits_{I}}%
%BeginExpansion
{\displaystyle \prod \limits_{I}}
%EndExpansion
z_{\mathcal{N},1}(\phi_{\mathcal{N},1}^{I},t) &
%TCIMACRO{\dprod \limits_{I}}%
%BeginExpansion
{\displaystyle \prod \limits_{I}}
%EndExpansion
z_{\mathcal{N},2}(\phi_{\mathcal{N},2}^{I},t) & ... &
%TCIMACRO{\dprod \limits_{I}}%
%BeginExpansion
{\displaystyle \prod \limits_{I}}
%EndExpansion
z_{\mathcal{N},\mathcal{M}}(\phi_{\mathcal{N},\mathcal{M}}^{I},t)
\end{array}
\right)  .
\end{equation}

In this paper, for simplify, we only consider the cases of
\[
\Delta_{i,(j-1,j)}^{I}=\Delta_{(j-1,j)}^{I}.
\]
As a result, the different vortex-membranes of the same level have the same
winding length. The hierarchy series of the particular type of composite
knot-crystal is given by
\begin{equation}
\{ \Delta_{(1,2)},\Delta_{(2,3)},...,\Delta_{(\mathcal{M}-1,\mathcal{M})}\}.
\end{equation}

\subsection{Classification}

To classify a knot-crystal, we introduce three indices: \emph{dimension }%
$d$\emph{, number of vortex-membranes }$\mathcal{N}$\emph{, level
}$\mathcal{M}$\emph{. }Each d-dimensional composite knot-crystal is denoted by
($\mathcal{N}$, $\mathcal{M}$).

The first index is dimension of knot-crystal, $d$. In this paper we focus on
3D knot-crystal ($d=3$) that is described by $\mathbf{Z(}\vec{x},t\mathbf{)}$
where $\vec{x}=(x,y,z)$. Fig.1 is an illustration of two types of 1D
knot-crystals in 3D space.

The second index is the number of vortex-membranes, $\mathcal{N}$. For a
composite knot-crystal with ($\mathcal{N}=1$, $\mathcal{M}$), the function is
\begin{equation}
\mathbf{z(}\vec{x},t\mathbf{)}=\left(
\begin{array}
[c]{ccc}%
r_{1}(\alpha_{1}e^{i\phi_{1}}+\beta_{1}e^{-i\phi_{1}})e^{i\omega_{1}%
t+i(\phi_{1})_{0}} & ... & r_{\mathcal{M}}(\alpha_{\mathcal{M}}e^{i\phi
_{\mathcal{M}}}+\beta_{1}e^{-i\phi_{\mathcal{M}}})e^{i\omega_{\mathcal{M}%
}t+i(\phi_{\mathcal{M}})_{0}}%
\end{array}
\right)
\end{equation}
where $\left \vert \alpha_{i}\right \vert ^{2}+\left \vert \beta_{i}\right \vert
^{2}=1$ and $r_{i}\gg r_{i-1}$. Fig.1(b) is an example of a 1D composite
knot-crystal from one vortex-membrane, of which the function is given by
\begin{equation}
\mathbf{z(}x,t\mathbf{)}=\left(
\begin{array}
[c]{cc}%
r_{1}e^{i\phi_{1}(x)+i\omega_{1}t+i(\phi_{1})_{0}} & r_{2}e^{i\phi
_{2}(x)+i\omega_{2}t+i(\phi_{2})_{0}}%
\end{array}
\right)
\end{equation}
where $r_{1}\ll r_{2},$ $\phi_{2}(x)=\Delta_{(1,2)}\cdot \phi_{1}(x)$ and
$\Delta_{(1,2)}$ is a positive number.

The third index is \emph{level }of knot-crystal, $\mathcal{M}$. To define the
concept of level, we introduce composite knot-crystal that corresponds to
polyatomic atom-crystal with a composite lattice. Fig.1(b) is an illustration
of a 2-level knot-crystal with ($\mathcal{N}=1$, $\mathcal{M}=2$)$.$ An
important composite knot-crystal is standard knot-crystal with ($\mathcal{N}%
=4$, $\mathcal{M}=3$), of which the hierarchy series is given by
\begin{equation}
\{ \Delta_{(1,2)},\text{ }\Delta_{(2,3)}\}
\end{equation}
where $\Delta_{(1,2)}=3$ and $\Delta_{(2,3)}\gg1$.

In addition to the three indices, $d$,\emph{ }$\mathcal{N}$,\emph{
}$\mathcal{M}$, to characterize a composite knot-crystal, one need to define
its tensor network state that denotes the entanglement pattern along different
directions. In Ref.\cite{kou}, we have studied a 3D simple knot-crystal with
two vortex-membranes ($\mathcal{N}=2$, $\mathcal{M}=1$) -- spin-orbital
coupling (SOC) knot-crystal that is characterized by the following tensor
network state
\begin{align}
\left \langle \mathbf{\sigma}^{X}\otimes \vec{1}\right \rangle  &  =\vec
{n}_{\sigma}^{X}=(1,0,0),\\
\left \langle \mathbf{\sigma}^{Y}\otimes \vec{1}\right \rangle  &  =\vec
{n}_{\sigma}^{Y}=(0,1,0),\nonumber \\
\left \langle \mathbf{\sigma}^{Z}\otimes \vec{1}\right \rangle  &  =\vec
{n}_{\sigma}^{Z}=(0,0,1).\nonumber
\end{align}

\subsection{Generalized spatial translation symmetry}

One of the most important properties of a composite knot-crystal is
generalized spatial translation symmetry.

It is obvious that the knot-crystals break continuous translation symmetry,
i.e.,
\begin{equation}
\mathbf{Z}^{\prime}\mathbf{(}\vec{x},t\mathbf{)}\neq \mathcal{T}(\delta
x^{I}\rightarrow0)\mathbf{Z(}\vec{x},t\mathbf{)}%
\end{equation}
where $\mathbf{Z(}\vec{x},t\mathbf{)}$ is function for a knot-crystal and
$\mathcal{T}(\delta x^{I})$ is translation operator for knot-crystal. The
knot-crystals have discrete translation symmetry as
\begin{align}
\mathbf{Z}^{\prime}\mathbf{(}\vec{x},t\mathbf{)}  &  =\mathcal{T}(\delta
x^{I}=2a^{I})\mathbf{Z(}\vec{x},t\mathbf{)}\\
&  =\mathbf{Z(}\vec{x},t\mathbf{)}.\nonumber
\end{align}
Here $a^{I}$ is the length that denotes the half pitch of largest windings of
vortex-membrane.\ However, we point out that all knot-crystals have
\emph{generalized spatial translation symmetry}.

To define the generalized translation symmetry for a knot-crystal, we do a
translation operation,\ under which all vortex-membranes shift a distance
$\left \vert \delta x^{I}\right \vert $ along the $x^{I}$-direction. i.e.,
\begin{align}
\mathbf{Z(}\vec{x},t\mathbf{)}  &  \rightarrow \mathbf{Z}^{\prime}%
\mathbf{(}\vec{x},t\mathbf{)}=\mathcal{T}(\delta x^{I})\mathbf{Z(}\vec
{x},t\mathbf{)}\\
&  =\left(
\begin{array}
[c]{cccc}%
z_{1,1}((\phi_{1,1}^{I}+\delta \phi_{1,1}^{I}),t) & z_{1,2}((\phi_{1,2}%
^{I}+\delta \phi_{1,2}^{I}),t) & ... & z_{1,\mathcal{M}}((\phi_{1,\mathcal{M}%
}^{I}+\delta \phi_{1,\mathcal{M}}^{I}),t)\\
z_{2,1}((\phi_{2,1}^{I}+\delta \phi_{2,1}^{I}),t) & z_{2,2}((\phi_{2,2}%
^{I}+\delta \phi_{2,2}^{I}),t) & ... & ...\\
... & ... & ... & ...\\
z_{\mathcal{N},1}((\phi_{\mathcal{N},1}^{I}+\delta \phi_{\mathcal{N},1}%
^{I}),t) & z_{\mathcal{N},2}((\phi_{\mathcal{N},2}^{I}+\delta \phi
_{\mathcal{N},2}^{I}),t) & ... & z_{\mathcal{N},\mathcal{M}}((\phi
_{\mathcal{N},\mathcal{M}}^{I}+\delta \phi_{\mathcal{N},\mathcal{M}}^{I}),t)
\end{array}
\right)  .\nonumber
\end{align}
Under the global generalized translation symmetry, we have
\begin{align}
z_{i,j}(\vec{x},t)  &  \rightarrow z_{i,j}^{\prime}(\vec{x},t)=\mathcal{T}%
(\delta x^{I})z_{i,j}(\vec{x},t)\\
&  =\mathcal{T}(\delta x^{I})z_{i,j}(\phi_{i,j}^{I},t)\nonumber \\
&  =z_{i,j}((\phi_{i,j}^{I}+\Delta \phi_{i,j}^{I}),t)\text{, }%
j=1,2,...\mathcal{M}\nonumber
\end{align}
where
\begin{equation}
\delta \phi_{i,j}^{I}=\Delta_{i,j}\cdot \delta \phi_{i,j-1}^{I}.
\end{equation}

We can define generalized spatial translation symmetry for each level of
composite knot-crystal, $\mathcal{T}_{l}(\delta x^{I}),$ i.e.,
\begin{align}
z_{i,j}(\vec{x},t)  &  \rightarrow z_{i,j}^{\prime}(\vec{x},t)=\mathcal{T}%
_{l}(\delta x^{I})z_{i,j}(\vec{x},t)\\
&  =z_{i,j}((\phi_{i,j}^{I}+\delta_{lj}\Delta \phi_{i,j}^{I}),t).\nonumber
\end{align}
For the case of $l=j$, we have $\delta_{lj}=1$; For the case of $l\neq j$, we
have $\delta_{lj}=0.$

\subsection{Examples of composite knot-crystals}

\subsubsection{1-level winding knot-crystal with ($\mathcal{N}=1$,
$\mathcal{M}=1$)}

Firstly, we consider the simplest knot-crystal -- 1-level winding knot-crystal
with ($\mathcal{N}=1$, $\mathcal{M}=1$)$.$

A 3D level winding knot-crystal with ($\mathcal{N}=1$, $\mathcal{M}=1$) is
just a pure state of Kelvin wave from a vortex-membrane in five dimensional
space $(x,y,z,\xi,\eta)$ that is described by
\begin{equation}
\mathrm{z}(\vec{x},t)=r_{0}(\alpha e^{i\phi(x)}+\beta e^{-i\phi(x)})e^{i\omega
t+i\phi_{0}}.
\end{equation}

To distinguish the travelling Kelvin wave and standing Kelvin wave, we have
introduced the spin network representation (a reduction representation of
tensor network state) of Kelvin waves\cite{kou}. Different spin network states
of Kelvin waves have different spin directions
\begin{align}
\left \langle \mathbf{\sigma}^{I}\right \rangle  &  =\left \langle \mathrm{z}%
\right]  (\mathbf{\sigma}^{I})\left[  \mathrm{z}\right \rangle \\
&  =\vec{n}_{\sigma}^{I},\text{ }I=x,y,z\nonumber
\end{align}
where $\mathbf{\sigma}^{I}$ is $2\times2$ Pauli matrices for helical degree of
freedom. Different knot-crystal with ($\mathcal{N}=1$, $\mathcal{M}=1$) is
characterized by different spin network states $\vec{n}_{\sigma}.$

\subsubsection{1-level knot-crystal with ($\mathcal{N}=2$, $\mathcal{M}=1$)}

We then review the properties of another simple knot-crystal -- three
dimensional 1-level knot-crystal with ($\mathcal{N}=2$, $\mathcal{M}=1$) (two
entangled vortex-membranes) in five dimensional space $(x,y,z,\xi,\eta)$ that
is described by $\mathbf{Z(}\vec{x},t\mathbf{)}=\left(
\begin{array}
[c]{c}%
\mathrm{z}_{\mathrm{A}}(\vec{x},t)\\
\mathrm{z}_{\mathrm{B}}(\vec{x},t)
\end{array}
\right)  $.

To distinguish the travelling Kelvin wave and standing Kelvin wave, we have
introduced the tensor network representation of Kelvin waves. Different tensor
states of Kelvin waves have different tensor network
\begin{align}
\left \langle \mathbf{\sigma}^{I}\otimes \mathbf{\tau}^{I}\right \rangle  &
=\left \langle \mathbf{Z}\right]  (\mathbf{\sigma}^{I}\otimes \mathbf{\tau}%
^{I})\left[  \mathbf{Z}\right \rangle \\
&  =\vec{n}_{\sigma}^{I}\otimes \vec{1},\text{ }I=x,y,z\nonumber
\end{align}
where $\mathbf{\sigma}^{I}$, $\mathbf{\tau}^{I}$ are $2\times2$ Pauli matrices
for helical and vortex degrees of freedom, respectively. An interesting type
of knot-crystal is spin-orbital coupling (SOC) knot-crystal that is
characterized by the tensor network states
\begin{align}
\left \langle \mathbf{\sigma}^{X}\otimes \vec{1}\right \rangle  &  =\vec
{n}_{\sigma}^{X}=(1,0,0),\\
\left \langle \mathbf{\sigma}^{Y}\otimes \vec{1}\right \rangle  &  =\vec
{n}_{\sigma}^{Y}=(0,1,0),\nonumber \\
\left \langle \mathbf{\sigma}^{Z}\otimes \vec{1}\right \rangle  &  =\vec
{n}_{\sigma}^{Z}=(0,0,1).\nonumber
\end{align}

There always exists leapfrogging motion for the two entangled vortex-membranes
with a fixed leapfrogging angular frequency $\omega^{\ast}$ and fixed distance
$r_{0}.$ As a result, along x-direction, the function of the Kelvin waves
becomes
\begin{equation}
\mathbf{Z}(z,t)=\frac{r_{0}}{\sqrt{2}}\left(
\begin{array}
[c]{c}%
1+e^{i\omega^{\ast}t}\\
1-e^{i\omega^{\ast}t}%
\end{array}
\right)  \cos(k\cdot x)e^{-i\omega_{0}t};
\end{equation}
along y-direction, the function becomes
\begin{align}
\mathbf{Z}(y,t)  &  =\frac{r_{0}}{\sqrt{2}}\left(
\begin{array}
[c]{c}%
1+e^{i\omega^{\ast}t}\\
1-e^{i\omega^{\ast}t}%
\end{array}
\right)  e^{-i\omega_{0}t}\nonumber \\
&  \cdot(e^{ik\cdot y}+ie^{-ik\cdot y});
\end{align}
along z-direction, the function becomes
\begin{equation}
\mathbf{Z}(z,t)=r_{0}\left(
\begin{array}
[c]{c}%
1+e^{i\omega^{\ast}t}\\
1-e^{i\omega^{\ast}t}%
\end{array}
\right)  e^{ik\cdot z}e^{-i\omega_{0}t}.
\end{equation}

\begin{figure}[ptb]
\includegraphics[clip,width=0.53\textwidth]{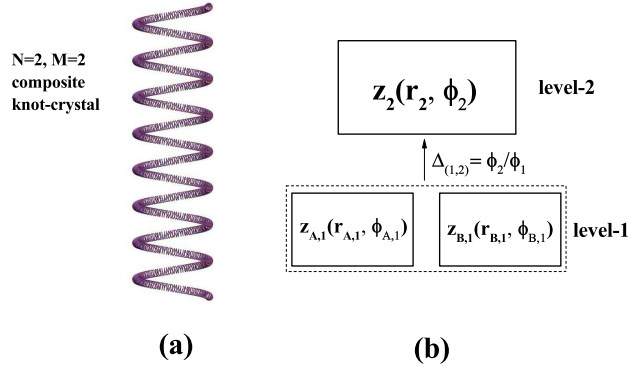}\caption{(a) An
illustration of a 1D composite knot-crystal with ($\mathcal{N}=2$,
$\mathcal{M}=2$) (a system of two entangled vortex-lines); (b) The hierarchy
structure of a 1D composite knot-crystal with ($\mathcal{N}=2$, $\mathcal{M}%
=2$)}%
\end{figure}

\subsubsection{$2$-level composite winding knot-crystal with ($\mathcal{N}=2$,
$\mathcal{M}=2$)}

A $2$-level composite winding knot-crystal is an object of two entangled
vortex-membranes $\mathrm{A}$ and $\mathrm{B}$. To generate a $2$-level
winding knot-crystal, we firstly tangle two symmetric vortex-membranes
$\mathrm{A}$ and $\mathrm{B}$ and get a knot-crystal. Next, we winding the
knot-crystal and get a $2$-level composite knot-crystal with ($\mathcal{N}=2$,
$\mathcal{M}=2$). See the illustration in Fig.2(a). In Fig.2(b), we show the
hierarchy structure of a composite knot-crystal with ($\mathcal{N}=2$,
$\mathcal{M}=2$).

\paragraph{The definition}

In general, the function of $2$-level composite knot-crystal with
($\mathcal{N}=2$, $\mathcal{M}=2$) is denoted by
\begin{equation}
\mathbf{Z(}x^{I},t\mathbf{)}=\left(
\begin{array}
[c]{cc}%
%TCIMACRO{\dprod \limits_{I}}%
%BeginExpansion
{\displaystyle \prod \limits_{I}}
%EndExpansion
z_{\mathrm{A},1}(\phi_{\mathrm{A},1}^{I},t) &
%TCIMACRO{\dprod \limits_{I}}%
%BeginExpansion
{\displaystyle \prod \limits_{I}}
%EndExpansion
z_{\mathrm{A},2}(\phi_{\mathrm{A},2}^{I},t)\\%
%TCIMACRO{\dprod \limits_{I}}%
%BeginExpansion
{\displaystyle \prod \limits_{I}}
%EndExpansion
z_{\mathrm{B},1}(\phi_{\mathrm{B},1}^{I},t) &
%TCIMACRO{\dprod \limits_{I}}%
%BeginExpansion
{\displaystyle \prod \limits_{I}}
%EndExpansion
z_{\mathrm{B},2}(\phi_{\mathrm{B},2}^{I},t)
\end{array}
\right)  .
\end{equation}
According to $\phi_{\mathrm{A},2}^{I}(x^{I})=\phi_{\mathrm{B},2}^{I}%
(x^{I})=\phi_{2}^{I}(x^{I},t)$, or
\begin{equation}%
%TCIMACRO{\dprod \limits_{I}}%
%BeginExpansion
{\displaystyle \prod \limits_{I}}
%EndExpansion
z_{\mathrm{A},2}(\phi_{\mathrm{A}_{1},2}^{I}(x^{I}),t)=%
%TCIMACRO{\dprod \limits_{I}}%
%BeginExpansion
{\displaystyle \prod \limits_{I}}
%EndExpansion
z_{\mathrm{B},2}(\phi_{\mathrm{B}_{1},2}^{I}(x^{I}),t)=\mathrm{z}_{2}(\phi
_{2}^{I}(x^{I},t),t),
\end{equation}
$\mathrm{z}_{2}(\vec{x},t)$ denotes the centre-membrane of
vortex-membrane-$\mathrm{A}$ and vortex-membrane-\textrm{B}. $\mathrm{z}%
_{\mathrm{A},1}(\vec{x},t)=%
%TCIMACRO{\dprod \limits_{I}}%
%BeginExpansion
{\displaystyle \prod \limits_{I}}
%EndExpansion
z_{\mathrm{A},1}(\phi_{\mathrm{A},1}^{I},t)$ denotes the
vortex-membrane-$\mathrm{A}$ and $\mathrm{z}_{\mathrm{B},1}(\vec{x},t)=%
%TCIMACRO{\dprod \limits_{I}}%
%BeginExpansion
{\displaystyle \prod \limits_{I}}
%EndExpansion
z_{\mathrm{B},1}(\phi_{\mathrm{B},1}^{I},t)$ denotes the
vortex-membrane-$\mathrm{B}$. The phase angles of $\mathrm{z}_{2}(x^{I},t)$
are $\phi_{2}^{I}(x^{I},t)$ and the winding radius of $\mathrm{z}_{2}(\vec
{x},t)$ is $r_{2}(\vec{x},t)$. The winding radii $r_{\mathrm{A/B},1}$ are the
local winding radii of vortex-membranes $\mathbf{z}_{\mathrm{A/B},1}(\vec
{x},t)$ around its center membrane; $\phi_{\mathrm{A/B},1}^{I}(x^{I},t)$ are
the phase angles of vortex-membranes-\textrm{A/B}. In particular, we consider
a perturbative condition,
\begin{equation}
r_{2}\gg r_{\mathrm{A/B},1}.
\end{equation}
The hierarchy series is just a number
\begin{equation}
\Delta_{(1,2)}=\frac{\phi_{\mathrm{A},1}^{I}(x^{I},t)}{\phi_{2}^{I}(x^{I}%
,t)}=\frac{\phi_{\mathrm{B},1}^{I}(x^{I},t)}{\phi_{2}^{I}(x^{I},t)}.
\end{equation}

\paragraph{Example}

In this paper, we focus on a particular type of $2$-level composite
knot-crystal with ($\mathcal{N}=2$, $\mathcal{M}=2$). $\left(
\begin{array}
[c]{c}%
\mathrm{z}_{\mathrm{A},1}(\vec{x},t)\\
\mathrm{z}_{\mathrm{B},1}(\vec{x},t)
\end{array}
\right)  $ becomes the function of an SOC knot-crystal with leapfrogging
motion. The tensor network state of $\left(
\begin{array}
[c]{c}%
\mathrm{z}_{\mathrm{A},1}(\vec{x},t)\\
\mathrm{z}_{\mathrm{B},1}(\vec{x},t)
\end{array}
\right)  $ is
\begin{align}
\left \langle \mathbf{\sigma}^{X}\otimes \vec{1}\right \rangle  &  =\vec
{n}_{\sigma}^{X}=(1,0,0),\\
\left \langle \mathbf{\sigma}^{Y}\otimes \vec{1}\right \rangle  &  =\vec
{n}_{\sigma}^{Y}=(0,1,0),\nonumber \\
\left \langle \mathbf{\sigma}^{Z}\otimes \vec{1}\right \rangle  &  =\vec
{n}_{\sigma}^{Z}=(0,0,1).\nonumber
\end{align}
$\mathrm{z}_{2}(\vec{x},t)$ is effective $\sigma_{z}$-type of knot-crystal
with leapfrogging motion. The tensor network state is
\begin{align}
\left \langle \mathbf{\sigma}^{X}\otimes \vec{1}\right \rangle  &  =\vec
{n}_{\sigma}^{X}=(1,0,0),\\
\left \langle \mathbf{\sigma}^{Y}\otimes \vec{1}\right \rangle  &  =\vec
{n}_{\sigma}^{Y}=(1,0,0),\nonumber \\
\left \langle \mathbf{\sigma}^{Z}\otimes \vec{1}\right \rangle  &  =\vec
{n}_{\sigma}^{Z}=(1,0,0).\nonumber
\end{align}
In this paper, the hierarchy series $\Delta_{(1,2)}$ is considered to be a
large positive integer number.

\paragraph{Twist-writhe locking condition}

We then introduce a topological constraint condition for a $2$-level composite
knot-crystal -- \emph{twist-writhe locking condition}.

To characterize a $2$-level composite knot-crystal with ($\mathcal{N}=2$,
$\mathcal{M}=2$), we introduce three types of 1D translation symmetry
projected topological invariable: linking-number, writhe-number, and
twist-number. So, there are three types of vectors for topological invariable
to describe entanglement between two vortex-membranes: a 1D linking-number
density-vector, a writhe-number density vector, a twist-number density vector.
We point out that there exists an important topological relationship between
these topological invariable -- the twist-writhe locking condition.

Firstly, we discuss the entanglement between two entangled vortex-membranes
for $2$-level composite knot-crystal with ($\mathcal{N}=2$, $\mathcal{M}=2$)
($\mathrm{z}_{\mathrm{A},1}(\vec{x},t)$ and $\mathrm{z}_{\mathrm{B},1}(\vec
{x},t)$), of which the vector of (translation symmetry projected) 1D linking
numbers $\vec{\zeta}_{\left(  \mathrm{AB}\right)  ,1D}=(\zeta_{\left(
\mathrm{AB}\right)  ,1D}^{x},\zeta_{\left(  \mathrm{AB}\right)  ,1D}^{y}%
,\zeta_{\left(  \mathrm{AB}\right)  ,1D}^{z})$\cite{kou} where
\begin{equation}
\zeta_{\left(  \mathrm{AB}\right)  ,1D}^{I}=\frac{1}{4\pi}\oint_{C_{x^{I}%
,\mathrm{A}}}\oint_{C_{x^{I},\mathrm{B}}}\frac{(\mathbf{s}_{\mathrm{A}}%
^{I}-\mathbf{s}_{\mathrm{B}}^{I})\cdot d\mathbf{s}_{\mathrm{A}}^{I}\times
d\mathbf{s}_{\mathrm{B}}^{I}}{|\mathbf{s}_{\mathrm{A}}^{I}-\mathbf{s}%
_{\mathrm{B}}^{I}|^{3}}.
\end{equation}
Here $\mathbf{s}_{\mathrm{A/B}}^{I}=\mathbf{r}\cdot \vec{e}^{I}$ is the spatial
vector of vortex-membranes along a given direction $\vec{e}^{I}$ ($I=x,y,z$).
We decompose the linking number $\zeta_{\left(  \mathrm{AB}\right)  ,1D}^{I}$
into the vector of writhe number $W_{\left(  \mathrm{AB}\right)  ,1D}%
^{I}=W_{1D,\mathrm{A}}^{I}+W_{1D,\mathrm{B}}^{I}$, and the vector of the twist
number $T_{\left(  \mathrm{AB}\right)  ,1D}^{I}=T_{1D,\mathrm{A}}%
^{I}+T_{1D,\mathrm{B}}^{I}$, i.e., $\zeta_{\left(  \mathrm{AB}\right)
,1D}^{I}=W_{\left(  \mathrm{AB}\right)  ,1D}^{I}+T_{\left(  \mathrm{AB}%
\right)  ,1D}^{I}$ where%
\begin{align}
W_{\left(  \mathrm{AB}\right)  ,1D}^{I}  &  =\frac{1}{4\pi}\oint
_{C_{x^{I},\mathrm{A}}}\oint_{C_{x^{I},\mathrm{A}}}\frac{(\mathbf{s}_{1}%
^{I}-\mathbf{s}_{2})\cdot d\mathbf{s}_{1}^{I}\times d\mathbf{s}_{2}%
}{|\mathbf{s}_{1}^{I}-\mathbf{s}_{2}|^{3}}\, \nonumber \\
&  +\frac{1}{4\pi}\oint_{C_{x^{I},\mathrm{B}}}\oint_{C_{x^{I},\mathrm{B}}%
}\frac{(\mathbf{s}_{1}^{I}-\mathbf{s}_{2})\cdot d\mathbf{s}_{1}^{I}\times
d\mathbf{s}_{2}}{|\mathbf{s}_{1}^{I}-\mathbf{s}_{2}|^{3}},\nonumber \\
T_{\left(  \mathrm{AB}\right)  ,1D}^{I}  &  =\frac{1}{2\pi}\oint
_{C_{x^{I},\mathrm{A}}}(\mathbf{N}^{I}\times(\mathbf{N}^{I})^{\prime})\cdot
d\mathbf{s}^{I}\nonumber \\
&  +\frac{1}{2\pi}\oint_{C_{x^{I},\mathrm{B}}}(\mathbf{N}^{I}\times
(\mathbf{N}^{I})^{\prime})\cdot d\mathbf{s}^{I}%
\end{align}
where a unit span-wise vector $\mathbf{N}^{I}=\hat{\mathbf{n}}^{I}\cos
\theta^{I}+\hat{\mathbf{b}}^{I}\sin \theta^{I}$ determines the twisting of the
vortex-membranes along $x^{I}$-direction. $\hat{\mathbf{n}}^{I}$ and
$\hat{\mathbf{b}}^{I}$ are local normal and bi-normal (unit vectors) of the
vortex-membrane in 5D fluid along $x^{I}$-direction, respectively ($\theta
^{I}$ is the corresponding mixing angle).

Because the linking number is a topological invariant for two entangled
vortex-membranes, we have a topological constraint condition -- twist-writhe
locking condition\cite{2,3},
\begin{equation}
\zeta_{\left(  \mathrm{AB}\right)  ,1D}^{I}=W_{\left(  \mathrm{AB}\right)
,1D}^{I}+T_{\left(  \mathrm{AB}\right)  ,1D}^{I}\equiv \mathrm{const},\text{
}(I=x,y,z).
\end{equation}
Theoretically under continuous deformation of the vortex-membranes, the writhe
number $W_{\left(  \mathrm{AB}\right)  ,1D}^{I}$ and the twist number
$T_{\left(  \mathrm{AB}\right)  ,1D}^{I}$ vary together, i.e.,
\begin{equation}
\delta W_{\left(  \mathrm{AB}\right)  ,1D}^{I}\equiv-\delta T_{\left(
\mathrm{AB}\right)  ,1D}^{I},\text{ }(I=x,y,z).
\end{equation}
From above discussion, we point out that for two entangled vortex-membranes
($2$-level composite knot-crystal with ($\mathcal{N}=2$, $\mathcal{M}=2$))
when the two vortex-membranes have additional global winding, finite $\delta
W_{\left(  \mathrm{AB}\right)  ,1D}^{I}$ leads to finite $-\delta T_{\left(
\mathrm{AB}\right)  ,1D}^{I}$ that is really additional entanglement between
two vortex-membranes; vice versa.

It was known that the vector of linking-number density operators for two
entangled vortex-membranes ($2$-level composite knot-crystal with
($\mathcal{N}=2$, $\mathcal{M}=2$)) are defined by
\begin{equation}
\mathbf{\hat{\rho}}_{\mathrm{linking}}=(-\frac{i}{2\pi r_{0}^{2}}\frac
{d}{dx^{x}},-\frac{i}{2\pi r_{0}^{2}}\frac{d}{dx^{y}},-\frac{i}{2\pi r_{0}%
^{2}}\frac{d}{dx^{z}}),
\end{equation}
respectively. For a $2$-level composite knot-crystal with ($\mathcal{N}=2$,
$\mathcal{M}=2$), the three 1D (spatial translation symmetry protected)
linking-numbers $\zeta_{1D}^{I}$ ($I=x,y,z$) are conserved.

For 3D $2$-level composite knot-crystal with ($\mathcal{N}=2$, $\mathcal{M}%
=2$) in 5D space, we define the (spatial translation symmetry projected) 1D
writhe density vector
\begin{equation}
\vec{\rho}_{\mathrm{writhe}}(\vec{x},t)=\frac{dW_{1D}^{I}}{dx^{I}}%
\end{equation}
and the (spatial translation symmetry projected) 1D twist density vector
\begin{equation}
\vec{\rho}_{\mathrm{twist}}(\vec{x},t)=\frac{dT_{1D}^{I}}{dx^{I}}.
\end{equation}
Due to the twist-writhe locking condition and spatial translation symmetry,
for $2$-level composite knot-crystal with ($\mathcal{N}=2$, $\mathcal{M}=2$),
we have the following equation
\begin{equation}
\rho_{\mathrm{linking}}^{I}(x,t)=\rho_{\mathrm{writhe}}^{I}(\vec{x}%
,t)+\rho_{\mathrm{twist}}^{I}(\vec{x},t),\text{ }(I=x,y,z).
\end{equation}

\subsubsection{$2$-level composite double-helix knot-crystal with
($\mathcal{N}=4$, $\mathcal{M}=2$)}

A $2$-level composite knot-crystal with ($\mathcal{N}=4$, $\mathcal{M}=2$) is
an object of four entangled vortex-membranes $\mathrm{A}_{1},$ $\mathrm{A}%
_{2}$ and $\mathrm{B}_{1},$ $\mathrm{B}_{2}$. To generate a $2$-level
composite knot-crystal with ($\mathcal{N}=4$, $\mathcal{M}=2$), we firstly
tangle two symmetric vortex-membranes $\mathrm{A}_{1}$ and $\mathrm{A}_{2}$
and get \textrm{A}-knot-crystal. Next, we tangle two symmetric
vortex-membranes $\mathrm{B}_{1}$ and $\mathrm{B}_{2}$ and get \textrm{B}%
-knot-crystal. Then, we tangle \textrm{A}-knot-crystal and \textrm{B}%
-knot-crystal into a composite knot-crystal. See the illustration in Fig.3(a).
In Fig.3(b), we show the hierarchy structure of a composite knot-crystal with
($\mathcal{N}=4$, $\mathcal{M}=2$).

\begin{figure}[ptb]
\includegraphics[clip,width=0.53\textwidth]{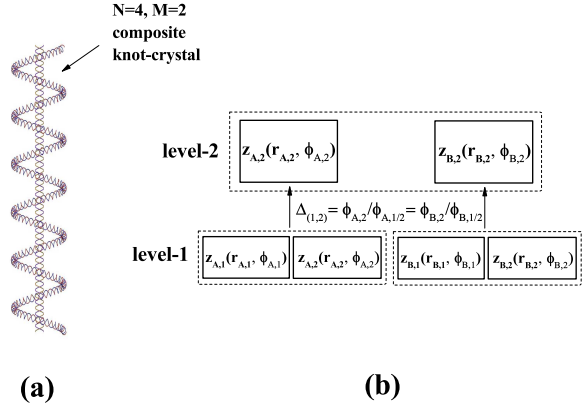}\caption{(a) An
illustration of a 1D composite knot-crystal with ($\mathcal{N}=4$,
$\mathcal{M}=2$) (a system of four entangled vortex-lines); (b) The hierarchy
structure of a 1D composite knot-crystal with ($\mathcal{N}=4$, $\mathcal{M}%
=2$)}%
\end{figure}

\paragraph{The definition}

In general, the function of $2$-level composite knot-crystal with
($\mathcal{N}=4$, $\mathcal{M}=2$) is denoted by
\begin{equation}
\mathbf{Z(}\vec{x},t\mathbf{)}=\left(
\begin{array}
[c]{cc}%
%TCIMACRO{\dprod \limits_{I}}%
%BeginExpansion
{\displaystyle \prod \limits_{I}}
%EndExpansion
z_{\mathrm{A}_{1},1}(\phi_{\mathrm{A}_{1},1}^{I},t) &
%TCIMACRO{\dprod \limits_{I}}%
%BeginExpansion
{\displaystyle \prod \limits_{I}}
%EndExpansion
z_{\mathrm{A}_{1},2}(\phi_{\mathrm{A}_{1},2}^{I},t)\\%
%TCIMACRO{\dprod \limits_{I}}%
%BeginExpansion
{\displaystyle \prod \limits_{I}}
%EndExpansion
z_{\mathrm{A}_{2},1}(\phi_{\mathrm{A}_{2},1}^{I},t) &
%TCIMACRO{\dprod \limits_{I}}%
%BeginExpansion
{\displaystyle \prod \limits_{I}}
%EndExpansion
z_{\mathrm{A}_{2},2}(\phi_{\mathrm{A}_{2},2}^{I},t)\\%
%TCIMACRO{\dprod \limits_{I}}%
%BeginExpansion
{\displaystyle \prod \limits_{I}}
%EndExpansion
z_{\mathrm{B}_{1},1}(\phi_{\mathrm{B}_{1},1}^{I},t) &
%TCIMACRO{\dprod \limits_{I}}%
%BeginExpansion
{\displaystyle \prod \limits_{I}}
%EndExpansion
z_{\mathrm{B}_{1},2}(\phi_{\mathrm{B}_{1},2}^{I},t)\\%
%TCIMACRO{\dprod \limits_{I}}%
%BeginExpansion
{\displaystyle \prod \limits_{I}}
%EndExpansion
z_{\mathrm{B}_{2},1}(\phi_{\mathrm{B}_{2},1}^{I},t) &
%TCIMACRO{\dprod \limits_{I}}%
%BeginExpansion
{\displaystyle \prod \limits_{I}}
%EndExpansion
z_{\mathrm{B}_{2},2}(\phi_{\mathrm{B}_{2},2}^{I},t)
\end{array}
\right)  .
\end{equation}
According to $\phi_{\mathrm{A}_{1},2}^{I}(x^{I})=\phi_{\mathrm{A}_{2},2}%
^{I}(x^{I})=\phi_{\mathrm{A},2}^{I}(x^{I},t)$ and $\phi_{\mathrm{B}_{1},2}%
^{I}(x^{I})=\phi_{\mathrm{B}_{2},2}^{I}(x^{I})=\phi_{\mathrm{B},2}^{I}%
(x^{I},t)$, or
\begin{equation}%
%TCIMACRO{\dprod \limits_{I}}%
%BeginExpansion
{\displaystyle \prod \limits_{I}}
%EndExpansion
z_{\mathrm{A}_{1},2}(\phi_{\mathrm{A}_{1},2}^{I}(x^{I}),t)=%
%TCIMACRO{\dprod \limits_{I}}%
%BeginExpansion
{\displaystyle \prod \limits_{I}}
%EndExpansion
z_{\mathrm{A}_{2},2}(\phi_{\mathrm{A}_{2},2}^{I}(x^{I}),t)=\mathrm{z}%
_{\mathrm{A},2}(\vec{x},t)
\end{equation}
and
\begin{equation}%
%TCIMACRO{\dprod \limits_{I}}%
%BeginExpansion
{\displaystyle \prod \limits_{I}}
%EndExpansion
z_{\mathrm{B}_{1},2}(\phi_{\mathrm{B}_{1},2}^{I}(x^{I}),t)=%
%TCIMACRO{\dprod \limits_{I}}%
%BeginExpansion
{\displaystyle \prod \limits_{I}}
%EndExpansion
z_{\mathrm{B}_{2},2}(\phi_{\mathrm{B}_{2},2}^{I}(x^{I}),t)=\mathrm{z}%
_{\mathrm{B},2}(\vec{x},t),
\end{equation}
$\mathbf{Z}_{2}(\vec{x},t)=\left(
\begin{array}
[c]{c}%
\mathrm{z}_{\mathrm{A},2}(\vec{x},t)\\
\mathrm{z}_{\mathrm{B},2}(\vec{x},t)
\end{array}
\right)  $ denote the centre-membranes of \textrm{A}-knot-crystal and
\textrm{B}-knot-crystal, respectively. Or, $\mathrm{z}_{\mathrm{A},2}(\vec
{x},t)$ denotes the global position of the entangled vortex-membranes
$\mathrm{A}_{1},$ $\mathrm{A}_{2}$ and $\mathrm{z}_{\mathrm{B},2}(\vec{x},t)$
denotes the global position of the entangled vortex-membranes $\mathrm{B}%
_{1},$ $\mathrm{B}_{2}$. $\mathbf{Z}_{\mathrm{A},1}^{\prime}(\vec
{x},t)=\left(
\begin{array}
[c]{c}%
z_{\mathrm{A}_{1},1}(\vec{x},t)\\
z_{\mathrm{A}_{2},1}(\vec{x},t)
\end{array}
\right)  =\left(
\begin{array}
[c]{c}%
%TCIMACRO{\dprod \limits_{I}}%
%BeginExpansion
{\displaystyle \prod \limits_{I}}
%EndExpansion
z_{\mathrm{A}_{1},1}(\phi_{\mathrm{A}_{1},1}^{I}(\phi_{\mathrm{A}_{1},2}%
^{I}(x^{I})),t)\\%
%TCIMACRO{\dprod \limits_{I}}%
%BeginExpansion
{\displaystyle \prod \limits_{I}}
%EndExpansion
z_{\mathrm{A}_{2},1}(\phi_{\mathrm{A}_{2},1}^{I}(\phi_{\mathrm{A}_{1},2}%
^{I}(x^{I})),t)
\end{array}
\right)  $ and $\mathbf{Z}_{\mathrm{B},1}^{\prime}(\vec{x},t)=\left(
\begin{array}
[c]{c}%
z_{\mathrm{B}_{1},1}(\vec{x},t)\\
z_{\mathrm{B}_{2},1}(\vec{x},t)
\end{array}
\right)  =\left(
\begin{array}
[c]{c}%
%TCIMACRO{\dprod \limits_{I}}%
%BeginExpansion
{\displaystyle \prod \limits_{I}}
%EndExpansion
z_{\mathrm{B}_{1},1}(\phi_{\mathrm{B}_{1},1}^{I}(\phi_{\mathrm{B}_{1},2}%
^{I}(x^{I})),t)\\%
%TCIMACRO{\dprod \limits_{I}}%
%BeginExpansion
{\displaystyle \prod \limits_{I}}
%EndExpansion
z_{\mathrm{B}_{2},1}(\phi_{\mathrm{B}_{2},1}^{I}(\phi_{\mathrm{B}_{1},2}%
^{I}(x^{I})),t)
\end{array}
\right)  $ denote local entanglement between two vortex-membranes
$\mathrm{A}_{1},$ $\mathrm{A}_{2}$ and $\mathrm{B}_{1},$ $\mathrm{B}_{2}$,
respectively. The phase angles of $\mathrm{z}_{2,\mathrm{A/B}}(x^{I},t)$ are
$\phi_{\mathrm{A/B},2}^{I}(x^{I},t)$ and the winding radii of $\mathrm{z}%
_{\mathrm{A/B},2}(\vec{x},t)$ are $r_{\mathrm{A/B},2}(\vec{x},t)$. The winding
radii $r_{\mathrm{A}_{i}\mathrm{/B}_{i},1}$ are the local winding radii of
vortex-membranes $\mathbf{Z}_{\mathrm{A}_{i}\mathrm{/B}_{i},1}^{\prime}%
(\vec{x},t)$ around its center; $\phi_{\mathrm{A}_{i}\mathrm{/B}_{i},1}%
^{I}(x^{I},t)$ are the angles of vortex-membranes $\mathbf{Z}_{\mathrm{A}%
_{i}\mathrm{/B}_{i},1}^{\prime}(\vec{x},t)$ around its center-membrane
\textrm{A/B}. Then, to characterize a knot on winding entangled
vortex-membranes, we need to two types phase angles -- one is $\phi
_{\mathrm{A/B},2}^{I}(\vec{x},t)$ that describe the winding position of the
knot on centre-membrane $\mathrm{z}_{\mathrm{A/B},2}(\vec{x},t),$ the other is
$\phi_{\mathrm{A}_{i}\mathrm{/B}_{i},1}^{I}(x^{I},t)$ that are the phase angle
of internal windings.

In particular, we consider a perturbative condition,
\begin{equation}
r_{2}=\sqrt{r_{\mathrm{A},2}^{2}(\vec{x},t)+r_{\mathrm{B},2}^{2}(\vec{x}%
,t)}\gg \left \vert r_{\mathrm{A}_{i}\mathrm{/B}_{i},1}(\vec{x},t)\right \vert .
\end{equation}
The hierarchy series is also a number
\begin{align}
\Delta_{(1,2)}  &  =\frac{\phi_{\mathrm{A}_{1},1}^{I}(x^{I},t)}{\phi
_{\mathrm{A},2}^{I}(x^{I},t)}=\frac{\phi_{\mathrm{B}_{1},1}^{I}(x^{I},t)}%
{\phi_{\mathrm{B},2}^{I}(x^{I},t)}\\
&  =\frac{\phi_{\mathrm{A}_{2},1}^{I}(x^{I},t)}{\phi_{\mathrm{A},2}^{I}%
(x^{I},t)}=\frac{\phi_{\mathrm{B}_{2},1}^{I}(x^{I},t)}{\phi_{\mathrm{B},2}%
^{I}(x^{I},t)}.\nonumber
\end{align}

\paragraph{Example}

In this paper, we focus on a particular type of $2$-level composite
knot-crystal with ($\mathcal{N}=4$, $\mathcal{M}=2$). $\mathbf{Z}_{2}(\vec
{x},t)$ becomes an effective SOC knot-crystal with leapfrogging motion, i.e.,
\begin{equation}
\mathbf{Z}_{2}(\vec{x},t)=\left(
\begin{array}
[c]{c}%
\mathrm{z}_{\mathrm{A},2}(\vec{x},t)\\
\mathrm{z}_{\mathrm{B},2}(\vec{x},t)
\end{array}
\right)  \mathbf{.}%
\end{equation}
The tensor network state of $\mathbf{Z}_{2}(\vec{x},t)$ is also
\begin{align}
\left \langle \mathbf{\sigma}^{X}\otimes \vec{1}\right \rangle  &  =\vec
{n}_{\sigma}^{X}=(1,0,0),\\
\left \langle \mathbf{\sigma}^{Y}\otimes \vec{1}\right \rangle  &  =\vec
{n}_{\sigma}^{Y}=(0,1,0),\nonumber \\
\left \langle \mathbf{\sigma}^{Z}\otimes \vec{1}\right \rangle  &  =\vec
{n}_{\sigma}^{Z}=(0,0,1).\nonumber
\end{align}
$\mathbf{Z}_{\mathrm{A},1}^{\prime}(\vec{x},t)$ and $\mathbf{Z}_{\mathrm{B}%
,1}^{\prime}(\vec{x},t)$ are effective SOC knot-crystal with leapfrogging
motion. The tensor network state is
\begin{align}
\left \langle \mathbf{\sigma}^{X}\otimes \vec{1}\right \rangle  &  =\vec
{n}_{\sigma}^{X}=(1,0,0),\\
\left \langle \mathbf{\sigma}^{Y}\otimes \vec{1}\right \rangle  &  =\vec
{n}_{\sigma}^{Y}=(1,0,0),\nonumber \\
\left \langle \mathbf{\sigma}^{Z}\otimes \vec{1}\right \rangle  &  =\vec
{n}_{\sigma}^{Z}=(1,0,0).\nonumber
\end{align}
The hierarchy series $\Delta_{(1,2)}$ is considered to be a positive integer
number $n$ (for example, $n=3$).

\paragraph{Twist-writhe locking condition}

We have discussed the twist-writhe locking relation for a composite
knot-crystal (vortex-membrane-\textrm{A} and vortex-membrane-\textrm{B} and
get
\begin{equation}
\zeta_{\left(  \mathrm{AB}\right)  ,1D}^{I}=W_{\left(  \mathrm{AB}\right)
,1D}^{I}+T_{\left(  \mathrm{AB}\right)  ,1D}^{I}\equiv \mathrm{const},\text{
}(I=x,y,z)
\end{equation}
where $\zeta_{\left(  \mathrm{AB}\right)  ,1D}^{I}$, $W_{\left(
\mathrm{AB}\right)  ,1D}^{I}$, and $T_{\left(  \mathrm{AB}\right)  ,1D}^{I}$
are linking number, writhe number, and twist number, respectively. In this
section, we consider the case of composite knot-crystal -- a composite system
with four entangled vortex-membranes ($\mathbf{Z(}\vec{x},t\mathbf{)}$) and
discuss the twist-writhe locking relation for it.

For the entanglement between $z_{\mathrm{A}_{1},1}(\vec{x},t)$ and
$z_{\mathrm{A}_{2},1}(\vec{x},t)$ or that between $z_{\mathrm{B}_{1},1}%
(\vec{x},t)$ and $z_{\mathrm{B}_{2},1}(\vec{x},t)$, the twist-writhe locking
relation is similar to that the entanglement between $\mathrm{z}%
_{\mathrm{A},2}(\vec{x},t)$ and $\mathrm{z}_{\mathrm{B},2}(\vec{x},t)$ in a
composite winding knot-crystal with ($\mathcal{N}=2$, $\mathcal{M}=2$):

There are three types of vectors for topological invariable to describe
entanglement between $z_{\mathrm{A}_{1},1}(\vec{x},t)$ and $z_{\mathrm{A}%
_{2},1}(\vec{x},t)$: a 1D linking-number density-vector $\vec{\zeta}_{\left(
\mathrm{A}_{1}\mathrm{A}_{2}\right)  ,1D}=(\zeta_{\left(  \mathrm{A}%
_{1}\mathrm{A}_{2}\right)  ,1D}^{x},\zeta_{\left(  \mathrm{A}_{1}%
\mathrm{A}_{2}\right)  ,1D}^{y},\zeta_{\left(  \mathrm{A}_{1}\mathrm{A}%
_{2}\right)  ,1D}^{z})$\cite{kou} where
\begin{equation}
\zeta_{\left(  \mathrm{A}_{1}\mathrm{A}_{2}\right)  ,1D}^{I}=\frac{1}{4\pi
}\oint_{C_{x^{I},\mathrm{A}_{1}}}\oint_{C_{x^{I},\mathrm{A}_{2}}}%
\frac{(\mathbf{s}_{\mathrm{A}_{1}}^{I}-\mathbf{s}_{\mathrm{A}_{2}}^{I})\cdot
d\mathbf{s}_{\mathrm{A}_{1}}^{I}\times d\mathbf{s}_{\mathrm{A}_{2}}^{I}%
}{|\mathbf{s}_{\mathrm{A}_{1}}^{I}-\mathbf{s}_{\mathrm{A}_{2}}^{I}|^{3}},
\end{equation}
a writhe-number density vector%
\begin{align}
W_{\left(  \mathrm{A}_{1}\mathrm{A}_{2}\right)  ,1D}^{I}  &  =\frac{1}{4\pi
}\oint_{C_{x^{I},\mathrm{A}_{1}}}\oint_{C_{x^{I},\mathrm{A}_{1}}}%
\frac{(\mathbf{s}_{1}^{I}-\mathbf{s}_{2})\cdot d\mathbf{s}_{1}^{I}\times
d\mathbf{s}_{2}}{|\mathbf{s}_{1}^{I}-\mathbf{s}_{2}|^{3}}\, \\
&  +\frac{1}{4\pi}\oint_{C_{x^{I},\mathrm{A}_{2}}}\oint_{C_{x^{I}%
,\mathrm{A}_{2}}}\frac{(\mathbf{s}_{1}^{I}-\mathbf{s}_{2})\cdot d\mathbf{s}%
_{1}^{I}\times d\mathbf{s}_{2}}{|\mathbf{s}_{1}^{I}-\mathbf{s}_{2}|^{3}%
},\nonumber
\end{align}
a twist-number density vector%
\begin{align}
T_{\left(  \mathrm{A}_{1}\mathrm{A}_{2}\right)  ,1D}^{I}  &  =\frac{1}{2\pi
}\oint_{C_{x^{I},\mathrm{A}_{1}}}(\mathbf{N}^{I}\times(\mathbf{N}^{I}%
)^{\prime})\cdot d\mathbf{s}^{I}\nonumber \\
&  +\frac{1}{2\pi}\oint_{C_{x^{I},\mathrm{A}_{2}}}(\mathbf{N}^{I}%
\times(\mathbf{N}^{I})^{\prime})\cdot d\mathbf{s}^{I};
\end{align}
There are three types of vectors for topological invariable to describe
entanglement between $z_{\mathrm{B}_{1},1}(\vec{x},t)$ and $z_{\mathrm{B}%
_{2},1}(\vec{x},t)$: a 1D linking-number density-vector $\vec{\zeta}_{\left(
\mathrm{B}_{1}\mathrm{B}_{2}\right)  ,1D}=(\zeta_{\left(  \mathrm{B}%
_{1}\mathrm{B}_{2}\right)  ,1D}^{x},\zeta_{\left(  \mathrm{B}_{1}%
\mathrm{B}_{2}\right)  ,1D}^{y},\zeta_{\left(  \mathrm{B}_{1}\mathrm{B}%
_{2}\right)  ,1D}^{z})$\cite{kou} where
\begin{equation}
\zeta_{\left(  \mathrm{B}_{1}\mathrm{B}_{2}\right)  ,1D}^{I}=\frac{1}{4\pi
}\oint_{C_{x^{I},\mathrm{B}_{1}}}\oint_{C_{x^{I},\mathrm{B}_{2}}}%
\frac{(\mathbf{s}_{\mathrm{B}_{1}}^{I}-\mathbf{s}_{\mathrm{B}_{2}}^{I})\cdot
d\mathbf{s}_{\mathrm{B}_{1}}^{I}\times d\mathbf{s}_{\mathrm{B}_{2}}^{I}%
}{|\mathbf{s}_{\mathrm{B}_{1}}^{I}-\mathbf{s}_{\mathrm{B}_{2}}^{I}|^{3}},
\end{equation}
a writhe-number density vector%
\begin{align}
W_{\left(  \mathrm{B}_{1}\mathrm{B}_{2}\right)  ,1D}^{I}  &  =\frac{1}{4\pi
}\oint_{C_{x^{I},\mathrm{B}_{1}}}\oint_{C_{x^{I},\mathrm{B}_{1}}}%
\frac{(\mathbf{s}_{1}^{I}-\mathbf{s}_{2})\cdot d\mathbf{s}_{1}^{I}\times
d\mathbf{s}_{2}}{|\mathbf{s}_{1}^{I}-\mathbf{s}_{2}|^{3}}\, \\
&  +\frac{1}{4\pi}\oint_{C_{x^{I},\mathrm{B}_{2}}}\oint_{C_{x^{I}%
,\mathrm{B}_{2}}}\frac{(\mathbf{s}_{1}^{I}-\mathbf{s}_{2})\cdot d\mathbf{s}%
_{1}^{I}\times d\mathbf{s}_{2}}{|\mathbf{s}_{1}^{I}-\mathbf{s}_{2}|^{3}%
},\nonumber
\end{align}
a twist-number density vector%
\begin{align}
T_{\left(  \mathrm{B}_{1}\mathrm{B}_{2}\right)  ,1D}^{I}  &  =\frac{1}{2\pi
}\oint_{C_{x^{I},\mathrm{B}_{1}}}(\mathbf{N}^{I}\times(\mathbf{N}^{I}%
)^{\prime})\cdot d\mathbf{s}^{I}\nonumber \\
&  +\frac{1}{2\pi}\oint_{C_{x^{I},\mathrm{B}_{2}}}(\mathbf{N}^{I}%
\times(\mathbf{N}^{I})^{\prime})\cdot d\mathbf{s}^{I}.
\end{align}

The twist-writhe locking condition for two entangled vortex-membranes
$\mathbf{Z}_{1,\mathrm{A}}^{\prime}(\vec{x},t)=\left(
\begin{array}
[c]{c}%
z_{\mathrm{A}_{1},1}(\vec{x},t)\\
z_{\mathrm{A}_{2},1}(\vec{x},t)
\end{array}
\right)  $ is given by%
\begin{equation}
\zeta_{\left(  \mathrm{A}_{1}\mathrm{A}_{2}\right)  ,1D}^{I}=W_{\left(
\mathrm{A}_{1}\mathrm{A}_{2}\right)  ,1D}^{I}+T_{\left(  \mathrm{A}%
_{1}\mathrm{A}_{2}\right)  ,1D}^{I}\equiv \mathrm{const},\text{ }(I=x,y,z)
\end{equation}
or
\begin{equation}
\delta W_{\left(  \mathrm{A}_{1}\mathrm{A}_{2}\right)  ,1D}^{I}\equiv-\delta
T_{\left(  \mathrm{A}_{1}\mathrm{A}_{2}\right)  ,1D}^{I},\text{ }(I=x,y,z).
\end{equation}
The twist-writhe locking condition for two entangled vortex-membranes
$\mathbf{Z}_{\mathrm{B},1}^{\prime}(\vec{x},t)=\left(
\begin{array}
[c]{c}%
z_{\mathrm{B}_{1},1}(\vec{x},t)\\
z_{\mathrm{B}_{2},1}(\vec{x},t)
\end{array}
\right)  $ is given by%
\begin{equation}
\zeta_{\left(  \mathrm{B}_{1}\mathrm{B}_{2}\right)  ,1D}^{I}=W_{\left(
\mathrm{B}_{1}\mathrm{B}_{2}\right)  ,1D}^{I}+T_{\left(  \mathrm{B}%
_{1}\mathrm{B}_{2}\right)  ,1D}^{I}\equiv \mathrm{const},\text{ }(I=x,y,z)
\end{equation}
or
\begin{equation}
\delta W_{\left(  \mathrm{B}_{1}\mathrm{B}_{2}\right)  ,1D}^{I}\equiv-\delta
T_{\left(  \mathrm{B}_{1}\mathrm{B}_{2}\right)  ,1D}^{I},\text{ }(I=x,y,z).
\end{equation}

\begin{figure}[ptb]
\includegraphics[clip,width=0.53\textwidth]{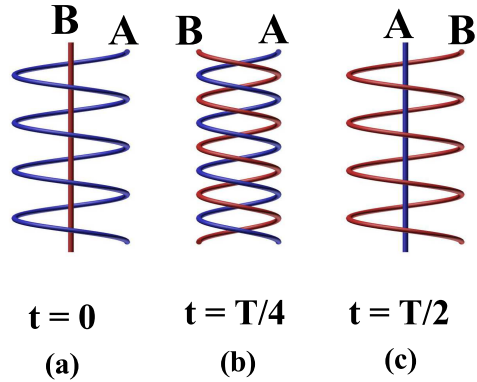}\caption{An illustration
of leapfrogging motion for two entangled vortex-lines}%
\end{figure}

In particular, there exist the following topological locking relationships%
\begin{equation}
W_{1D,\mathrm{A}}^{I}\Longrightarrow \delta W_{\mathrm{A}_{1},1D}^{I}=\delta
W_{\mathrm{A}_{2},1D}^{I}=W_{1D,\mathrm{A}}^{I},
\end{equation}
and%
\begin{equation}
W_{1D,\mathrm{B}}^{I}\Longrightarrow \delta W_{\mathrm{B}_{1},1D}^{I}=\delta
W_{\mathrm{B}_{2},1D}^{I}=W_{1D,\mathrm{B}}^{I}.
\end{equation}
Because $W_{1D,\mathrm{A}}^{I}$ or $W_{1D,\mathrm{B}}^{I}$ is time-dependent
due to leapfrogging motion, $W_{\mathrm{A}_{1},1D}^{I},$ $W_{\mathrm{A}%
_{2},1D}^{I},$ $W_{\mathrm{B}_{1},1D}^{I}$, $W_{\mathrm{B}_{2},1D}^{I}$ change
with time. See the illustration in Fig.4.

According to twist-writhe locking conditions, $\delta W_{\left(
\mathrm{A}_{1}\mathrm{A}_{2}\right)  ,1D}^{I}\equiv-\delta T_{\left(
\mathrm{A}_{1}\mathrm{A}_{2}\right)  ,1D}^{I}$ and $\delta W_{\left(
\mathrm{B}_{1}\mathrm{B}_{2}\right)  ,1D}^{I}\equiv-\delta T_{\left(
\mathrm{B}_{1}\mathrm{B}_{2}\right)  ,1D}^{I}$, we have
\begin{equation}
\delta T_{\left(  \mathrm{A}_{1}\mathrm{A}_{2}\right)  ,1D}^{I}%
=-W_{1D,\mathrm{A}}^{I},\text{ }(I=x,y,z),
\end{equation}
and
\begin{equation}
\delta T_{\left(  \mathrm{B}_{1}\mathrm{B}_{2}\right)  ,1D}^{I}%
=-W_{1D,\mathrm{B}}^{I},\text{ }(I=x,y,z).
\end{equation}

From above discussion, we point out that for uniformly entangled
vortex-membranes when the two vortex-membranes have additional global winding,
finite $\delta W_{\left(  \mathrm{AB}\right)  ,1D}^{I}$ leads to finite
$-\delta T_{\left(  \mathrm{AB}\right)  ,1D}^{I}$ that is really additional
entanglement between two vortex-membranes; vice versa. That means a global
winding of \textrm{A}-knot-crystal and \textrm{B}-knot-crystal leads to
additional internal twisting between vortex-membranes $\mathrm{A}_{1}$,
$\mathrm{A}_{2}$ and vortex-membranes $\mathrm{B}_{1},$ $\mathrm{B}_{2}$.

\subsubsection{$3$-level composite knot-crystal with ($\mathcal{N}=4$,
$\mathcal{M}=3$)}

A $3$-level composite knot-crystal with ($\mathcal{N}=4$, $\mathcal{M}=3$) is
an object of four entangled vortex-membranes $\mathrm{A}_{1},$ $\mathrm{A}%
_{2}$ and $\mathrm{B}_{1},$ $\mathrm{B}_{2}$. We firstly tangle two symmetric
vortex-membranes $\mathrm{A}_{1}$ and $\mathrm{A}_{2}$ and get \textrm{A}%
-knot-crystal and tangle two symmetric vortex-membranes $\mathrm{B}_{1}$ and
$\mathrm{B}_{2}$ and get \textrm{B}-knot-crystal. Then, we tangle
\textrm{A}-knot-crystal and \textrm{B}-knot-crystal into a 2-level composite
double-helix knot-crystal. Finally, we wind the 2-level composite double-helix
knot-crystal and get a $3$-level composite knot-crystal with ($\mathcal{N}=4$,
$\mathcal{M}=3$). See the illustration in Fig.5(a). In Fig.5(b), we show the
hierarchy structure of a composite knot-crystal with ($\mathcal{N}=4$,
$\mathcal{M}=3$).

\begin{figure}[ptb]
\includegraphics[clip,width=0.53\textwidth]{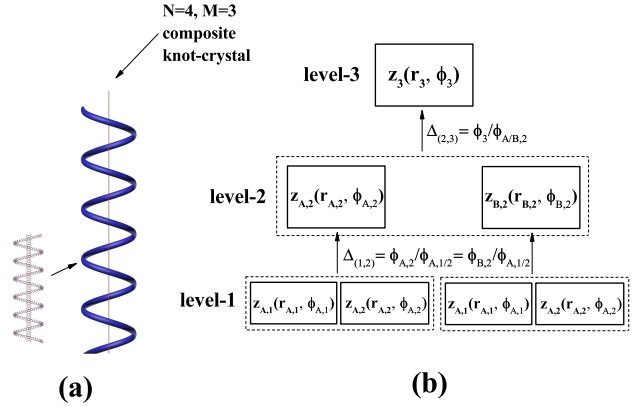}\caption{(a) An
illustration of a 1D composite knot-crystal with ($\mathcal{N}=4$ and
$\mathcal{M}=3$) (a system of four entangled vortex-lines); (b) The hierarchy
structure}%
\end{figure}

\paragraph{The definition}

In general, the function of $3$-level composite knot-crystal with
($\mathcal{N}=4$, $\mathcal{M}=3$) is denoted by
\begin{equation}
\mathbf{Z(}\vec{x},t\mathbf{)}=%
\begin{array}
[c]{ccc}%
%TCIMACRO{\dprod \limits_{I}}%
%BeginExpansion
{\displaystyle \prod \limits_{I}}
%EndExpansion
z_{\mathrm{A}_{1},1}(\phi_{\mathrm{A}_{1},1}^{I},t) &
%TCIMACRO{\dprod \limits_{I}}%
%BeginExpansion
{\displaystyle \prod \limits_{I}}
%EndExpansion
z_{\mathrm{A}_{1},2}(\phi_{\mathrm{A}_{1},2}^{I},t) &
%TCIMACRO{\dprod \limits_{I}}%
%BeginExpansion
{\displaystyle \prod \limits_{I}}
%EndExpansion
z_{\mathrm{A}_{1},3}(\phi_{\mathrm{A}_{1},3}^{I},t)\\%
%TCIMACRO{\dprod \limits_{I}}%
%BeginExpansion
{\displaystyle \prod \limits_{I}}
%EndExpansion
z_{\mathrm{A}_{2},1}(\phi_{\mathrm{A}_{2},1}^{I},t) &
%TCIMACRO{\dprod \limits_{I}}%
%BeginExpansion
{\displaystyle \prod \limits_{I}}
%EndExpansion
z_{\mathrm{A}_{2},2}(\phi_{\mathrm{A}_{2},2}^{I},t) &
%TCIMACRO{\dprod \limits_{I}}%
%BeginExpansion
{\displaystyle \prod \limits_{I}}
%EndExpansion
z_{\mathrm{A}_{2},3}(\phi_{\mathrm{A}_{2},3}^{I},t)\\%
%TCIMACRO{\dprod \limits_{I}}%
%BeginExpansion
{\displaystyle \prod \limits_{I}}
%EndExpansion
z_{\mathrm{B}_{1},1}(\phi_{\mathrm{B}_{1},1}^{I},t) &
%TCIMACRO{\dprod \limits_{I}}%
%BeginExpansion
{\displaystyle \prod \limits_{I}}
%EndExpansion
z_{\mathrm{B}_{1},2}(\phi_{\mathrm{B}_{1},2}^{I},t) &
%TCIMACRO{\dprod \limits_{I}}%
%BeginExpansion
{\displaystyle \prod \limits_{I}}
%EndExpansion
z_{\mathrm{B}_{1},3}(\phi_{\mathrm{B}_{1},3}^{I},t)\\%
%TCIMACRO{\dprod \limits_{I}}%
%BeginExpansion
{\displaystyle \prod \limits_{I}}
%EndExpansion
z_{\mathrm{B}_{2},1}(\phi_{\mathrm{B}_{2},1}^{I},t) &
%TCIMACRO{\dprod \limits_{I}}%
%BeginExpansion
{\displaystyle \prod \limits_{I}}
%EndExpansion
z_{\mathrm{B}_{2},2}(\phi_{\mathrm{B}_{2},2}^{I},t) &
%TCIMACRO{\dprod \limits_{I}}%
%BeginExpansion
{\displaystyle \prod \limits_{I}}
%EndExpansion
z_{\mathrm{B}_{2},3}(\phi_{\mathrm{B}_{2},3}^{I},t)
\end{array}
.
\end{equation}
According to $\phi_{\mathrm{A},3}^{I}(x^{I})=\phi_{\mathrm{B},3}^{I}%
(x^{I})=\phi_{3}^{I}(x^{I},t)$, or
\begin{align}%
%TCIMACRO{\dprod \limits_{I}}%
%BeginExpansion
{\displaystyle \prod \limits_{I}}
%EndExpansion
z_{\mathrm{A}_{1},3}(\phi_{\mathrm{A}_{1},3}^{I},t)  &  =%
%TCIMACRO{\dprod \limits_{I}}%
%BeginExpansion
{\displaystyle \prod \limits_{I}}
%EndExpansion
z_{\mathrm{A}_{2},3}(\phi_{\mathrm{A}_{2},3}^{I},t)\\
&  =%
%TCIMACRO{\dprod \limits_{I}}%
%BeginExpansion
{\displaystyle \prod \limits_{I}}
%EndExpansion
z_{\mathrm{B}_{1},3}(\phi_{\mathrm{B}_{1},3}^{I},t)=%
%TCIMACRO{\dprod \limits_{I}}%
%BeginExpansion
{\displaystyle \prod \limits_{I}}
%EndExpansion
z_{\mathrm{B}_{2},3}(\phi_{\mathrm{B}_{2},3}^{I},t)\nonumber \\
&  =\mathrm{z}_{3}(x^{I},t),\nonumber
\end{align}
$\mathrm{z}_{3}(\vec{x},t)$ denotes the centre-membranes of \textrm{A}%
-knot-crystal and \textrm{B}-knot-crystal. The phase angles of $\mathrm{z}%
_{3}(x^{I},t)$ are $\phi_{3}^{I}(x^{I},t)$ and the winding radii of
$\mathrm{z}_{3}(\vec{x},t)$ are $r_{3}(\vec{x},t)$. According to
$\phi_{\mathrm{A}_{1},2}^{I}(x^{I})=\phi_{\mathrm{A}_{2},2}^{I}(x^{I}%
)=\phi_{\mathrm{A},2}^{I}(x^{I},t)$ and $\phi_{\mathrm{B}_{1},2}^{I}%
(x^{I})=\phi_{\mathrm{B}_{2},2}^{I}(x^{I})=\phi_{\mathrm{B},2}^{I}(x^{I},t)$,
or
\begin{equation}%
%TCIMACRO{\dprod \limits_{I}}%
%BeginExpansion
{\displaystyle \prod \limits_{I}}
%EndExpansion
z_{\mathrm{A}_{1},2}(\phi_{\mathrm{A}_{1},2}^{I}(x^{I}),t)=%
%TCIMACRO{\dprod \limits_{I}}%
%BeginExpansion
{\displaystyle \prod \limits_{I}}
%EndExpansion
z_{\mathrm{A}_{2},2}(\phi_{\mathrm{A}_{2},2}^{I}(x^{I}),t)=\mathrm{z}%
_{\mathrm{A},2}(\vec{x},t)
\end{equation}
and
\begin{equation}%
%TCIMACRO{\dprod \limits_{I}}%
%BeginExpansion
{\displaystyle \prod \limits_{I}}
%EndExpansion
z_{\mathrm{B}_{1},2}(\phi_{\mathrm{B}_{1},2}^{I}(x^{I}),t)=%
%TCIMACRO{\dprod \limits_{I}}%
%BeginExpansion
{\displaystyle \prod \limits_{I}}
%EndExpansion
z_{\mathrm{B}_{2},2}(\phi_{\mathrm{B}_{2},2}^{I}(x^{I}),t)=\mathrm{z}%
_{\mathrm{B},2}(\vec{x},t),
\end{equation}
$\mathbf{Z}_{2}(\vec{x},t)=\left(
\begin{array}
[c]{c}%
\mathrm{z}_{\mathrm{A},2}(\vec{x},t)\\
\mathrm{z}_{\mathrm{B},2}(\vec{x},t)
\end{array}
\right)  $ also denote the centre-membranes of \textrm{A}-knot-crystal and
\textrm{B}-knot-crystal, respectively. $\mathrm{z}_{2,\mathrm{A}}(\vec{x},t)$
denotes the global position of the entangled vortex-membranes $\mathrm{A}%
_{1},$ $\mathrm{A}_{2}$ and $\mathrm{z}_{\mathrm{B},2}(\vec{x},t)$ denotes the
global position of the entangled vortex-membranes $\mathrm{B}_{1},$
$\mathrm{B}_{2}$. $\mathbf{Z}_{\mathrm{A},1}^{\prime}(\vec{x},t)=\left(
\begin{array}
[c]{c}%
z_{\mathrm{A}_{1},1}(\vec{x},t)\\
z_{\mathrm{A}_{2},1}(\vec{x},t)
\end{array}
\right)  =\left(
\begin{array}
[c]{c}%
%TCIMACRO{\dprod \limits_{I}}%
%BeginExpansion
{\displaystyle \prod \limits_{I}}
%EndExpansion
z_{\mathrm{A}_{1},1}(\phi_{\mathrm{A}_{1},1}^{I}(\phi_{\mathrm{A}_{1},2}%
^{I}(x^{I})),t)\\%
%TCIMACRO{\dprod \limits_{I}}%
%BeginExpansion
{\displaystyle \prod \limits_{I}}
%EndExpansion
z_{\mathrm{A}_{2},1}(\phi_{\mathrm{A}_{2},1}^{I}(\phi_{\mathrm{A}_{1},2}%
^{I}(x^{I})),t)
\end{array}
\right)  $ and $\mathbf{Z}_{\mathrm{B},1}^{\prime}(\vec{x},t)=\left(
\begin{array}
[c]{c}%
z_{\mathrm{B}_{1},1}(\vec{x},t)\\
z_{\mathrm{B}_{2},1}(\vec{x},t)
\end{array}
\right)  =\left(
\begin{array}
[c]{c}%
%TCIMACRO{\dprod \limits_{I}}%
%BeginExpansion
{\displaystyle \prod \limits_{I}}
%EndExpansion
z_{\mathrm{B}_{1},1}(\phi_{\mathrm{B}_{1},1}^{I}(\phi_{\mathrm{B}_{1},2}%
^{I}(x^{I})),t)\\%
%TCIMACRO{\dprod \limits_{I}}%
%BeginExpansion
{\displaystyle \prod \limits_{I}}
%EndExpansion
z_{\mathrm{B}_{2},1}(\phi_{\mathrm{B}_{2},1}^{I}(\phi_{\mathrm{B}_{1},2}%
^{I}(x^{I})),t)
\end{array}
\right)  $ denote local entanglement between two vortex-membranes
$\mathrm{A}_{1},$ $\mathrm{A}_{2}$ and $\mathrm{B}_{1},$ $\mathrm{B}_{2}$,
respectively. The phase angles of $\mathrm{z}_{\mathrm{A/B},2}(x^{I},t)$ are
$\phi_{\mathrm{A/B},2}^{I}(x^{I},t)$ and the winding radii of $\mathrm{z}%
_{\mathrm{A/B},2}(\vec{x},t)$ are $r_{\mathrm{A/B},2}(\vec{x},t)$. The winding
radii $r_{\mathrm{A}_{i}\mathrm{/B}_{i},1}$ are the local winding radii of
vortex-membranes $\mathbf{Z}_{\mathrm{A}_{i}\mathrm{/B}_{i},1}^{\prime}%
(\vec{x},t)$ around its center; $\phi_{\mathrm{A}_{i}\mathrm{/B}_{i},1}%
^{I}(x^{I},t)$ are the angles of vortex-membranes $\mathbf{Z}_{\mathrm{A}%
_{i}\mathrm{/B}_{i},1}^{\prime}(\vec{x},t)$ around its center-membrane
\textrm{A/B}.

In particular, we consider a perturbative condition,
\begin{equation}
r_{3}\gg r_{2}=\sqrt{r_{\mathrm{A},2}^{2}(\vec{x},t)+r_{\mathrm{B},2}^{2}%
(\vec{x},t)}\gg \left \vert r_{\mathrm{A}_{i}\mathrm{/B}_{i},1}(\vec
{x},t)\right \vert .
\end{equation}
The hierarchy series is
\[
\{ \Delta_{(1,2)},\Delta_{(2,3)}\}
\]
where
\begin{align}
\Delta_{(1,2)}  &  =\frac{\phi_{\mathrm{A}_{1},1}^{I}(x^{I},t)}{\phi
_{\mathrm{A},2}^{I}(x^{I},t)}=\frac{\phi_{\mathrm{B}_{1},1}^{I}(x^{I},t)}%
{\phi_{\mathrm{B},2}^{I}(x^{I},t)}\\
&  =\frac{\phi_{\mathrm{A}_{2},1}^{I}(x^{I},t)}{\phi_{\mathrm{A},2}^{I}%
(x^{I},t)}=\frac{\phi_{\mathrm{B}_{2},1}^{I}(x^{I},t)}{\phi_{\mathrm{B},2}%
^{I}(x^{I},t)}\nonumber
\end{align}
and
\begin{equation}
\Delta_{(2,3)}=\frac{\phi_{\mathrm{A},2}^{I}(x^{I},t)}{\phi_{3}^{I}(x^{I}%
,t)}=\frac{\phi_{\mathrm{B},2}^{I}(x^{I},t)}{\phi_{3}^{I}(x^{I},t)}.
\end{equation}

\paragraph{Example}

In this paper, we focus on a particular type of $3$-level composite
knot-crystal with ($\mathcal{N}=4$, $\mathcal{M}=3$).

$\mathrm{z}_{3}(\vec{x},t)$ is $\sigma_{z}$-type Kelvin wave. The spin network
state is
\begin{align}
\left \langle \mathbf{\sigma}^{X}\otimes \vec{1}\right \rangle  &  =\vec
{n}_{\sigma}^{X}=(1,0,0),\\
\left \langle \mathbf{\sigma}^{Y}\otimes \vec{1}\right \rangle  &  =\vec
{n}_{\sigma}^{Y}=(1,0,0),\nonumber \\
\left \langle \mathbf{\sigma}^{Z}\otimes \vec{1}\right \rangle  &  =\vec
{n}_{\sigma}^{Z}=(1,0,0).\nonumber
\end{align}
$\mathbf{Z}_{2}(\vec{x},t)$ becomes an SOC knot-crystal with leapfrogging
motion, i.e.,
\begin{equation}
\mathbf{Z}_{2}(\vec{x},t)=\left(
\begin{array}
[c]{c}%
\mathrm{z}_{2,\mathrm{A}}(\vec{x},t)\\
\mathrm{z}_{2,\mathrm{B}}(\vec{x},t)
\end{array}
\right)  \mathbf{.}%
\end{equation}
The tensor network states of $\mathbf{Z}_{2}(\vec{x},t)$ are
\begin{align}
\left \langle \mathbf{\sigma}^{X}\otimes \vec{1}\right \rangle  &  =\vec
{n}_{\sigma}^{X}=(1,0,0),\\
\left \langle \mathbf{\sigma}^{Y}\otimes \vec{1}\right \rangle  &  =\vec
{n}_{\sigma}^{Y}=(0,1,0),\nonumber \\
\left \langle \mathbf{\sigma}^{Z}\otimes \vec{1}\right \rangle  &  =\vec
{n}_{\sigma}^{Z}=(0,0,1).\nonumber
\end{align}
$\mathbf{Z}_{\mathrm{A},1}^{\prime}(\vec{x},t)$ and $\mathbf{Z}_{\mathrm{B}%
,1}^{\prime}(\vec{x},t)$ are SOC knot-crystal with leapfrogging motion. The
tensor states are also
\begin{align}
\left \langle \mathbf{\sigma}^{X}\otimes \vec{1}\right \rangle  &  =\vec
{n}_{\sigma}^{X}=(1,0,0),\\
\left \langle \mathbf{\sigma}^{Y}\otimes \vec{1}\right \rangle  &  =\vec
{n}_{\sigma}^{Y}=(1,0,0),\nonumber \\
\left \langle \mathbf{\sigma}^{Z}\otimes \vec{1}\right \rangle  &  =\vec
{n}_{\sigma}^{Z}=(1,0,0).\nonumber
\end{align}
The hierarchy number between level-1 and level-2 $\Delta_{(1,2)}$ is
considered to be a positive integer number $n$ (for example, $n=3$). The
hierarchy series $\Delta_{(2,3)}$ is considered to be a very large number, but
not necessary an integer number, i.e., $\Delta_{(2,3)}\gg \Delta_{(1,2)}$.

For the $3$-level composite knot-crystal with ($\mathcal{N}=4$, $\mathcal{M}%
=3$), there exists complex twist-writhe locking condition. One can use the
above approach to discuss the twist-writhe locking condition for an $3$-level
composite knot-crystal with ($\mathcal{N}=4$, $\mathcal{M}=3$).

\section{Zero-lattice and zeroes}

\subsection{Projection of vortex-membranes}

There are two types projections on vortex-membranes: the projection for single
vortex-membrane and that for entangled vortex-membranes. We call the
projection for single vortex-membrane \emph{W-type projection} and that for
entangled vortex-membranes \emph{T-type projection}. See the illustration in Fig.6.

\begin{figure}[ptb]
\includegraphics[clip,width=0.53\textwidth]{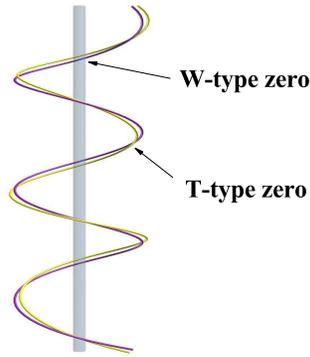}\caption{An illustration
of a W-type zero from W-type projection and a T-type zero from T-type
projection}%
\end{figure}

\subsubsection{Projection for single vortex-membrane}

Firstly, we discuss the projection for single vortex-membrane, of which the
function is
\begin{equation}
\mathrm{z}(\vec{x},t)=\xi(\vec{x},t)+i\eta(\vec{x},t).
\end{equation}
To locally characterize the windings of a helical vortex-membrane, we define
\emph{the projection} via a projection angle $\theta$ on $\{ \xi(\vec{x},t),$
$\eta(\vec{x},t)\}$ by $\hat{P}_{\theta}\left(
\begin{array}
[c]{c}%
\xi(\vec{x},t)\\
\eta(\vec{x},t)
\end{array}
\right)  =\left(
\begin{array}
[c]{c}%
\xi_{\theta}(\vec{x},t)\\
\left[  \eta_{\theta}(\vec{x},t)\right]  _{0}%
\end{array}
\right)  $ where $\xi_{\theta}(\vec{x},t)=\xi(\vec{x},t)\sin \theta-\eta
(\vec{x},t)\cos \theta$ is variable and $\left[  \eta_{\theta}(\vec
{x},t)\right]  _{0}=\xi(\vec{x},t)\cos \theta+\eta(\vec{x},t)\sin \theta$ is
constant. Thus, the projected helical vortex-membrane is described by the
function $\xi(\vec{x},t).$ A crossing between a helical vortex-membrane and a
straight one ($\mathrm{z}(\vec{x},t)=0$) in its center corresponds to a
solution of the equation
\begin{equation}
\hat{P}_{\theta}[\mathrm{z}(\vec{x},t)]=0,
\end{equation}
that is $\xi(\vec{x},t)=0.$ We call the equation zero equation and its
solution zero solution.

\subsubsection{Projection for entangled vortex-membranes}

Next, we discuss the projection for entangled vortex-membranes. For two
entangled vortex-membranes described by $\mathrm{z}_{\mathrm{A/B}}(\vec
{x},t)=\xi_{\mathrm{A/B}}(\vec{x},t)+i\eta_{\mathrm{A/B}}(\vec{x},t),$ the
projection along a given direction $\theta$ in 5D space is defined by
\begin{equation}
\hat{P}_{\theta}\left(
\begin{array}
[c]{c}%
\xi_{\mathrm{A/B}}(\vec{x},t)\\
\eta_{\mathrm{A/B}}(\vec{x},t)
\end{array}
\right)  =\left(
\begin{array}
[c]{c}%
\xi_{\mathrm{A/B},\theta}(\vec{x},t)\\
\left[  \eta_{\mathrm{A/B},\theta}(\vec{x},t)\right]  _{0}%
\end{array}
\right)
\end{equation}
where $\xi_{\mathrm{A/B},\theta}(\vec{x},t)=\xi_{\mathrm{A/B}}(\vec{x}%
,t)\cos \theta+\eta_{\mathrm{A/B}}(\vec{x},t)\sin \theta$ is variable and
$\left[  \eta_{\mathrm{A/B},\theta}(\vec{x},t)\right]  _{0}=\xi_{\mathrm{A/B}%
}(\vec{x},t)\sin \theta-\eta_{\mathrm{A/B}}(\vec{x},t)\cos \theta$ is constant.
So the projected vortex-membrane is described by the function $\xi
_{\mathrm{A/B},\theta}(\vec{x},t).$ For two projected vortex-membranes
described by $\xi_{\mathrm{A},\theta}(\vec{x},t)$ and $\xi_{\mathrm{B},\theta
}(\vec{x},t),$ a zero is solution of the equation
\begin{align}
\hat{P}_{\theta}[\mathrm{z}_{\mathrm{A}}(\vec{x},t)]  &  \equiv \xi
_{\mathrm{A},\theta}(\vec{x},t)\\
&  =\hat{P}_{\theta}[\mathrm{z}_{\mathrm{B}}(\vec{x},t)]\equiv \xi
_{\mathrm{B},\theta}(\vec{x},t).\nonumber
\end{align}

\subsection{Zero-lattice}

We then introduce two types of zero-lattices by the two types of projections.

\subsubsection{W-type zero-lattice from W-type projection on a helical
vortex-membrane}

Firstly, we consider the zero-lattice from W-type projection on a helical vortex-membrane.

The function of a 1D helical vortex-line in a 3D fluid is
\begin{equation}
\mathrm{z}(x,t)=\xi(x,t)+i\eta(x,t)=r_{0}e^{\pm ik_{0}\cdot x-i\omega
_{0}t+i\phi_{0}}%
\end{equation}
where $r_{0}$ is the winding radius of vortex-line that is set be constant,
$k_{0}=\frac{\pi}{a}>0$ and $a$ is a fixed length that denotes the half pitch
of the windings. $\phi_{0}$ is a constant angle. $\pm$ denotes two possible
chiralities: left-hand with clockwise winding, or right-hand with
counterclockwise winding.

For a helical vortex-line, from the zero solution $\xi(x,t)=0,$ we get the
zero solutions to be $\pm \bar{x}(t)=a\cdot X-\frac{a}{\pi}\omega_{0}t$ where
$X$ is an integer along $x$-direction and $\theta=-\frac{\pi}{2}+\phi_{0}$.
From the projection of a helical vortex-membrane, we have a crystal of
crossings. Because the winding-number of helical vortex-line is half of
crossing number, each crossing corresponds to a piece of helical vortex-line
with half winding-number. We call the object with half winding-number a knot.
As a result, the system can be regarded as a crystal of knots. It is obvious
that the global rotation doesn't change the winding-number density. For a
helical vortex-membrane with $\omega_{0}\neq0,$ we have a finite velocity of
the system, $\frac{a}{\pi}\omega_{0}.$

For an arbitrary 1D Kelvin wave with different spin states,\ the zero solution
doesn't change i.e., $\left \langle \mathbf{\sigma}^{Z}\right \rangle =\vec
{n}_{\sigma}^{Z}=(0,0,1)\rightarrow \vec{n}_{\sigma}=(n_{x},n_{y},n_{x})$ with
$\left \vert \vec{n}_{\sigma}\right \vert =1$\cite{kou}.

Therefore, in the following parts, we call the crystal with discrete lattice
sites described by the integer numbers $X$ to be "\emph{zero-lattice}%
"\cite{kou1}.

For an 3D SOC Kelvin wave of single vortex-membrane, we can use similar W-type
projection to obtain a 3D W-type zero-lattice.

\subsubsection{T-type zero-lattice from T-type projection on two entangled
vortex-membranes}

For two entangled vortex-membranes, there exists leapfrogging motion. So we
call it leapfrogging knot-crystal. A leapfrogging knot-crystal (two entangled
vortex-membranes) is described by%
\begin{equation}
\mathbf{Z}_{\mathrm{KC}}\mathbf{(}\vec{x},t\mathbf{)}=\left(
\begin{array}
[c]{c}%
\mathrm{z}_{\mathrm{A}}(\vec{x},t)\\
\mathrm{z}_{\mathrm{B}}(\vec{x},t)
\end{array}
\right)  =\left(
\begin{array}
[c]{c}%
r_{\mathrm{A}}\\
r_{\mathrm{B}}%
\end{array}
\right)  e^{i\vec{k}\cdot \vec{x}-i\omega_{0}t+i\omega^{\ast}t/2},
\end{equation}
where $r_{\mathrm{A}}=r_{0}\cos(\frac{\omega^{\ast}t}{2})$ and $r_{\mathrm{B}%
}=-r_{0}i\sin(\frac{\omega^{\ast}t}{2}).$ According to the knot-equation
$\hat{P}_{\theta}[z_{\mathrm{A}}(\vec{x})]=\hat{P}_{\theta}[z_{\mathrm{B}%
}(\vec{x})],$ we have
\begin{equation}
\bar{x}_{\mathrm{F},0}^{I}=a\cdot X^{I}+\frac{a}{\pi}\omega_{0}t
\end{equation}
where $\bar{x}_{\mathrm{F},0}^{I}=\vec{x}\cdot \vec{e}^{I}$ is the coordination
on the axis along a given direction $\vec{e}^{I}$ and $X^{I}$ is an integer
number. As a result, we also have a periodic distribution of zeroes (knots)
that is a T-type zero-lattice. 

\subsubsection{Generalized spatial translation symmetry for zero-lattices}

For both types of zero-lattice, owing to the generalized spatial translation
symmetry for the vortex-membranes there exist corresponding generalized
spatial translation symmetries. For example, for a helical vortex-membrane, by
doing a spatial translation operation $\mathcal{T}(\Delta x)=e^{i\Delta
x\cdot \hat{k}^{I}},$ we have
\begin{align}
\mathrm{z}(\vec{x},t)  &  \longrightarrow \mathrm{z}(\vec{x}+\Delta \vec
{x},t)\nonumber \\
&  =e^{\pm i(\vec{k}\cdot \Delta \vec{x})}(\vec{x},t).
\end{align}
Under the spatial transformation, the zero-lattices shift, i.e.,
\begin{align}
\bar{x}^{I}(t)  &  \rightarrow(\bar{x}^{I}(t))^{\prime}=\bar{x}^{I}(t)+\Delta
x^{I}\\
&  =a\cdot X^{I}-\frac{a}{\pi}\omega_{0}t+\Delta x^{I}.\nonumber
\end{align}
However after changing the projection angle, $\theta \rightarrow \theta
+\frac{\pi}{a}\Delta x^{I}$, the zero-lattice is invariant,
\begin{equation}
\bar{x}^{I}(t)\rightarrow(\bar{x}^{I}(t))^{\prime}=\bar{x}^{I}(t).
\end{equation}

Therefore, for the zero-lattices, the generalized spatial translation
operation is also a combination of a\ continuum spatial translation operation
and a global phase rotation operation.

\subsection{Zeroes and knots}

From the point view of "information", each zero becomes the element of a
zero-lattice. Thus, the information of vortex-membranes is characterized by
the distribution of zeroes. For the case of an extra zero, we have a knot; for
the case of missing zero, we have an anti-knot. According to the existence of
two types of zero-lattices, there are two types of zeroes: W-type zero and
T-type zero. Obviously, a W-type zero is an element object of W-type
zero-lattice and a T-type zero is an element object of T-type zero-lattice.

\subsubsection{W-type zero and W-type knot}

The element of W-type zero-lattice is W-type zero that corresponds to a
crossing of a helical vortex-membrane and a straight line in its center. In
the following parts, we call a knot with half winding-number corresponding to
W-type zero \emph{W-type knot}.

From point view of information, a knot is an information unit with fixed
geometric properties that is always anti-phase changing along arbitrary
direction $\vec{e}$. When there exists a knot, the periodic boundary condition
of Kelvin waves along arbitrary direction is changed into anti-periodic
boundary condition. Based on the projected vortex-membranes, we define a knot
by a monotonic function $F_{\theta}(x)=\xi_{\theta}(x)$ with
\begin{equation}
\mathrm{sgn}\text{ }[F_{\theta}(x\rightarrow-\infty)\cdot F_{\theta
}(x\rightarrow \infty)]=-1
\end{equation}
where $x$ denotes the position along the given direction $\vec{e}$. So the
sign-switching character can be labeled by winding number $w_{1D}$. The
winding number $w_{1D}$ for a knot along given direction is $\pm \frac{1}{2}.$
On the other hand, from the topological character of a knot, there must exist
a point, each knot corresponds to a zero between two vortex-membranes along
the given direction.

\begin{figure}[ptb]
\includegraphics[clip,width=0.45\textwidth]{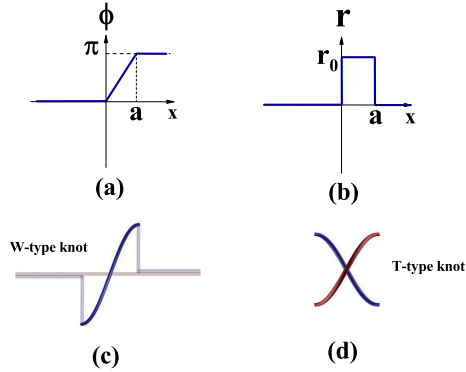}\caption{(a) The phase
angle of a knot ; (b) The radius of a knot; (c) A illustration of a W-type
knot; (d) An illustration of a T-type knot.}%
\end{figure}

The inset in Fig.7 is an illustration of a 1D unified W-type knot. Its
function is given by
\begin{equation}
\mathrm{z}(x)=r(x)e^{i\phi(x)}%
\end{equation}
where%
\begin{equation}
r(x)\rightarrow \left \{
\begin{array}
[c]{c}%
0,\text{ }x\in(-\infty,x_{0}]\\
r_{0},\text{ }x\in(x_{0},x_{0}+a]\\
0,\text{ }x\in(x_{0}+a,\infty)
\end{array}
\right \}
\end{equation}
and
\begin{equation}
\phi(x)=\left \{
\begin{array}
[c]{c}%
\phi_{0}\mp \frac{\pi}{2},\text{ }x\in(-\infty,x_{0}]\\
\phi_{0}\mp \frac{\pi}{2}\pm k_{0}(x-x_{0}),\text{ }x\in(x_{0},x_{0}+a]\\
\phi_{0}\pm \frac{\pi}{2},\text{ }x\in(x_{0}+a,\infty)
\end{array}
\right \}
\end{equation}
where $+$ denotes a clockwise winding and $-$ denotes a counterclockwise
winding. There is a linear relationship between $\phi(x)$ and $x$ as
$\phi(x)\propto x-x_{0}$ in the winding region of $x_{0}<x\leq x_{0}+a$. Thus,
we obtain an anti-periodic boundary condition for the system,
\begin{equation}
\phi(x\rightarrow \infty)-\phi(x\rightarrow-\infty)=\pm \pi.
\end{equation}
Under projection, we have the knot equation as
\begin{equation}
x-x_{0}=(\pm \frac{\pi}{2}\pm \frac{\pi}{2}+\theta-\phi_{0})/k_{0}.
\end{equation}
In Fig.7(a), the relation between the phase and the coordination of a unified
W-type knot is shown. It is obvious that there always exists a single knot
solution for a unified W-type knot.

\subsubsection{T-type zero and T-type knot}

The element of the projected two entangled vortex-membranes is T-type zero
that corresponds to a crossing of the two projected vortex-membranes. In the
following parts, we call a knot with half twisting-number corresponding to
T-type zero \emph{T-type knot}.

Based on the projected vortex-membranes, we define a knot by a monotonic
function $F_{\theta}(x)=\xi_{\mathrm{A},\theta}(x)-\xi_{\mathrm{B},\theta}(x)$
with
\begin{equation}
\mathrm{sgn}\text{ }[F_{\theta}(x\rightarrow-\infty)\cdot F_{\theta
}(x\rightarrow \infty)]=-1
\end{equation}
where $x$ denotes the position along the given direction $\vec{e}$. From the
topological character of a knot, each knot corresponds to a zero between two
vortex-membranes along the given direction. See the illustration in Fig.7(d).

A knot (a zero) has four degrees of freedom: two spin degrees of freedom
$\uparrow$ or $\downarrow$ from the helicity degrees of freedom, the other two
vortex degrees of freedom from the vortex degrees of freedom that characterize
the vortex-membranes, $\mathrm{A}$ or $\mathrm{B}$. For example, for an
up-spin knot on vortex-membrane-\textrm{A}, the function is defined by
$\mathrm{z}_{\mathrm{knot},\uparrow,\mathrm{A}}(x,t)=r_{\uparrow,\mathrm{A}%
}(x)\exp[i\phi_{\mathrm{knot},\uparrow,\mathrm{A}}(x,t)]$ where
\begin{equation}
\phi_{\mathrm{knot},\uparrow,\mathrm{A}}(x)=\left \{
\begin{array}
[c]{c}%
-\phi_{0}+\frac{\pi}{2},\text{ }x\in(-\infty,x_{0}]\\
-\phi_{0}+\frac{\pi}{2}+k_{0}(x-x_{0}),\text{ }x\in(x_{0},x_{0}+a]\\
-\phi_{0}-\frac{\pi}{2},\text{ }x\in(x_{0}+a,\infty)
\end{array}
\right \}
\end{equation}
and%
\begin{equation}
r_{\uparrow,\mathrm{A}}(x)\rightarrow \left \{
\begin{array}
[c]{c}%
0,\text{ }x\in(-\infty,x_{0}]\\
r_{0},\text{ }x\in(x_{0},x_{0}+a]\\
0,\text{ }x\in(x_{0}+a,\infty)
\end{array}
\right \}
\end{equation}
where $x=\vec{x}\cdot \vec{e}$ is the coordination on the axis along a given
direction $\vec{e}$ and $\phi_{0}$ is an arbitrary constant angle,
$k_{0}=\frac{\pi}{a}.$

\subsection{Examples:}

\subsubsection{2-level composite knot-crystal with ($\mathcal{N}=2$,
$\mathcal{M}=2$)}

For 2-level composite knot-crystal with ($\mathcal{N}=2$, $\mathcal{M}=2$),
the function of the centre-membranes of vortex-membrane-$\mathrm{A}$ and
vortex-membrane-$\mathrm{B,}$ $\mathrm{z}_{2}(\vec{x},t)$; the function of the
vortex-membrane-$\mathrm{A}$ is $\mathrm{z}_{\mathrm{A},1}(\vec{x},t)$ and the
function of the vortex-membrane-$\mathrm{B}$ is $\mathrm{z}_{\mathrm{B}%
,1}(\vec{x},t)$.

The level-2 W-type zero-lattice of the composite winding knot-crystal with
($\mathcal{N}=2$, $\mathcal{M}=2$) is obtained by level-2 W-type projection of
centre-membranes of vortex-membrane-$\mathrm{A}$ and
vortex-membrane-$\mathrm{B},$ i.e.,
\begin{equation}
\hat{P}_{\theta_{2}}[\mathrm{z}_{2}(\vec{x},t)]=0,
\end{equation}
that is $\xi_{2}(\vec{x},t)=0.$ The solution of zero-lattice is given by
$\bar{x}^{I}(t)=a_{2}\cdot X^{I}-\frac{a_{2}}{\pi}\omega_{0}t$ where $X^{I}$
is an integer along $x^{I}$-direction and $\theta_{2}=-\frac{\pi}{2}%
+\phi_{2,0}$. $a_{2}$ is a fixed length that denotes the half pitch of the
windings of $\mathrm{z}_{2}(\vec{x},t)$.

The level-1 T-type zero-lattice of the 2-level composite knot-crystal with
($\mathcal{N}=2$, $\mathcal{M}=2$) is obtained by level-1 T-type projection of
vortex-membrane-$\mathrm{A}$ and vortex-membrane-\textrm{B}, i.e.,
\begin{equation}
\hat{P}_{\theta_{1}}[\mathrm{z}_{\mathrm{A},1}(\vec{x},t)]=\hat{P}_{\theta
_{2}}[\mathrm{z}_{\mathrm{B},1}(\vec{x},t)],
\end{equation}
that is $\xi_{\mathrm{A},1}(\vec{x},t)=\xi_{\mathrm{B},1}(\vec{x},t).$

In particular, there exists an intrinsic relationship between the number of
T-type zeroes and level-2 W-type zeroes,%
\begin{equation}
\zeta_{\left(  \mathrm{A,B}\right)  ,1D}^{I}=W_{\left(  \mathrm{A,B}\right)
,1D}^{I}+T_{\left(  \mathrm{A,B}\right)  ,1D}^{I}\equiv \mathrm{const},\text{
}(I=x,y,z)
\end{equation}
where $\zeta_{\left(  \mathrm{A,B}\right)  ,1D}^{I},$ $W_{\left(
\mathrm{A,B}\right)  ,1D}^{I},$ and $T_{\left(  \mathrm{A,B}\right)  ,1D}^{I}$
are linking number, writhe number, and twist number, respectively. Here, the
number of level-1 T-type zeroes and level-2 W-type zeroes are equal to be the
writhe number $W_{\left(  \mathrm{A,B}\right)  ,1D}^{I}$ and the twist number
$T_{\left(  \mathrm{A,B}\right)  ,1D}^{I}$, respectively. Because the sum of
the number of level-2 T-type zeroes and level-1 W-type zeroes is invariant,
when one winds two entangled vortex-membranes (a 1-level knot-crystal with
($\mathcal{N}=2$, $\mathcal{M}=1$)) into a 2-level composite knot-crystal with
($\mathcal{N}=2$, $\mathcal{M}=2$)$,$ some level-1 T-type of zeroes are
replaced by level-2 W-type zeroes. We call it \emph{substitution effect of
level-1 T-type of zeroes by level-2 W-type zeroes}.

\subsubsection{2-level composite knot-crystal with ($\mathcal{N}=4$,
$\mathcal{M}=2$)}

For 2-level composite knot-crystal with ($\mathcal{N}=4$, $\mathcal{M}=2$)$,$
the functions $\mathbf{Z}_{2}(\vec{x},t)=\left(
\begin{array}
[c]{c}%
\mathrm{z}_{\mathrm{A},2}(\vec{x},t)\\
\mathrm{z}_{\mathrm{B},2}(\vec{x},t)
\end{array}
\right)  $ are described by another knot-crystal that characterizes the
centre-membrane of \textrm{A}-knot-crystal by $\mathrm{z}_{\mathrm{A},2}%
(\vec{x},t)$ and that of \textrm{B}-knot-crystal by $\mathrm{z}_{\mathrm{B}%
,2}(\vec{x},t)$, respectively. The \textrm{A}-knot-crystal (the entangled
vortex-membranes $\mathrm{A}_{1},$ $\mathrm{A}_{2}$) around the $\mathrm{z}%
_{\mathrm{A},2}(\vec{x},t)$ is described by
\begin{equation}
\mathbf{Z}_{\mathrm{A},1}^{\prime}(\vec{x},t)=\left(
\begin{array}
[c]{c}%
z_{\mathrm{A}_{1},1}(\vec{x},t)\\
z_{\mathrm{A}_{2},1}(\vec{x},t)
\end{array}
\right)  =\left(
\begin{array}
[c]{c}%
%TCIMACRO{\dprod \limits_{I}}%
%BeginExpansion
{\displaystyle \prod \limits_{I}}
%EndExpansion
z_{\mathrm{A}_{1},1}(\phi_{\mathrm{A}_{1},1}^{I}(\phi_{\mathrm{A}_{1},2}%
^{I}(x^{I})),t)\\%
%TCIMACRO{\dprod \limits_{I}}%
%BeginExpansion
{\displaystyle \prod \limits_{I}}
%EndExpansion
z_{\mathrm{A}_{2},1}(\phi_{\mathrm{A}_{2},1}^{I}(\phi_{\mathrm{A}_{1},2}%
^{I}(x^{I})),t)
\end{array}
\right)
\end{equation}
and \textrm{B}-knot-crystal (the entangled vortex-membranes $\mathrm{B}_{1},$
$\mathrm{B}_{2}$) around $\mathrm{z}_{\mathrm{B},2}(\vec{x},t)$ is described
by
\begin{equation}
\mathbf{Z}_{\mathrm{B},1}^{\prime}(\vec{x},t)=\left(
\begin{array}
[c]{c}%
z_{\mathrm{B}_{1},1}(\vec{x},t)\\
z_{\mathrm{B}_{2},1}(\vec{x},t)
\end{array}
\right)  =\left(
\begin{array}
[c]{c}%
%TCIMACRO{\dprod \limits_{I}}%
%BeginExpansion
{\displaystyle \prod \limits_{I}}
%EndExpansion
z_{\mathrm{B}_{1},1}(\phi_{\mathrm{B}_{1},1}^{I}(\phi_{\mathrm{B}_{1},2}%
^{I}(x^{I})),t)\\%
%TCIMACRO{\dprod \limits_{I}}%
%BeginExpansion
{\displaystyle \prod \limits_{I}}
%EndExpansion
z_{\mathrm{B}_{2},1}(\phi_{\mathrm{B}_{2},1}^{I}(\phi_{\mathrm{B}_{1},2}%
^{I}(x^{I})),t)
\end{array}
\right)  .
\end{equation}

After projection, there are five zero-lattices: level-2 W-type zero-lattice
for \textrm{A}-knot-crystal, level-2 W-type zero-lattice for \textrm{B}%
-knot-crystal, level-2 T-type zero-lattice between \textrm{A}-knot-crystal and
\textrm{B}-knot-crystal, level-1 T-type zero-lattice between two entangled
vortex-membrane-$\mathrm{A}_{1},$ $\mathrm{A}_{2}$ for \textrm{A}%
-knot-crystal, level-1 T-type zero-lattice between two entangled
vortex-membrane-$\mathrm{B}_{1},$ $\mathrm{B}_{2}$ for \textrm{B}-knot-crystal.

The level-2 W-type zero-lattice for \textrm{A}-knot-crystal is obtained by
level-2 W-type projection of vortex-membrane-$\mathrm{A}$, i.e.,
\begin{equation}
\hat{P}_{\theta_{2}}[\mathrm{z}_{\mathrm{A},2}(\vec{x},t)]=0,
\end{equation}
that is $\xi_{\mathrm{A},2}(\vec{x},t)=0.$ The solution of the zero-lattice is
obtained as
\begin{equation}
\bar{x}_{2\mathrm{W,A},0}^{I}=a_{2}\cdot X^{I}+\frac{na_{2}}{\pi}\omega
_{0}t+\phi_{2\mathrm{W,A},0}%
\end{equation}
where $X^{I}$ is an integer number along $x^{I}$-direction. $\phi
_{2\mathrm{W,A},0}$ is a constant phase angle and$\ a_{2}$ is a fixed length
that denotes the half pitch of the half windings of $\mathrm{z}_{\mathrm{A}%
,2}(\vec{x},t)$.

The level-2 W-type zero-lattice for \textrm{B}-knot-crystal is obtained by
level-2 W-type projection of vortex-membrane-\textrm{B}, i.e.,
\begin{equation}
\hat{P}_{\theta_{2}}[\mathrm{z}_{\mathrm{B},2}(\vec{x},t)]=0,
\end{equation}
that is $\xi_{\mathrm{B},2}(\vec{x},t)=0.$ The solution of the zero-lattice is
obtained as
\begin{equation}
\bar{x}_{2\mathrm{W,B},0}^{I}=a_{2}\cdot X^{I}+\frac{na_{2}}{\pi}\omega
_{0}t+\phi_{2\mathrm{W,B},0}%
\end{equation}
where $X^{I}$ is an integer number along $x^{I}$-direction. $\phi
_{2\mathrm{W,B},0}$ is a constant phase angle and$\ a_{2}$ is a fixed length
that denotes the half pitch of the windings of $\mathrm{z}_{\mathrm{B},2}%
(\vec{x},t)$.

The level-2 T-type zero-lattice is obtained by level-2 T-type projection
between \textrm{A}-knot-crystal and \textrm{B}-knot-crystal, i.e.,
\begin{equation}
\hat{P}_{\theta_{2}}[\mathrm{z}_{\mathrm{A},2}(\vec{x},t)]=\hat{P}_{\theta
}[\mathrm{z}_{\mathrm{B},2}(\vec{x},t)],
\end{equation}
that is $\xi_{\mathrm{A},2}(\vec{x},t)=\xi_{\mathrm{B},2}(\vec{x},t).$ The
solution of the zero-lattice is obtained as
\begin{equation}
\bar{x}_{2\mathrm{T,}0}^{I}=a_{2}\cdot X^{I}+\frac{na_{2}}{\pi}\omega
_{0}t+\phi_{2\mathrm{T},0}%
\end{equation}
where $X^{I}$ is an integer number along $x^{I}$-direction. $\phi
_{2\mathrm{T,}0}$ is a constant phase angle and$\ a_{2}$ is a fixed length
that denotes the half pitch of the twistings of $\mathrm{z}_{\mathrm{A/B}%
,2}(\vec{x},t)$.

The level-1 T-type zero-lattice between two entangled
vortex-membrane-$\mathrm{A}_{1},$ $\mathrm{A}_{2}$ for \textrm{A}-knot-crystal
is obtained by level-1 T-type projection of $\mathrm{A}$-knot-crystal,
\begin{equation}
\hat{P}_{\theta_{1}}[z_{\mathrm{A}_{1},1}(\vec{x},t)]=\hat{P}_{\theta
}[z_{\mathrm{A}_{2},1}(\vec{x},t)],
\end{equation}
i.e.,that is $\xi_{\mathrm{A}_{1},1}(\vec{x},t)=\xi_{\mathrm{A}_{2},1}(\vec
{x},t)$. The solution of the zero-lattice is obtained as
\begin{equation}
\bar{x}_{1\mathrm{T,A},0}^{I}=a_{1}\cdot X^{I}+\frac{na_{1}}{\pi}\omega
_{0}t+\phi_{1\mathrm{T,A},0}%
\end{equation}
where $X^{I}$ is an integer number along $x^{I}$-direction. $\phi
_{1\mathrm{T,A},0}$ is a constant phase angle and$\ a_{1}$ is a fixed length
that denotes the half pitch of the windings of $\mathrm{z}_{\mathrm{A}_{1}%
,1}(\vec{x},t)$.

The level-1 T-type zero-lattice between two entangled
vortex-membrane-$\mathrm{B}_{1},$ $\mathrm{B}_{2}$ for \textrm{B}-knot-crystal
is obtained by level-1 T-type projection of $\mathrm{B}$-knot-crystal,
\begin{equation}
\hat{P}_{\theta_{1}}[z_{\mathrm{B}_{1},1}(\vec{x},t)]=\hat{P}_{\theta
}[z_{\mathrm{B}_{2},1}(\vec{x},t)],
\end{equation}
i.e.,that is $\xi_{\mathrm{B}_{1},1}(\vec{x},t)=\xi_{\mathrm{B}_{2},1}(\vec
{x},t)$. The solution of the zero-lattice is obtained as
\begin{equation}
\bar{x}_{1\mathrm{T,B},0}^{I}=a_{1}\cdot X^{I}+\frac{na_{1}}{\pi}\omega
_{0}t+\phi_{1\mathrm{T,B},0}%
\end{equation}
where $X^{I}$ is an integer number along $x^{I}$-direction. $\phi
_{1\mathrm{T,B},0}$ is a constant phase angle and$\ a_{1}$ is a fixed length
that denotes the half pitch of the windings of $\mathrm{z}_{\mathrm{B}_{1}%
,1}(\vec{x},t)$.

In particular, there exist two intrinsic relationships between the number of
T-type zeroes and level-2 W-type zeroes. One is twist-writhe locking relation
for \textrm{A}-knot-crystal%
\begin{equation}
\zeta_{\left(  \mathrm{A}_{1}\mathrm{,A}_{2}\right)  ,1D}^{I}=W_{\left(
\mathrm{A}_{1}\mathrm{,A}_{2}\right)  ,1D}^{I}+T_{\left(  \mathrm{A}%
_{1}\mathrm{,A}_{2}\right)  ,1D}^{I}\equiv \mathrm{const},\text{ }(I=x,y,z)
\end{equation}
where $\zeta_{\left(  \mathrm{A}_{1}\mathrm{,A}_{2}\right)  ,1D}^{I},$
$W_{\left(  \mathrm{A}_{1}\mathrm{,A}_{2}\right)  ,1D}^{I},$ and $T_{\left(
\mathrm{A}_{1}\mathrm{,A}_{2}\right)  ,1D}^{I}$ are linking number, writhe
number, and twist number between vortex-membranes $\mathrm{A}_{1}\mathrm{,}$
$\mathrm{A}_{2}$, respectively. Here, the number of level-1 T-type zeroes and
level-2 W-type zeroes are equal to be the writhe number $W_{\left(
\mathrm{A}_{1}\mathrm{,A}_{2}\right)  ,1D}^{I}$ and the twist number
$T_{\left(  \mathrm{A}_{1}\mathrm{,A}_{2}\right)  ,1D}^{I}$, respectively. The
other is twist-writhe locking relation for \textrm{B}-knot-crystal%
\begin{equation}
\zeta_{\left(  \mathrm{B}_{1}\mathrm{,B}_{2}\right)  ,1D}^{I}=W_{\left(
\mathrm{B}_{1}\mathrm{,B}_{2}\right)  ,1D}^{I}+T_{\left(  \mathrm{B}%
_{1}\mathrm{,B}_{2}\right)  ,1D}^{I}\equiv \mathrm{const},\text{ }(I=x,y,z)
\end{equation}
where $\zeta_{\left(  \mathrm{B}_{1}\mathrm{,B}_{2}\right)  ,1D}^{I},$
$W_{\left(  \mathrm{B}_{1}\mathrm{,B}_{2}\right)  ,1D}^{I},$ and $T_{\left(
\mathrm{B}_{1}\mathrm{,B}_{2}\right)  ,1D}^{I}$ are linking number, writhe
number, and twist number between vortex-membranes $\mathrm{B}_{1}\mathrm{,}$
$\mathrm{B}_{2}$, respectively. Here, the number of level-1 T-type zeroes and
level-2 W-type zeroes are equal to be the writhe number $W_{\left(
\mathrm{B}_{1}\mathrm{,B}_{2}\right)  ,1D}^{I}$ and the twist number
$T_{\left(  \mathrm{B}_{1}\mathrm{,B}_{2}\right)  ,1D}^{I}$, respectively.

\subsubsection{3-level composite knot-crystal with ($\mathcal{N}=4$,
$\mathcal{M}=3$)}

To generate a 3-level composite knot-crystal with ($\mathcal{N}=4$,
$\mathcal{M}=3$), we symmetrically wind a 2-level double-helix knot-crystal
with ($\mathcal{N}=4$, $\mathcal{M}=3$) along different spatial directions. As
a result, there exists an additional W-type zero-lattice by W-type projection
on the center membrane of \textrm{A}-knot-crystal and \textrm{B}-knot-crystal
-- the level-3 W-type zero-lattice. We use $\mathrm{z}_{3}(\vec{x},t)$ to
denote the centre-membranes of \textrm{A}-knot-crystal and \textrm{B}-knot-crystal.

So the level-3 W-type zero-lattice of the 3-level composite knot-crystal with
($\mathcal{N}=4$, $\mathcal{M}=3$) is obtained by level-3 W-type projection of
centre-membranes of $\mathrm{z}_{3}(\vec{x},t),$ i.e.,
\begin{equation}
\hat{P}_{\theta_{3}}[\mathrm{z}_{3}(\vec{x},t)]=0,
\end{equation}
that is $\xi_{3}(\vec{x},t)=0.$ The solution of zero-lattice is given by
\begin{equation}
\bar{x}_{3\mathrm{W},0}^{I}(t)=a_{3}\cdot X^{I}-\frac{a_{3}}{\pi}\omega
_{0}t+\phi_{3\mathrm{W},0}%
\end{equation}
where $X^{I}$ is an integer along $x^{I}$-direction and $\theta_{3}=-\frac
{\pi}{2}+\phi_{3\mathrm{W},0}$. $\phi_{3\mathrm{W},0}$ is a constant phase
angle and$\ a_{3}$ is a fixed length that denotes the half pitch of the
windings of $\mathrm{z}_{3}(\vec{x},t)$.

Owing to the existence the additional level-3 W-type zero-lattice, there
exists an additional intrinsic relationship between the number of level-2
T-type zeroes and level-3 W-type zeroes,%
\begin{equation}
\zeta_{\left(  \mathrm{A,B}\right)  ,1D}^{I}=W_{\left(  \mathrm{A,B}\right)
,1D}^{I}+T_{\left(  \mathrm{A,B}\right)  ,1D}^{I}\equiv \mathrm{const},\text{
}(I=x,y,z)
\end{equation}
where $\zeta_{\left(  \mathrm{A,B}\right)  ,1D}^{I},$ $W_{\left(
\mathrm{A,B}\right)  ,1D}^{I},$ and $T_{\left(  \mathrm{A,B}\right)  ,1D}^{I}$
are linking number, writhe number, and twist number, respectively. Here, the
number of level-2 T-type zeroes and level-3 W-type zeroes are equal to be the
writhe number $W_{\left(  \mathrm{A,B}\right)  ,1D}^{I}$ and the twist number
$T_{\left(  \mathrm{A,B}\right)  ,1D}^{I}$, respectively. Because the sum of
the number of level-2 T-type zeroes and level-3 W-type zeroes is invariant,
when one winds two entangled vortex-membrane (a 2-level double-helix
knot-crystal with ($\mathcal{N}=4$, $\mathcal{M}=3$)) into a 3-level winding
knot-crystal with ($\mathcal{N}=4$, $\mathcal{M}=3$), some level-2 T-type of
zeroes are replaced by level-3 W-type zeroes. this is \emph{the substitution
effect of level-2 T-type of zeroes by level-3 W-type zeroes}.

\section{Emergent quantum field theory for 1-level knot-crystal with
($\mathcal{N}=1$, $\mathcal{M}=1$)}

From the above discussion, the winding-number density becomes a physical
quantity to characterize local windings of a helical vortex-membrane (1-level
knot-crystal with ($\mathcal{N}=1$, $\mathcal{M}=1$)) and the element of the
projected helical vortex-membrane is knot with half winding-number that
corresponds to a crossing of a helical vortex-membrane and a straight line in
its center. We then discuss the local fluctuations of a perturbative helical
vortex-membrane with an extra knot and develop a theory to characterize its dynamics.

\subsection{Knot}

A knot of 1-level knot-crystal with ($\mathcal{N}=1$, $\mathcal{M}=1$) is an
anti-phase changing along arbitrary direction $\vec{e}$. As a result, there
exists a point, each knot corresponds to a zero between two vortex-membranes.

A W-type knot with a W-type zero is a half-winding of 1-level knot-crystal
with ($\mathcal{N}=1$, $\mathcal{M}=1$). Fig.7(c) is an example of a W-type
knot. We then introduce the operation for the 1D W-type knot
\begin{equation}
\hat{U}(\phi(x))=\mathrm{\hat{\digamma}}_{\mathrm{knot}}(r_{0})\exp
[i\phi(x)\cdot \hat{K}]
\end{equation}
on a constant complex field $z_{0}=0$ (we use $[0]$ to denote the flat
vortex-line) to generate a single W-type knot, i.e.,
\begin{equation}
\hat{U}(\phi(x))\cdot \lbrack0]=\mathrm{z}(x)=r(x)e^{i\phi(x)}.
\end{equation}
$\mathrm{\hat{\digamma}}_{\mathrm{knot}}(r_{0})$ is an expanding operator by
shifting $0$ to $r_{0}$ in the winding region of a knot (for example,
$x\in(x_{0},x_{0}+a]$). Here $\hat{K}=-i\frac{d}{d\phi}$ is knot number
operator and%
\begin{equation}
r(x)\rightarrow \left \{
\begin{array}
[c]{c}%
0,\text{ }x\in(-\infty,x_{0}]\\
r_{0},\text{ }x\in(x_{0},x_{0}+a]\\
0,\text{ }x\in(x_{0}+a,\infty)
\end{array}
\right \}
\end{equation}
and
\begin{equation}
\phi(x)=\left \{
\begin{array}
[c]{c}%
\phi_{0}\mp \frac{\pi}{2},\text{ }x\in(-\infty,x_{0}]\\
\phi_{0}\mp \frac{\pi}{2}\pm k_{0}(x-x_{0}),\text{ }x\in(x_{0},x_{0}+a]\\
\phi_{0}\pm \frac{\pi}{2},\text{ }x\in(x_{0}+a,\infty)
\end{array}
\right \}
\end{equation}
where $+$ denotes a clockwise winding and $-$ denotes a counterclockwise
winding. In Fig.7(a), the relation between the phase and the coordination of a
1D unified W-type knot is shown. It is obvious that there always exists a
single knot solution for a unified W-type knot.

The knot number for 1D W-type knot can be obtained by the following equation
\begin{equation}
\left \langle \hat{K}\right \rangle =\frac{1}{\pi \left \vert z_{0}\right \vert
^{2}}\int \mathrm{z}^{\ast}(\phi(x))\cdot \hat{K}\cdot \mathrm{z}(\phi(x))d\phi
\end{equation}
where $\mathrm{z}^{\ast}(x)$ is a complex conjugation of $\mathrm{z}(x)$. In
physics, $\left \langle \hat{K}\right \rangle $ measures the total phase
changing for a knot with half winding-number that can be regarded as an
anti-phase domain wall along given direction in a 1D complex field
$\mathrm{z}(x)$, i.e., $\left \langle \hat{K}\right \rangle =1.$ The knot
density (the density of crossings) and the density of winding-numbers are
defined by $\rho_{\mathrm{knot}}^{I}=\left \langle \frac{\hat{K}}{\Delta
x}\right \rangle $ and $\rho_{\mathrm{wind}}=2\rho_{\mathrm{knot}}=\left \langle
\frac{2\hat{K}}{\Delta x}\right \rangle ,$ respectively.

We call the extended object unified W-type knot. The definition can be
generalized to d-dimensional W-type knot. The knot number for d-dimensional
W-type knot along $x^{I}$-direction can be obtained by the following equation
\begin{equation}
\left \langle \hat{K}^{I}\right \rangle =\frac{1}{\pi \left \vert z_{0}\right \vert
^{2}}\int \mathrm{z}^{\ast}(\phi(x^{I}))\cdot \hat{K}\cdot \mathrm{z}(\phi
(x^{I}))d\phi^{I}.
\end{equation}
In physics, $\left \langle \hat{K}^{I}\right \rangle $ measures the total phase
changing for a knot with half winding-number that can be regarded as an
anti-phase domain wall along given direction in a 1D complex field
$\mathrm{z}(x^{I})$, i.e., $\left \langle \hat{K}^{I}\right \rangle =1.$ The
knot density (the density of crossings) and the density of winding-numbers are
defined by $\rho_{\mathrm{knot}}^{I}=\left \langle \frac{\hat{K}^{I}}{\Delta
x^{I}}\right \rangle $ and $\rho_{\mathrm{wind}}^{I}=2\rho_{\mathrm{knot}}%
^{I}=\left \langle \frac{2\hat{K}^{I}}{\Delta x^{I}}\right \rangle ,$
respectively. The total density of a knot is defined by
\begin{equation}
\rho_{\mathrm{knot}}(\vec{x},t)=%
%TCIMACRO{\dprod \limits_{I}}%
%BeginExpansion
{\displaystyle \prod \limits_{I}}
%EndExpansion
\rho_{\mathrm{knot}}^{I}(x^{I},t).
\end{equation}

\paragraph{Fragmentized W-type knot}

However, because W-type knot comes from the winding of a helical
vortex-membrane, it is not a rigid object. Instead, it can split and be
fragmentized. We then introduce the concept of "fragmentized W-type knot" by
breaking a W-type knot into $N$ pieces ($N\rightarrow \infty$), each of which
is an identical $\frac{1}{N}$-knot with $\frac{\pi}{N}$ phase-changing. The
function of Kelvin wave with a fragmentized W-type knot (a composite object of
$N$ identical $\frac{1}{N}$-knot) is
\begin{equation}
\left[  z(\phi)\right]  _{\mathrm{fragment,N}}=%
%TCIMACRO{\dprod \limits_{i=1}^{N}}%
%BeginExpansion
{\displaystyle \prod \limits_{i=1}^{N}}
%EndExpansion
\hat{U}(\Delta \phi=\frac{\pi}{N},\left(  x_{0}\right)  _{i})z_{0}%
\end{equation}
where $N$ identical $\frac{1}{N}$-knots are at $\left(  x_{0}\right)  _{1},$
$\left(  x_{0}\right)  _{2},$..., $\left(  x_{0}\right)  _{N}$. For each
$\frac{1}{N}$-knot, the knot number $\left \langle \hat{K}\right \rangle $ is
$\frac{1}{N}$\ and the corresponding phase changing is $\Delta \phi=\frac{\pi
}{N}$. Thus, for a fragmentized knot, there also exists only a single knot
solution and the knot number is conserved. At the limit of $N\rightarrow
\infty,$ we have a uniform distribution of the $N$ identical $\frac{1}{N}$-knots.

\subsection{Emergent quantum mechanics}

Knots can be regarded as quantum particles that obey emergent quantum
mechanics and that the distribution of fragmentized knots is determined by
Schr\"{o}dinger equation.

The function of the Kelvin wave with a fragmentized knot describes the
distribution of the $N$ identical $\frac{1}{N}$-knots and plays the role of
the wave function in emergent quantum mechanics as
\begin{equation}
\frac{\left[  z(\phi)\right]  _{\mathrm{fragment}}}{\left \vert z_{0}%
\right \vert }=\sqrt{\Omega(\vec{x},t)}e^{i\Delta \phi(\vec{x},t)}%
\Longleftrightarrow \psi(\vec{x},t),
\end{equation}
and $\Omega(\vec{x},t)=\rho_{\mathrm{knot}}(\vec{x},t)\Longleftrightarrow
n_{\mathrm{knot}}(\vec{x},t)$ where the function of the Kelvin wave with a
fragmentized knot $\frac{\left[  z(\phi)\right]  _{\mathrm{fragment}}%
}{\left \vert z_{0}\right \vert }$ becomes the wave function $\psi(\vec{x},t)$
in emergent quantum mechanics; the angle $\Delta \phi(\vec{x},t)$ becomes the
quantum phase angle of the wave function; $\Omega(\vec{x},t)$ is equal to the
knot density $\rho_{\mathrm{knot}}$ and thus becomes the probability density
for finding a knot $n_{\mathrm{knot}}(\vec{x})$. The geometrical Kelvin waves
turn into "probability wave" for finding knots and the functions of Kelvin
waves become the wave functions.

In emergent quantum mechanics, the projected energy and the projected momentum
become operators for a fragmentized knot.

The projected momentum of a fragmentized knot on helical vortex-membrane with
an excited Kelvin wave $\psi(\vec{x},t)=\frac{1}{\sqrt{V}}e^{-i\Delta \omega
t+i\Delta \vec{k}\cdot \vec{x}}$ is defined to be
\begin{equation}
p_{\mathrm{knot}}=\hbar_{\mathrm{eff}}\Delta k
\end{equation}
where the effective Planck constant $\hbar_{\mathrm{eff}}$ is obtained as the
projected angular momentum of a knot
\begin{equation}
\hbar_{\mathrm{eff}}=J_{\mathrm{knot}}=\frac{1}{2}\rho_{0}\kappa V_{P}%
r_{0}^{2}.
\end{equation}
Given the superposition principle of Kelvin waves, a generalized wave function
is $\psi(\vec{x},t)=%
%TCIMACRO{\dsum \nolimits_{p}}%
%BeginExpansion
{\displaystyle \sum \nolimits_{p}}
%EndExpansion
c_{p}\exp(\frac{-iE_{\mathrm{knot}}t+i\vec{p}_{\mathrm{knot}}\cdot \vec{x}%
}{\hslash_{\mathrm{eff}}}).$ For an arbitrary wave function $\psi(\vec{x},t)$,
we have $\left \langle \vec{p}_{\mathrm{knot}}\right \rangle =\int \vec
{p}_{\mathrm{knot}}\Omega(\vec{x})dx=\int \psi^{\ast}(\vec{x},t)(i\hbar
_{\mathrm{eff}}\frac{d}{d\vec{x}})\psi(\vec{x},t)dV.$ This result indicates
that the projected momentum for a fragmentized knot becomes operator $\vec
{p}_{\mathrm{knot}}\rightarrow \hat{p}_{\mathrm{knot}}=-i\hbar_{\mathrm{eff}%
}\frac{d}{d\vec{x}}.$ For a plane Kelvin wave $\psi(\vec{x},t)=\frac{1}%
{\sqrt{V}}e^{-i\Delta \omega t+i\Delta \vec{k}\cdot \vec{x}},$ the projected
energy of a fragmentized knot is
\begin{equation}
E_{\mathrm{knot}}=J_{\mathrm{knot}}\Delta \omega=\hbar_{\mathrm{eff}}%
\Delta \omega.
\end{equation}
Using a similar approach, one can see that the projected energy for a
fragmentized knot becomes operator $E_{\mathrm{knot}}\rightarrow \hat
{H}_{\mathrm{knot}}=i\hbar_{\mathrm{eff}}\frac{d}{dt}.$

\subsection{Quantum Fermionic lattice model for knots}

Because the knots on 1-level knot-crystal with ($\mathcal{N}=1$,
$\mathcal{M}=1$) has only one chirality (we assume right-hand chirality), the
effective quantum model is that of Weyl fermions\cite{we}.

First, we discuss the statistics of a knot. In quantum mechanics, particles
with wave functions antisymmetric under exchange are called Fermions. To
illustrate the Fermi statistics of knots, we define the Fermionic operator for
knots with right-hand chirality as $c^{\dagger}(\vec{x})=\hat{U}(\phi(\vec
{x})).$ It is obvious that the wave function's antisymmetry by exchanging two
knots is a result of the $\pi$-phase changing nature of knots, i.e.,
\begin{equation}
\hat{U}(\phi^{\prime}(\vec{x}^{\prime}))\cdot \hat{U}(\phi(\vec{x}))=-\hat
{U}(\phi(\vec{x}))\cdot \hat{U}(\phi^{\prime}(\vec{x}^{\prime})).
\end{equation}
As a result, knots obey Fermi statistics, $\{c^{\dagger}(\vec{x}),c^{\dagger
}(\vec{x}^{\prime})\}=0$.

Next, we derive the effective quantum field theory for Weyl fermions. A knot
has two spin degrees of freedom $\uparrow$ or $\downarrow$ from the helicity
degrees of freedom. The basis to define the microscopic structure of a knot is
given by $\left \vert \uparrow \right \rangle ,$ $\left \vert \downarrow
\right \rangle .$ We define operator of knot states by the region of the phase
angle of a knot: for the case of $\phi_{0}\operatorname{mod}(2\pi)\in
(-\pi,0],$ we have $c^{\dagger}\left \vert 0\right \rangle $; For the case of
$\phi_{0}\operatorname{mod}(2\pi)\in(0,\pi],$ we have $(c^{\dagger}\left \vert
0\right \rangle )^{\dagger}$.

To characterize the energy cost from global winding, we use an effective
Hamiltonian to describe the coupling between 2-knot states along $x^{I}%
$-direction on 3D SOC knot-crystal
\begin{equation}
Jc_{i}^{\dagger}T^{I}c_{i+e^{I}}%
\end{equation}
with the annihilation operator of knots at the site $i,$ $c_{i}=\left(
\begin{array}
[c]{c}%
c_{\uparrow,i}\\
c_{\downarrow,i}%
\end{array}
\right)  $. $J$ is the coupling constant between two nearest-neighbor knots.
According to the generalized translation symmetry, the transfer matrices
$T^{I}$ along $x^{I}$-direction are defined by
\begin{equation}
T^{I}=e^{ia(\hat{k}^{I}\cdot \sigma^{I})}%
\end{equation}
Then, we get the total kinetic term as
\begin{equation}
J\sum_{\left \langle i,j\right \rangle }c_{i}^{\dagger}T^{I}c_{i+e^{I}}+h.c.
\end{equation}
where $\left \langle {i,j}\right \rangle $ denotes the nearest-neighbor knots.
Fig.8(a) is an illustration of entanglement pattern of a 2D SOC knot-crystal
with ($\mathcal{N}=1$, $\mathcal{M}=1$) and Fig.8(d) is an illustration of a
knot that changes entanglement along x and y directions. In Fig.8(a), each
circle denotes a zero.

\begin{figure}[ptb]
\includegraphics[clip,width=0.53\textwidth]{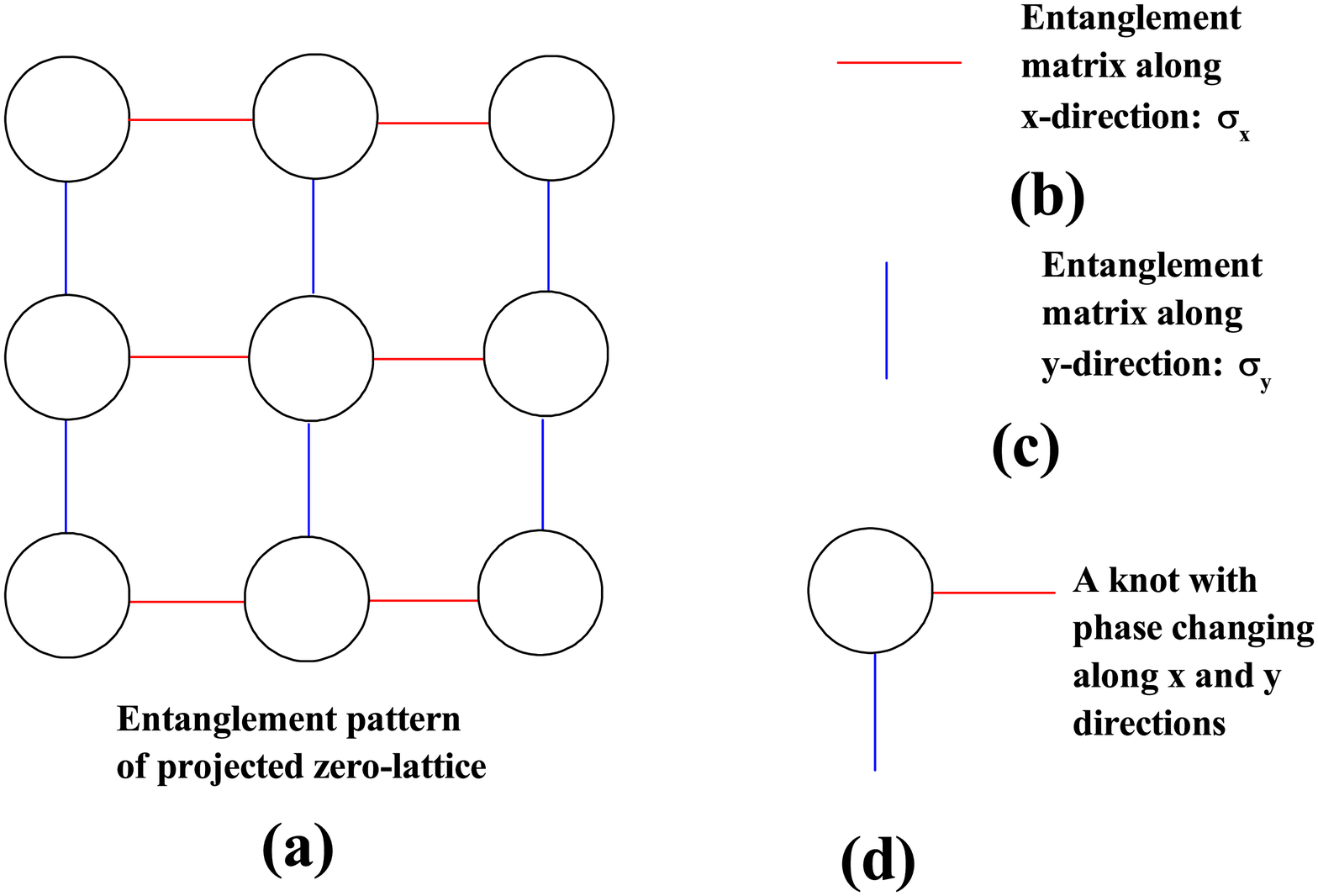}\caption{(a) An
illustration of entanglement pattern of a 2D SOC knot-crystal with
($\mathcal{N}=1$, $\mathcal{M}=1$). Each circle denotes a zero; (b) An
illustration of entanglement matrix along x-direction; (c) An illustration of
entanglement matrix along y-direction; (d) (b) An illustration of knot
changing entanglement along x and y directions.}%
\end{figure}

We then use path-integral formulation to characterize the effective
Hamiltonian for a knot-crystal as
\begin{equation}
\int \mathcal{D}\psi^{\dagger}(t,\vec{x})\mathcal{D}\psi(t)e^{i\mathcal{S}%
/\hbar}%
\end{equation}
where $\mathcal{S}=\int \mathcal{L}dt$ and $\mathcal{L}=i%
%TCIMACRO{\dsum \limits_{i}}%
%BeginExpansion
{\displaystyle \sum \limits_{i}}
%EndExpansion
\psi_{i}^{\dagger}\partial_{t}\psi_{i}-\mathcal{H}_{\mathrm{coupling}}$. To
describe the knot states on 3D knot-crystal, we have introduced a
two-component fermion field as $\psi(t,\vec{x})=\left(
\begin{array}
[c]{c}%
\psi_{\uparrow}(t,\vec{x})\\
\psi_{\downarrow}(t,\vec{x})
\end{array}
\right)  $ where $\uparrow,\downarrow$ label two spin degrees of freedom that
denote the two possible winding directions along a given direction $\vec{e}$.

In continuum limit, we have%
\begin{equation}
\mathcal{H}_{\mathrm{coupling}}=2aJ%
%TCIMACRO{\dsum \limits_{k}}%
%BeginExpansion
{\displaystyle \sum \limits_{k}}
%EndExpansion
\psi_{k}^{\dagger}[\sigma_{x}\cos k_{x}+\sigma_{y}\cos k_{y}+\sigma_{z}\cos
k_{z}]\psi_{k}%
\end{equation}
where the dispersion of knots is
\begin{equation}
E_{\mathrm{A/B,}k}\simeq c_{\mathrm{eff}}[(\vec{k}-\vec{k}_{0})\cdot
\vec{\sigma}],
\end{equation}
where $\vec{k}_{0}=(\frac{\pi}{2},\frac{\pi}{2},\frac{\pi}{2})$ and
$c_{\mathrm{eff}}=2aJ$ is the velocity. In the following part we ignore
$\vec{k}_{0}$.

From above equation, in the limit $\left \vert \vec{k}\right \vert \rightarrow0$
we derive low energy effective Hamiltonian as
\begin{align}
\mathcal{H}_{\mathrm{3D}}  &  \simeq2aJ%
%TCIMACRO{\dsum \limits_{k}}%
%BeginExpansion
{\displaystyle \sum \limits_{k}}
%EndExpansion
\psi_{k}^{\dagger}(\vec{\sigma}\cdot \vec{k})\psi_{k}\\
&  =c_{\mathrm{eff}}\int \psi^{\dagger}(\vec{\sigma}\cdot \hat{k})\psi d^{3}x.
\end{align}
We then re-write the effective Hamiltonian to be
\begin{equation}
\mathcal{H}_{\mathrm{3D}}=\int(\psi^{\dagger}\hat{H}_{\mathrm{3D}}\psi)d^{3}x
\end{equation}
and
\begin{equation}
\hat{H}_{\mathrm{3D}}=c_{\mathrm{eff}}\vec{\sigma}\cdot \vec{p}_{\mathrm{knot}}%
\end{equation}
where $\vec{p}=\hbar_{\mathrm{knot}}\vec{k}$ is the momentum operator.
$c_{\mathrm{eff}}$ play the role of light speed where $a$ is a fixed length
that denotes the half pitch of the windings on the knot-crystal.

The Schr\"{o}dinger equation for knot becomes
\begin{equation}
i\hbar_{\mathrm{eff}}\frac{d\psi(\vec{x},t)}{dt}=\hat{H}_{\mathrm{3D}}%
\psi(\vec{x},t).
\end{equation}
With help of Schr\"{o}dinger equation, we can predict the spacial distribution
of fragmentized$\  \frac{1}{N}$-knots by varying the rotating velocity,
$\omega_{0}\rightarrow \omega_{0}+\Delta \omega$. In the following parts, we set
$\hbar_{\mathrm{knot}}=1$ and $c_{\mathrm{eff}}=1$.

\section{Emergent quantum field theory for 1-level knot-crystal with
($\mathcal{N}=2$, $\mathcal{M}=1$)}

The dynamic of T-type knot on a 1-level SOC knot-crystal with ($\mathcal{N}%
=2$, $\mathcal{M}=1$) has been developed in Ref.\cite{kou}. In Ref.\cite{kou},
the low energy effective Hamiltonian of knot has been obtained as
\begin{equation}
\mathcal{H}_{\mathrm{3D}}=\int(\Psi^{\dagger}\hat{H}_{\mathrm{3D}}\Psi)d^{3}x
\end{equation}
and
\begin{equation}
\hat{H}_{\mathrm{3D}}=c_{\mathrm{eff}}\vec{\Gamma}\cdot \vec{p}_{\mathrm{knot}%
}+m_{\mathrm{knot}}c_{\mathrm{eff}}^{2}\Gamma^{5}%
\end{equation}
where
\begin{align}
\Gamma^{5}  &  =\tau^{x}\otimes \vec{1}\mathbf{,}\text{ }\Gamma^{1}=\tau
^{z}\otimes \sigma^{x},\\
\Gamma^{2}  &  =\tau^{z}\otimes \sigma^{y},\text{ }\Gamma^{3}=\tau^{z}%
\otimes \sigma^{z}.\nonumber
\end{align}
$\vec{p}=\hbar_{\mathrm{knot}}\vec{k}$ is the momentum operator.
$\Psi^{\dagger}=(\psi_{\mathrm{A},\uparrow}^{\ast},\psi_{\mathrm{B},\uparrow
},\psi_{\mathrm{A},\downarrow}^{\ast},\psi_{\mathrm{B},\downarrow})$ is the
annihilation operator of four-component fermions. $m_{\mathrm{knot}%
}c_{\mathrm{eff}}^{2}=2\hbar_{\mathrm{knot}}\omega^{\ast}$ plays role of the
mass of knots. In the following parts, we set $\hbar_{\mathrm{knot}}=1$ and
$c_{\mathrm{eff}}=1$.

The low energy effective Lagrangian of 3D SOC knot-crystal is
\begin{align}
\mathcal{L}_{\mathrm{3D}}  &  =i\Psi^{\dagger}\partial_{t}\Psi-\mathcal{H}%
_{\mathrm{3D}}\\
&  =\bar{\Psi}(i\gamma^{\mu}\hat{\partial}_{\mu}-m_{\mathrm{knot}}%
)\Psi \nonumber
\end{align}
where $\bar{\Psi}=\Psi^{\dagger}\gamma^{0},$ $\gamma^{\mu}$ are the reduced
Gamma matrices,
\begin{equation}
\gamma^{1}=\gamma^{0}\Gamma^{1},\text{ }\gamma^{2}=\gamma^{0}\Gamma^{2},\text{
}\gamma^{3}=\gamma^{0}\Gamma^{3},
\end{equation}
and
\begin{align}
\gamma^{0}  &  =\Gamma^{5}=\tau_{x}\otimes \vec{1},\\
\gamma^{5}  &  =i\gamma^{0}\gamma^{1}\gamma^{2}\gamma^{3}.\nonumber
\end{align}

\section{Emergent quantum field theory for 2-level knot-crystal with
($\mathcal{N}=2$, $\mathcal{M}=2$)}

For 2-level winding knot-crystal with ($\mathcal{N}=2$, $\mathcal{M}=2$),
there are two types of zero-lattices: level-2 W-type zero-lattice and level-1
T-type zero-lattice. There are two types of zeroes or knots: level-2 W-type
zero (knot) and level-1 T-type zero (knot) In principle, the level-2 W-type
knot becomes Weyl fermion and the level-1 T-type knot becomes Dirac fermion.
In addition, we will show that the two types of knots are coupled by
$\mathrm{SU}_{\mathrm{weak}}\mathrm{(2)}$ gauge field.

Due to the writhe-twist locking condition, there exists an intrinsic
relationship between the number of T-type zeroes and level-2 W-type zeroes,%
\begin{equation}
\zeta_{\left(  \mathrm{A,B}\right)  ,1D}^{I}=W_{\left(  \mathrm{A,B}\right)
,1D}^{I}+T_{\left(  \mathrm{A,B}\right)  ,1D}^{I}\equiv \mathrm{const},\text{
}(I=x,y,z)
\end{equation}
where $\zeta_{\left(  \mathrm{A,B}\right)  ,1D}^{I},$ $W_{\left(
\mathrm{A,B}\right)  ,1D}^{I},$ and $T_{\left(  \mathrm{A,B}\right)  ,1D}^{I}$
are linking number, writhe number (the number of level-1 W-type zeroes), and
twist number (the number of level-2 W-type zeroes), respectively. So, the
topological invariable of level-2 W-type knot and that of level-1 T-type knot
are same -- the linking number between two vortex-membranes. According to the
substitution effect of level-1 T-type of zeroes by level-2 W-type zeroes, the
properties of W-type knot and those of left-hand T-type knot are same. This
symmetry is characterized by $\mathrm{SU}_{\mathrm{weak}}\mathrm{(2)}$ gauge
symmetry -- an $\mathrm{SU(2)}$ gauge symmetry between Weyl fermions and Dirac fermions.

\subsection{Knots on 2-level knot-crystal with ($\mathcal{N}=2$,
$\mathcal{M}=2$)}

In this paper, we only consider the knot dynamics in the limit of
$\Delta_{\{1,2\}}=\frac{a_{2}}{a_{1}}\gg1$. Now, owing to the existence of two
types zeroes (the level-2 W-type zeroes and the level-1 T-type zeroes), there
exist two types of knots: the level-2 W-type knots and the level-1 T-type
knots. So, a knot is defined by changing half linking number, i.e.,
\begin{equation}
\Delta \zeta_{\left(  \mathrm{A,B}\right)  ,1D}^{I}=\pm \frac{1}{2}.
\end{equation}
For a level-2 W-type knot, the changing of half writhe number leads to the
changing half linking number, i.e.,
\begin{equation}
\Delta \zeta_{\left(  \mathrm{A,B}\right)  ,1D}^{I}=\Delta W_{\left(
\mathrm{A,B}\right)  ,1D}^{I}=\pm \frac{1}{2};
\end{equation}
For a level-1 T-type knot, the changing of half twist number leads to the
changing half linking number, i.e.,
\begin{equation}
\Delta \zeta_{\left(  \mathrm{A,B}\right)  ,1D}^{I}=T_{\left(  \mathrm{A,B}%
\right)  ,1D}^{I}=\pm \frac{1}{2}.
\end{equation}
According to the substitution effect of level-1 T-type of zeroes by level-2
W-type zeroes, the property of the level-2 W-type knots is same to that of
level-1 T-type knots.

\subsection{Low energy effective model for two types of knots}

Firstly, we consider the low energy effective model for level-1 T-type knots.

Fig.9(a) is an illustration of entanglement pattern of a 2D SOC knot-crystal
with ($\mathcal{N}=2$, $\mathcal{M}=2$). Each circle denotes a level-2 W-type
zero. The number of dotted lines that connect the two circles is considered to
a very larger number (here the number is $6$). Fig.9(b) is an illustration of
level-2 W-type knot and Fig.9(c) is an illustration of level-1 T-type knot.

\begin{figure}[ptb]
\includegraphics[clip,width=0.53\textwidth]{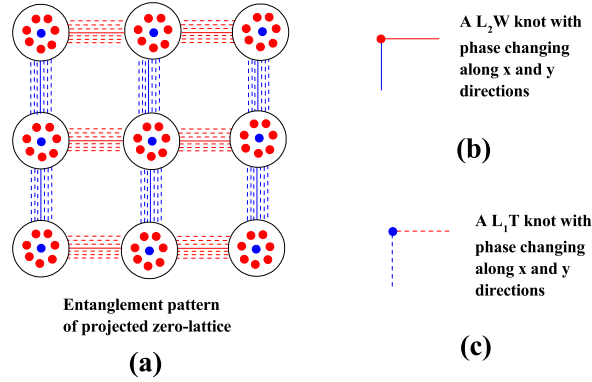}\caption{(a) An
illustration of entanglement pattern of a 2D SOC knot-crystal with
($\mathcal{N}=2$, $\mathcal{M}=2$). Each circle denotes a level-2 W-type zero.
The number of dotted lines that connect the two circles is considered to a
very larger number (here the number is $6$); (b) An illustration of level-2
W-type knot; (c) An illustration of level-1 T-type knot.}%
\end{figure}

In the limit of $\Delta_{\{1,2\}}=\frac{a_{2}}{a_{1}}\gg1,$ the winding angles
for $\mathrm{z}_{2}(\vec{x},t)$ become dynamic coordinate. Along a given
direction $\vec{e}$, the unit cell of the level-1 T-type knots turns into
$a_{2}.$ The position is determined by two kinds of values: $\vec{X}$ are
numbers of zeroes for type-2 W-type zero-lattice, i.e.,
\begin{equation}
\vec{X}=(X,Y,Z)=\frac{1}{\pi}\vec{\Phi}_{2}-\frac{1}{\pi}\vec{\Phi}%
_{2}\operatorname{mod}\pi
\end{equation}
and $\vec{\varphi}_{2}$ denote internal winding angles
\begin{equation}
\vec{\varphi}_{2}=(\varphi_{2,x},\varphi_{2,y},\varphi_{2,z})=\vec{\Phi}%
_{2}\operatorname{mod}\pi
\end{equation}
with $\varphi_{2,x},\varphi_{2,y},\varphi_{2,z}\in(0,\pi]$.

Therefore, on level-2 W-type zero-lattice, the effective Hamiltonian for
level-1 T-type knots turns into
\begin{align}
\hat{H}_{\mathrm{3D}}  &  =\vec{\Gamma}\cdot \vec{p}_{\mathrm{knot}%
}+m_{\mathrm{knot}}\Gamma^{5}\nonumber \\
&  =\vec{\Gamma}\cdot \vec{p}_{X,\mathrm{knot}}+\vec{\Gamma}\cdot \vec
{p}_{\varphi,\mathrm{knot}}+m_{\mathrm{knot}}\Gamma^{5}%
\end{align}
where $\vec{p}_{X}=\frac{1}{a_{2}}i\frac{d}{d\vec{X}}$ and $\vec{p}_{\varphi
}=\frac{1}{a_{2}}i\frac{d}{d\vec{\varphi}}$. Because of $\varphi_{j}\in
(0,\pi]$, quantum number of $\vec{p}_{\varphi}$ is angular momentum $\vec
{L}_{\varphi}$ and the energy spectra are $\frac{1}{a_{2}}\left \vert \vec
{L}_{\varphi}\right \vert .$ If we focus on the low energy physics $E\ll
\frac{1}{a_{2}}$ (or $\vec{L}_{\varphi}=0$), we may get the low energy
effective Hamiltonian as
\begin{equation}
\hat{H}_{\mathrm{3D}}\simeq \vec{\Gamma}\cdot \vec{p}_{X,\mathrm{knot}%
}+m_{\mathrm{knot}}\Gamma^{5}.
\end{equation}
The low energy effective Hamiltonian for level-1 T-type knots indicates that
the knot-pieces of level-1 T-type knots have a uniform distribution inside the
unit cell of level-2 W-type zero-lattice.

Next, we consider the low energy effective model for level-2 W-type knots.

The level-2 W-type knots change half linking number between two entangled
vortex-membranes by winding globally. However, the level-2 W-type knots on a
2-level knot-crystal with ($\mathcal{N}=2$, $\mathcal{M}=2$) are similar to
the W-type knots on a 1-level knot-crystal with ($\mathcal{N}=1$,
$\mathcal{M}=1$) and have one chirality. We may set a W-type knot to be left-hand.

As a result, in the limit of $\Delta_{\{1,2\}}=\frac{a_{2}}{a_{1}}\gg1$, every
left-hand W-type knots can be regarded as a replacement of a left-hand T-type
knot on a uniform knot-crystal. As a result, in long-wave length limit the
left-hand W-type knot has same properties and we cannot distinguish a
left-hand W-type knot and a left-hand T-type knot. If we focus on the low
energy physics $E\ll \frac{1}{a_{2}}$ (or $\vec{L}_{\varphi}=0$), we may get
the low energy effective Hamiltonian of left-hand W-type knot as
\begin{equation}
\hat{H}_{\mathrm{3D}}\simeq \vec{\sigma}\cdot \vec{p}_{X,\mathrm{knot}}.
\end{equation}
The low energy effective Hamiltonian for level-2 W-type knots indicates that
the knot-pieces of level-2 W-type knots also have a uniform distribution
inside the unit cell of level-2 W-type zero-lattice.

Finally, the effective Lagrangian of the two types of knots becomes$\left(
\begin{array}
[c]{c}%
\psi_{\mathrm{L}_{2}\mathrm{W}}\\
\psi_{\mathrm{L}_{1}\mathrm{T,L}}%
\end{array}
\right)  $%
\begin{align}
\mathcal{L}_{\mathrm{fermion}}(x)  &  =\bar{\psi}_{\mathrm{L}_{2}\mathrm{W}%
}(\vec{X})i\gamma^{\mu}\partial_{\mu}\psi_{\mathrm{L}_{2}\mathrm{W}}(\vec
{X})\nonumber \\
&  +\bar{\psi}_{\mathrm{L}_{1}\mathrm{T}}(\vec{X})i\gamma^{\mu}\partial_{\mu
}\psi_{\mathrm{L}_{1}\mathrm{T}}(\vec{X})+m_{\mathrm{L}_{1}\mathrm{T}}%
\bar{\psi}_{\mathrm{L}_{1}\mathrm{T}}(\vec{X})\psi_{\mathrm{L}_{1}\mathrm{T}%
}(\vec{X}).
\end{align}
There is no mass term for the left-hand level-2 W-type knots, i.e.,
$m_{\mathrm{L}_{2}\mathrm{W}}=0$.

\subsection{$\mathrm{SU}_{\mathrm{weak}}\mathrm{(2)}$ gauge symmetry}

An $\mathrm{SU}_{\mathrm{weak}}\mathrm{(2)}$ gauge field that couples the
level-2 W-type knots and the level-1 T-type knots characterizes the
interaction by exchanging the fluctuations of linking number on level-2 W-type
zero-lattice. The non-diagonal fluctuations of $\mathrm{SU}_{\mathrm{weak}%
}\mathrm{(2)}$ gauge theory come from the density fluctuations of linking
number and the diagonal fluctuations of $\mathrm{SU}_{\mathrm{weak}%
}\mathrm{(2)}$ gauge theory come from the phase fluctuations of the zeroes
(T-type or W-type) inside the unit cell of level-2 W-type zero-lattice. Fig.10
is an illustration of the theoretical structure of emergent $\mathrm{SU}%
_{\mathrm{weak}}\mathrm{(2)}$ non-Abelian gauge symmetry and non-Abelian gauge fields

\begin{figure}[ptb]
\includegraphics[clip,width=0.53\textwidth]{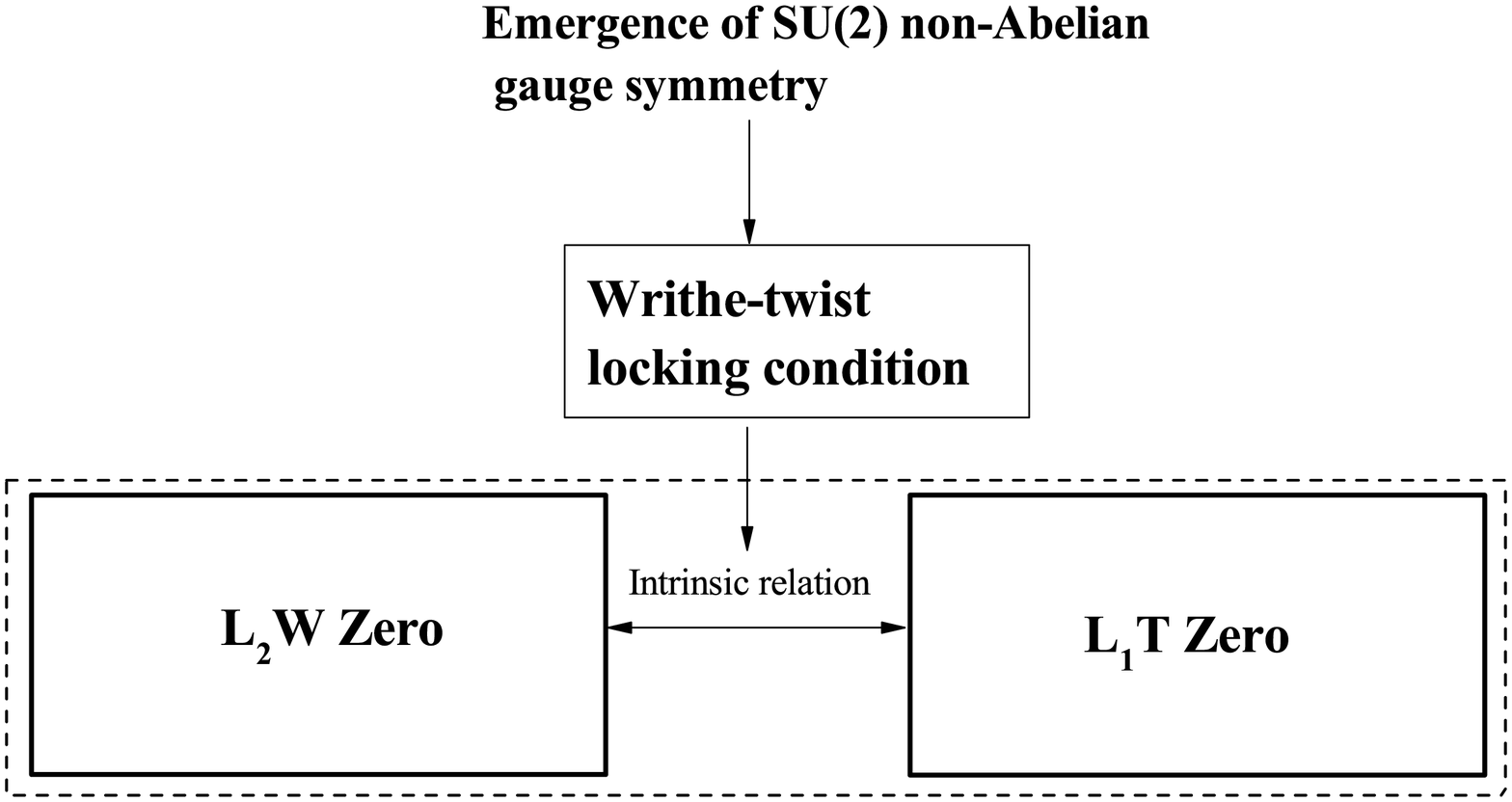}\caption{An illustration
of the theoretical structure of emergent $\mathrm{SU}_{\mathrm{weak}%
}\mathrm{(2)}$ non-Abelian gauge symmetry and non-Abelian gauge fields}%
\end{figure}

Because in long-wave length limit the left-hand level-2 W-type knot and
left-hand level-1 T-type knot have same properties, we cannot distinguish a
right-hand W-type knot and a right-hand T-type knot. In addition, a left-hand
W-type knot can change into a left-hand T-type knot, vice versa. As a result,
to characterize the symmetry of the two types of knots, a left-hand level-2
W-type knot $\psi_{\mathrm{L}_{2}\mathrm{W}}$ and a left-hand level-2 T-type
knot $\psi_{\mathrm{L}_{1}\mathrm{T,L}}$ make up a $\mathrm{SU}_{\mathrm{weak}%
}\mathrm{(2)}$ spinor\cite{ly}
\begin{equation}
\psi=\left(
\begin{array}
[c]{c}%
\psi_{\mathrm{L}_{2}\mathrm{W}}\\
\psi_{\mathrm{L}_{1}\mathrm{T,L}}%
\end{array}
\right)  .
\end{equation}
These considerations lead us to assign the left-handed components of the knots
to doublets of $\mathrm{SU}_{\mathrm{weak}}\mathrm{(2)}$
\begin{equation}
\psi_{\mathrm{L}}=\frac{1}{2}(1+\gamma_{5})\left(
\begin{array}
[c]{c}%
\psi_{\mathrm{L}_{2}\mathrm{W}}\\
\psi_{\mathrm{L}_{1}\mathrm{T,L}}%
\end{array}
\right)  \label{leptrepr1}%
\end{equation}
In physics, there exists mixed knot state as $\alpha(\vec{x},t)\psi
_{\mathrm{L}_{2}\mathrm{W}}+\beta(\vec{x},t)\psi_{\mathrm{L}_{1}\mathrm{T,L}}$
with $\alpha^{2}(\vec{x},t)+\beta^{2}(\vec{x},t)=1$.

The right-handed components are assigned to singlets of $\mathrm{SU}%
_{\mathrm{weak}}\mathrm{(2)}$ that has level-1 T-type knot and we have
\begin{equation}
\psi_{\mathrm{L}_{1}\mathrm{T,R}}=\frac{1}{2}(1-\gamma_{5})\psi_{\mathrm{L}%
_{1}\mathrm{T}}.
\end{equation}
As a result, we have
\begin{align}
\psi_{\mathrm{L}}  &  \rightarrow U_{\mathrm{SU_{\mathrm{weak}}(2)}}(\vec
{X})\psi_{\mathrm{L}},\text{ }\nonumber \\
\psi_{\mathrm{L}_{1}\mathrm{T,R}}  &  \rightarrow \psi_{\mathrm{L}%
_{1}\mathrm{T,R}}%
\end{align}
where $U_{\mathrm{SU_{\mathrm{weak}}(2)}}(\vec{X})=e^{i\vec{\tau}\vec{\theta
}(\vec{X})}$ and $\vec{\tau}$ is the three Pauli matrices.

So, $\mathrm{SU}_{\mathrm{weak}}\mathrm{(2)}$ gauge symmetry comes from the
local symmetry of two types of knots in the unit cell of level-2 W-type
zero-lattice. The Lagrangian density $\mathcal{L}_{\mathrm{YM}}(\mathrm{SU}%
_{\mathrm{weak}}\mathrm{(2)})$ is invariant under the gauge transformations
with an $x$-dependent $\Theta$. To characterize the non-Abelian gauge
symmetry, C. N. Yang and R. L. Mills in 1954 introduced \textquotedblleft
Yang-Mills field\textquotedblright, $\mathcal{A}_{\mu}(x)=\sum_{a=1}^{3}%
A_{\mu}^{a}(x)\tau^{a}\rightarrow A_{\mu}^{a}\tau^{a}$, $a=1,2,3$ that belong
to the adjoint representation of $\mathrm{SU}_{\mathrm{weak}}\mathrm{(2)}%
$\cite{yang}. In the paper of \cite{yang}, neutrons and protons had been
considered to the doublets of $\mathrm{SU}_{\mathrm{weak}}\mathrm{(2)}$.
However, in this paper, a left-hand level-2 W-type knot $\psi_{\mathrm{L}%
_{2}\mathrm{W}}$ and a left-hand level-2 T-type knot $\psi_{\mathrm{L}%
_{1}\mathrm{T,L}}$ are the doublets of $\mathrm{SU}_{\mathrm{weak}%
}\mathrm{(2)}$.

The non-Abelian gauge symmetry is represented by%
\begin{align}
\psi &  =\left(
\begin{array}
[c]{c}%
\psi_{\mathrm{L}_{2}\mathrm{W}}\\
\psi_{\mathrm{L}_{1}\mathrm{T,L}}%
\end{array}
\right) \\
&  \rightarrow \psi^{\prime}(\vec{X})=\left(
\begin{array}
[c]{c}%
\psi_{\mathrm{L}_{2}\mathrm{W}}^{\prime}(\vec{X})\\
\psi_{\mathrm{L}_{1}\mathrm{T,L}}^{\prime}(\vec{X})
\end{array}
\right)  =U_{\mathrm{SU_{\mathrm{weak}}(2)}}(\vec{X})\left(
\begin{array}
[c]{c}%
\psi_{\mathrm{L}_{2}\mathrm{W}}\\
\psi_{\mathrm{L}_{1}\mathrm{T,L}}%
\end{array}
\right)
\end{align}
and
\begin{align}
W_{\mu}(\vec{X})  &  \rightarrow U_{\mathrm{SU_{\mathrm{weak}}(2)}}(\vec
{X})W_{\mu}(\vec{X})\left(  U_{\mathrm{SU_{\mathrm{weak}}(2)}}(\vec
{X})\right)  ^{-1}\nonumber \\
&  +\frac{i}{g}\left(  \partial_{\mu}U_{\mathrm{SU_{\mathrm{weak}}(2)}}%
(\vec{X})\right)  \left(  U_{\mathrm{SU_{\mathrm{weak}}(2)}}(\vec{X})\right)
^{-1}.
\end{align}
The gauge strength is defined by $W_{\mu \nu}$ as%
\begin{align}
W_{\mu \nu}(\vec{X})  &  =\partial_{\mu}W_{v}(\vec{X})-\partial_{\nu}W_{\mu
}(\vec{X})\nonumber \\
&  -ig\left[  W_{\mu}(\vec{X}),\text{ }W_{v}(\vec{X})\right]
\end{align}
or
\begin{align}
W_{\mu \nu}(\vec{X})  &  =(\partial_{\mu}A_{\nu}^{a}(\vec{X})-\partial_{\nu
}A_{\mu}^{a}(\vec{X})\nonumber \\
&  +gf^{abc}A_{\mu}^{b}(\vec{X})A_{\nu}^{c}(\vec{X}))t^{a}%
\end{align}
where $g$ is coupling constant of $\mathrm{SU_{\mathrm{weak}}(2)}$ gauge
field. The Lagrangian of Yang-Mills field can only be written as:
\begin{equation}
\mathcal{L}_{\mathrm{YM}}(\mathrm{SU_{\mathrm{weak}}(2)})=-\frac{1}%
{2}\mathrm{Tr}W_{\mu \nu}W^{\mu \nu}+{g}\mathrm{Tr}W_{\mu}{j}_{w}^{\mu}%
\end{equation}
where the weak current is%
\begin{equation}
{j}_{w\text{ }-}^{\mu}=i\bar{\psi}_{\mathrm{L}_{1}\mathrm{T,L}}\gamma_{\mu
}\psi_{\mathrm{L}_{2}\mathrm{W}},\text{ }{j}_{w\,+}^{\mu}=i\bar{\psi
}_{\mathrm{L}_{2}\mathrm{W}}\gamma_{\mu}\psi_{\mathrm{L}_{1}\mathrm{T,L}}.
\end{equation}

We write down the effective Lagrangian of $\mathrm{SU}_{\mathrm{weak}%
}\mathrm{(2)}$ gauge theory%
\begin{align}
\mathcal{L}  &  =\mathrm{Tr}\bar{\psi}_{L}i\gamma^{\mu}(\partial_{\mu
}-igW_{\mu})\psi_{L}+\bar{\psi}_{R}i\gamma^{\mu}\partial_{\mu}\psi
_{R}\nonumber \\
&  -\mathrm{Tr}\frac{1}{2}W_{\mu \nu}W^{\mu \nu}%
\end{align}
where $W_{\mu}$ denotes the gauge fields associated to $\mathrm{SU}%
_{\mathrm{weak}}\mathrm{(2)}$ respectively, of which the corresponding field
strengths are $W_{\mu \nu}$. Because linking number of composite knot-crystal
can only be changed for left-hand knots, the charged $W$'s couple only to the
left-handed components of the two types of knots.

\subsection{Higgs mechanism and spontaneous symmetry breaking}

In this part we focus on leapfrogging motion of the composite knot-crystal, of
which the wave functions of level-1 T-type knots become time-dependent with
fixed angular velocity $\omega^{\ast}$
\begin{equation}
\psi_{\mathrm{L}_{1}\mathrm{T,A}}(\vec{X},t)\longleftrightarrow \psi
_{\mathrm{L}_{1}\mathrm{T,B}}^{\ast}(\vec{X},t).
\end{equation}
We then consider the fluctuating angular velocity of leapfrogging motion
\begin{equation}
\omega^{\ast}\rightarrow \omega(\vec{X},t)
\end{equation}
that plays the role of Higgs field $\Phi(\vec{X},t)$\cite{higgs}, i.e.,%
\[
\omega(\vec{X},t)\longleftrightarrow \Phi(\vec{X},t)/2.
\]
And the condensation of Higgs field $\left \langle \Phi(\vec{X},t)\right \rangle
\neq0$ corresponds to a finite leapfrogging angular velocity of the
knot-crystal $\omega^{\ast}\neq0$. It is the leapfrogging motion that gives
masses to knots. Due to chirality, without leapfrogging motion left-hand
level-2 W-type knot doesn't obtain mass.

We then study the properties of leapfrogging field $\omega(\vec{X},t)$ (that
is really the Higgs field $\Phi(\vec{X},t)/2$). The effect of leapfrogging
motion is to change $\psi_{\mathrm{L}_{1}\mathrm{T,L}}(\vec{X},t)$ (that is
$\psi_{\mathrm{L}_{1}\mathrm{T,A}}$) to $\psi_{\mathrm{L}_{1}\mathrm{T,R}%
}(\vec{X},t)$ (that is $\psi_{\mathrm{L}_{1}\mathrm{T,B}}^{\ast}$) and there
appears an extra term in Hamiltonian as
\begin{equation}
\psi^{\dagger}\omega(\vec{X},t)\tau_{x}\psi=\psi_{\mathrm{L}_{1}\mathrm{T,}%
R}^{\dagger}\omega(\vec{X},t)\psi_{\mathrm{L}_{1}\mathrm{T,}L}.
\end{equation}
On the other hand, due to
\begin{align}
\psi_{\mathrm{L}}(\vec{X})  &  \rightarrow U_{\mathrm{SU_{\mathrm{weak}}(2)}%
}\psi_{\mathrm{L}}(\vec{X}),\text{ }\nonumber \\
\psi_{\mathrm{R}}(\vec{X})  &  \rightarrow \psi_{\mathrm{R}}(\vec{X}),
\end{align}
$\omega(\vec{X},t)$ must be an $\mathrm{SU}_{\mathrm{weak}}\mathrm{(2)}$
complex doublet as
\begin{align}
\omega(\vec{X},t)  &  =\left(
\begin{array}
[c]{c}%
\phi^{+}\\
\phi^{0}%
\end{array}
\right)  ,\text{ }\nonumber \\
\omega(\vec{X},t)  &  \rightarrow U_{\mathrm{SU_{\mathrm{weak}}(2)}}%
\omega(\vec{X},t).
\end{align}

Next, we write down an effective Lagrangian of the leapfrogging field
$\omega(\vec{X},t).$ Because the leapfrogging field $\omega(\vec{X},t)$ is an
$\mathrm{SU}_{\mathrm{weak}}\mathrm{(2)}$ complex doublet, we get the kinetic
term of leapfrogging field $\omega(\vec{X},t)$ that is
\begin{equation}
|(\partial_{\mu}-ig\frac{\vec{\tau}}{2}\cdot \vec{W}_{\mu})\omega(\vec
{X},t)|^{2}.
\end{equation}
To obtain the finite leapfrogging angular velocity, we add a phenomenological
potential term $V(\omega(\vec{X},t)).$ Finally, by adding Yukawa coupling
between the leapfrogging field and fermions, the full Lagrangian of
leapfrogging field $\omega(\vec{X},t)$ is given by%
\begin{align}
\mathcal{L}_{\mathrm{Higgs}}  &  =|(\partial_{\mu}-ig\frac{\vec{\tau}}{2}%
\cdot \vec{W}_{\mu})\omega(\vec{X},t)|^{2}-V(\omega(\vec{X},t))\nonumber \\
&  +2\bar{\psi}\omega(\vec{X},t)\psi+h.c..
\end{align}

A finite leapfrogging angular velocity is given by minimizing $\omega(\vec
{X},t),$ of which the expected value is $\omega^{\ast}$. Then, the weak gauge
symmetry is spontaneously broken, we get a finite leapfrogging angular
velocity:
\begin{equation}
\left \langle \omega(\vec{X},t)\right \rangle =\left(
\begin{array}
[c]{c}%
0\\
\omega^{\ast}%
\end{array}
\right)  +\delta \omega(\vec{X},t).
\end{equation}
The finite leapfrogging angular velocity plays the role of Higgs condensation.
Under Higgs condensation, there exists Higgs mechanism. The Higgs mechanism
breaks the original gauge symmetry according to $\mathrm{SU}_{\mathrm{weak}%
}\mathrm{(2)}\rightarrow \mathrm{Z2}$. As a result, the $\mathrm{SU}%
_{\mathrm{weak}}\mathrm{(2)}$ gauge fields obtain masses from the following
terms\cite{wein}
\begin{equation}
\frac{1}{2}(\omega^{\ast})^{2}g^{2}(W_{\mu}W^{\mu}).
\end{equation}
The mass for the charged vector bosons $W_{\mu}\ $is $m_{W}=\omega^{\ast}g.$

Before considering Higgs condensation or $\omega^{\ast}=0,$ the low energy
effective Lagrangian density is
\begin{align}
\mathcal{L}_{\mathrm{SM}}  &  =\mathrm{Tr}\bar{\psi}_{L}i\gamma^{\mu}%
(\partial_{\mu}-ig\frac{\vec{\tau}}{2}\cdot \vec{W}_{\mu})\psi_{L}+\bar{\psi
}_{R}i\gamma^{\mu}\partial_{\mu}\psi_{R}\nonumber \\
&  -\mathrm{Tr}\frac{1}{2}W_{\mu \nu}W^{\mu \nu}+|(\partial_{\mu}-ig\frac
{\vec{\tau}}{2}\cdot \vec{W}_{\mu})\omega|^{2}\nonumber \\
&  -V(\omega)+2\bar{\psi}_{L}\omega \psi_{R}+h.c..
\end{align}
After considering the Higgs condensation $\omega^{\ast}\neq0,$ we have the low
energy effective Lagrangian as
\begin{align}
\mathcal{L}_{\mathrm{SM}}  &  =\mathrm{Tr}\bar{\psi}_{L}i\gamma^{\mu}%
(\partial_{\mu}-ig\frac{\vec{\tau}}{2}\cdot \vec{W}_{\mu})\psi_{L}+\bar{\psi
}_{R}i\gamma^{\mu}\partial_{\mu}\psi_{R}+m_{T}\bar{\psi}_{T}\psi_{T}\\
&  -\mathrm{Tr}\frac{1}{2}\mathrm{Tr}W_{\mu \nu}W^{\mu \nu}+\frac{1}{2}%
(\omega^{\ast})^{2}g^{2}\mathrm{Tr}W_{\mu}W^{\mu}+{g}\mathrm{Tr}W_{\mu}{j}%
_{w}^{\mu}\nonumber \\
&  +|\partial_{\mu}\omega|^{2}+m_{\mathrm{Higgs}}\left \vert \omega \right \vert
^{2}+...\nonumber
\end{align}
\ A finite leapfrogging angular velocity creates a mass term for the level-1
T-type knots, leaving the W-types massless,
\begin{equation}
m_{T}=2\omega^{\ast},\text{ }m_{W}=\omega^{\ast}g.
\end{equation}
In addition, the leapfrogging field also has mass $m_{\mathrm{Higgs}}.$ This
is an $\mathrm{SU}_{\mathrm{weak}}\mathrm{(2)}$ gauge theory with Higgs
mechanism due to spontaneous symmetry breaking.

\section{Emergent quantum field theory for 2-level composite knot-crystal with
($\mathcal{N}=4$, $\mathcal{M}=2$)}

In this part, we will derive the low energy effective model for knots on the
composite knot-crystal with ($\mathcal{N}=4$, $\mathcal{M}=2$). The composite
knots correspond to elementary particles in particle physics, including
electron and quarks.

\subsection{Composite knots -- definition and classification}

For 2-level composite knot-crystal with ($\mathcal{N}=4$, $\mathcal{M}=2$),
there are five zero-lattices: level-2 W-type zero-lattice for \textrm{A}%
-knot-crystal, level-2 W-type zero-lattice for \textrm{B}-knot-crystal,
level-2 T-type zero-lattice between \textrm{A}-knot-crystal and \textrm{B}%
-knot-crystal, level-1 T-type zero-lattice between two entangled
vortex-membrane-$\mathrm{A}_{1},$ $\mathrm{A}_{2}$ for \textrm{A}%
-knot-crystal, level-1 T-type zero-lattice between two entangled
vortex-membrane-$\mathrm{B}_{1},$ $\mathrm{B}_{2}$ for \textrm{B}-knot-crystal.

The knots for the composite knot-crystal are defined to correspond to the
level-2 T-type zeroes by level-2 T-type projection between \textrm{A}%
-knot-crystal and \textrm{B}-knot-crystal. The level-2 T-type knots have four
degrees of freedom, two spin degrees of freedom and two vortex degrees of
freedom. For knot on a composite knot-crystal, the linking number $\zeta
_{1D}^{I}$ along given direction is changed by $\pm \frac{1}{2},$ i.e.,
$\delta \zeta_{1D}^{I}=\pm \frac{1}{2}$. A knot (an anti-knot) removes (or adds)
a projected zero of level-2 T-type zero-lattice that corresponds to removes
(or adds) half of "lattice unit" on the level-2 T-type zero-lattice according
to $\Delta x_{i}=\pm a_{2}.$ So the size of the knot is
\begin{equation}
a_{2}\cdot a_{2}...a_{2}.
\end{equation}

However, by trapping different internal zeroes, there exist different types of
level-2 T-type knots. We use the following number series to label different
types of knots,
\begin{equation}
\lbrack n_{\mathrm{L}_{2}},\text{ }n_{\mathrm{L}_{1}}]
\end{equation}
where $n_{\mathrm{L}_{2}}$ is the half linking number of level-2 entangled
knot-crystals that is equal to the number of level-2 T-type zeroes
$n_{\mathrm{L}_{2}\mathrm{T}}$ as
\begin{equation}
n_{\mathrm{L}_{2}}=n_{\mathrm{L}_{2}\mathrm{T}},
\end{equation}
$n_{\mathrm{L}_{1}}$ is the half linking number of level-1 entangled
vortex-membranes that is equal to the sum of the number of level-1 T-type
zeroes $n_{2\mathrm{T}}$ and the number of level-2 W-type zeroes
$n_{\mathrm{L}_{2}\mathrm{W}}$ $,$
\begin{equation}
n_{\mathrm{L}_{1}}=n_{\mathrm{L}_{1}\mathrm{T}}+n_{\mathrm{L}_{2}\mathrm{W}}.
\end{equation}
For different types of knots, due to trapping half linking number of two
entangled knot-crystals, we must have
\begin{equation}
n_{\mathrm{L}_{2}\mathrm{T}}\equiv1.
\end{equation}
So the classification of the knot type is based on the internal half linking
number $n_{\mathrm{L}_{1}}$ (the linking number of level-1 entangled
vortex-membranes). For a 2-level composite knot-crystal with ($\mathcal{N}=4,$
$\mathcal{M}=2$) with an integer hierarchy number $\Delta_{\{1,2\}}=n$, there
are $n$ different types of knots with $[1,$ $0],$ $[1,$ $1],$ ... $[1,$
$n-1].$ The composite knots with $[1,$ $n]$ correspond to electrons in
particle physics and the composite knots with $[1,$ $n_{k}]$ ($n_{k}$ is an
integer number, $0<n_{k}<n$)\ correspond to quarks.

An object with level-2 half linking number is a knot and an object with
level-1 half linking number is an \emph{internal-knot}. For a free level-2
composite knot with changing $\Delta n_{\mathrm{L}_{2}}=1,$ the effective
Planck constant is
\begin{equation}
\hbar_{\mathrm{L}_{2}\mathrm{T}}=\frac{1}{2}\rho_{0}\kappa V_{P}r_{2}^{2};
\end{equation}
For a free level-1 internal knot with changing $\Delta n_{\mathrm{L}_{1}}=1$,
the effective Planck constant is%
\begin{equation}
\hbar_{\mathrm{L}_{1}\mathrm{T}}=\frac{1}{2n}\rho_{0}\kappa V_{P}r_{1}^{2}.
\end{equation}
As a result, the effective Planck constant for different knots dependents on
the number of internal zeroes $n_{k},$
\begin{align}
\hbar_{\lbrack1,n_{k}]}  &  =\hbar_{\mathrm{L}_{2}\mathrm{T}}-(n-n_{k}%
)\hbar_{\mathrm{L}_{1}\mathrm{T}}\\
&  =\frac{1}{2}\rho_{0}\kappa V_{P}r_{2}^{2}-\frac{(n-n_{k})}{2n}\rho
_{0}\kappa V_{P}r_{1}^{2}.\nonumber
\end{align}
In this paper, we focus on the case of weak coupling limit ($r_{1}\ll r_{2}$)
and have%
\begin{equation}
\hbar_{\lbrack1,n_{k}]}\simeq \hbar_{\mathrm{L}_{2}\mathrm{T}}.
\end{equation}

In general, the quantum state of a composite knot with $[1,$ $n_{i}]$ is
denoted by
\begin{equation}
\psi_{\mathrm{L}_{2}\mathrm{T},n_{i},\tau,\sigma}^{\dagger}(\vec
{x},t)\left \vert \mathrm{vac}\right \rangle
\end{equation}
where $\tau=\mathrm{A}/\mathrm{B}$ labels the vortex index, $\sigma
=\uparrow/\downarrow$ labels spin-degrees of freedom. A knot with $[1,$
$n_{i}]$ is a composite object with a knot and $n_{i}$ internal knots. Because
the knot with $[1,$ $n]$ corresponds to electron, its quantum state is denoted
by%
\begin{equation}
\psi_{\mathrm{L}_{2}\mathrm{T},n,\tau,\sigma}^{\dagger}(\vec{x},t)\left \vert
\mathrm{vac}\right \rangle =e_{\tau,\sigma}^{\dagger}(\vec{x},t)\left \vert
\mathrm{vac}\right \rangle
\end{equation}
There is $n$ internal knots inside electron. Except for electrons, other types
knots correspond to quarks with $n_{i}$ internal knots, of which quantum
states are denoted by%
\begin{equation}
\psi_{\mathrm{L}_{2}\mathrm{T},n_{i}\neq n,\tau,\sigma}^{\dagger}(\vec
{x},t)\left \vert \mathrm{vac}\right \rangle =q_{\tau,\sigma}^{\dagger}(\vec
{x},t)\left \vert \mathrm{vac}\right \rangle .
\end{equation}

We then show several 2-level composite knot-crystals with ($\mathcal{N}=4$,
$\mathcal{M}=2$).

For the case of $n=1$, there exists one type of knot without internal
additional zero ($n_{k}=0$) that is just electron.

For the case of $n=2$, except for electrons, there exists one type of quark
with one internal additional zero (or an extra internal knot), $n_{k}%
=n_{\mathrm{quark}}=1$. Due to the existence of two "sites" inside the knots,
the internal additional zero has two degenerate internal states that are
described by
\begin{equation}
\psi_{\mathrm{L}_{2}\mathrm{T},n_{i}=1,\tau,\sigma}^{\dagger}(\vec
{x},t)\left \vert \mathrm{vac}\right \rangle =q_{\tau,\sigma}^{\dagger}(\vec
{x},t)=\left \vert \mathrm{vac}\right \rangle
\end{equation}
or%
\begin{equation}
\left(
\begin{array}
[c]{c}%
\psi_{\mathrm{L}_{2}\mathrm{T},n_{i}=1,1,\tau,\sigma}^{\dagger}(\vec
{x},t)\left \vert \mathrm{vac}\right \rangle \\
\psi_{\mathrm{L}_{2}\mathrm{T},n_{i}=1,2,\tau,\sigma}^{\dagger}(\vec
{x},t)\left \vert \mathrm{vac}\right \rangle
\end{array}
\right)  =\left(
\begin{array}
[c]{c}%
q_{1,\tau,\sigma}(\vec{x},t)\left \vert \mathrm{vac}\right \rangle \\
q_{2,\tau,\sigma}(\vec{x},t)\left \vert \mathrm{vac}\right \rangle
\end{array}
\right)  .
\end{equation}

For the case of $n=3$, except for electrons, there exists two types of quarks,
u-quark and d-quark. The d-quarks are composite knots with one internal
additional zero (or an extra internal knot), $n_{k}=n_{\mathrm{d-quark}}=1$,
of which there are three degenerate states described by%
\begin{equation}
\psi_{\mathrm{L}_{2}\mathrm{T},n_{i}=1,\tau,\sigma}^{\dagger}(\vec
{x},t)\left \vert \mathrm{vac}\right \rangle =d_{\tau,\sigma}^{\dagger}(\vec
{x},t)=\left \vert \mathrm{vac}\right \rangle
\end{equation}
or
\begin{equation}
\left(
\begin{array}
[c]{c}%
\psi_{\mathrm{L}_{2}\mathrm{T},n_{i}=1,1,\tau,\sigma}^{\dagger}(\vec
{x},t)\left \vert \mathrm{vac}\right \rangle \\
\psi_{\mathrm{L}_{2}\mathrm{T},n_{i}=1,2,\tau,\sigma}^{\dagger}(\vec
{x},t)\left \vert \mathrm{vac}\right \rangle \\
\psi_{\mathrm{L}_{2}\mathrm{T},n_{i}=1,3,\tau,\sigma}^{\dagger}(\vec
{x},t)\left \vert \mathrm{vac}\right \rangle
\end{array}
\right)  =\left(
\begin{array}
[c]{c}%
d_{1,\tau,\sigma}(\vec{x},t)\left \vert \mathrm{vac}\right \rangle \\
d_{2,\tau,\sigma}(\vec{x},t)\left \vert \mathrm{vac}\right \rangle \\
d_{3,\tau,\sigma}(\vec{x},t)\left \vert \mathrm{vac}\right \rangle
\end{array}
\right)  .
\end{equation}
The three degenerate states of d-quarks are called red d-quark, blue d-quark
and green d-quark, respectively. The u-quarks are composite knots with $2$
internal-zeroes (or two extra internal knots), $n_{k}=n_{\mathrm{u-quark}}=2$,
of which the three degenerate states are described by%
\begin{equation}
\psi_{\mathrm{L}_{2}\mathrm{T},n_{i}=2,\tau,\sigma}^{\dagger}(\vec
{x},t)\left \vert \mathrm{vac}\right \rangle =u_{\tau,\sigma}^{\dagger}(\vec
{x},t)=\left \vert \mathrm{vac}\right \rangle
\end{equation}
or
\begin{equation}
\left(
\begin{array}
[c]{c}%
\psi_{\mathrm{L}_{2}\mathrm{T},n_{i}=2,1,\tau,\sigma}^{\dagger}(\vec
{x},t)\left \vert \mathrm{vac}\right \rangle \\
\psi_{\mathrm{L}_{2}\mathrm{T},n_{i}=2,2,\tau,\sigma}^{\dagger}(\vec
{x},t)\left \vert \mathrm{vac}\right \rangle \\
\psi_{\mathrm{L}_{2}\mathrm{T},n_{i}=2,3,\tau,\sigma}^{\dagger}(\vec
{x},t)\left \vert \mathrm{vac}\right \rangle
\end{array}
\right)  =\left(
\begin{array}
[c]{c}%
u_{1,\tau,\sigma}(\vec{x},t)\left \vert \mathrm{vac}\right \rangle \\
u_{2,\tau,\sigma}(\vec{x},t)\left \vert \mathrm{vac}\right \rangle \\
u_{3,\tau,\sigma}(\vec{x},t)\left \vert \mathrm{vac}\right \rangle
\end{array}
\right)  .
\end{equation}
The three degenerate states of u-quarks are called red u-quark, blue u-quark
and green u-quark, respectively.

\subsection{Emergent Dirac model}

In this part we discuss the emergent quantum field theory for composite knots
without taking into account the dynamics of internal zeroes. For different
types of knots on 2-level knot-crystal with ($\mathcal{N}=4$, $\mathcal{M}%
=2$), there also exist two types of energy costs -- the kinetic term and the
mass term from leapfrogging motion. As a result, the low energy effective
model is similar to that of knot on 1-level knot-crystal with ($\mathcal{N}%
=2$, $\mathcal{M}=1$)$.$

For simplicity, we take the case of $n=3$ as an example. By introducing
operator representation, we use the traditional effective Hamiltonian to
describe the dynamics of composite knot-crystal as
\begin{equation}
\mathcal{\hat{H}}_{\mathrm{coupling}}=\frac{J}{2}%
%TCIMACRO{\dsum \nolimits_{\left\langle ij\right\rangle }}%
%BeginExpansion
{\displaystyle \sum \nolimits_{\left \langle ij\right \rangle }}
%EndExpansion
\psi_{i}^{\dagger}T_{ij}\psi_{j}+h.c.
\end{equation}
where $T_{ij}$ is translation operator from $i$-site to $j$-site,
\begin{equation}
\psi_{i}^{\dagger}=\left(
\begin{array}
[c]{c}%
e_{i}^{\dagger}\\
u_{i}^{\dagger}\\
d_{i}^{\dagger}%
\end{array}
\right)
\end{equation}
and
\begin{equation}
u_{i}^{\dagger}=\left(
\begin{array}
[c]{c}%
u_{1,i}^{\dagger}\\
u_{2,i}^{\dagger}\\
u_{3,i}^{\dagger}%
\end{array}
\right)  ,\text{ }d_{i}^{\dagger}=\left(
\begin{array}
[c]{c}%
d_{1,i}^{\dagger}\\
d_{2,i}^{\dagger}\\
d_{3,i}^{\dagger}%
\end{array}
\right)  .
\end{equation}
The seven types of knots generated by $e_{i}^{\dagger},$ $u_{1,i}^{\dagger}%
,$\ $u_{2,i}^{\dagger},$\ $u_{3,i}^{\dagger},$\ $d_{1,i}^{\dagger},$%
\ $d_{2,i}^{\dagger},$\ $d_{3,i}^{\dagger}$\ have four degrees of freedom, two
spin degrees of freedom and two vortex (or chiral) degrees of freedom.

After considering the leapfrogging motion, we write down the low energy
effective Lagrangian as
\begin{equation}
\mathcal{L}=i\int \Psi^{\dagger}\partial_{t}\Psi d^{3}x-\mathcal{\hat{H}%
}_{\mathrm{3D}}%
\end{equation}
where
\begin{align}
\mathcal{\hat{H}}_{\mathrm{3D}}  &  \simeq \int \Psi^{\dagger}[(\tau_{z}%
\otimes \vec{\sigma}\otimes \mathbf{1})\cdot \hat{k}]\Psi d^{3}x\\
&  +2\omega^{\ast}\int \Psi^{\dagger}[\tau_{x}\otimes \vec{1}\otimes
\mathbf{1]}\Psi d^{3}x\nonumber
\end{align}
where $\mathbf{1}$ is a 7-by-7 matrix and $\vec{1}$ is 2-by-2 unit matrix.

Finally, the Lagrangian of Dirac fermion of composite knot-crystal is derived
by
\begin{align}
\mathcal{L}_{\mathrm{fermion}}(x)  &  ={\bar{e}}i\gamma^{\mu}\partial_{\mu
}e+{\bar{d}}i\gamma^{\mu}\partial_{\mu}d+{\bar{u}}i\gamma^{\mu}\partial_{\mu
}u\nonumber \\
&  +m_{e}{\bar{e}}e+m_{d}{\bar{d}}d+m_{u}{\bar{u}}u.
\end{align}
Due to leapfrogging process, all fermionic elementary particles have finite
mass,
\begin{equation}
m_{e}=m_{d}=m_{u}=2\omega^{\ast}.
\end{equation}

\subsection{Emergent U(1) gauge field}

Gauge field is an important concept in physics. By understanding
electromagnetism, Maxwell introduced the concept of the electromagnetic field
and believed that the propagation of light required a medium for the waves,
named the luminiferous aether. Quantum gauge field theory, particular the
quantum electrodynamics (QED) is an extremely successful theory of
electromagnetic interaction that agrees with experiments very well. The key
point of QED is the existence of U(1) gauge symmetry.

For 1-level knot-crystal with ($\mathcal{N}=2$, $\mathcal{M}=1$) and $n$
internal knots, there are $n$ types of composite knots that are topological
excitations in a composite knot-crystal. Except for the topological
excitations (the knots), there exist collective excitations of internal
twistings -- gauge fields. The phase fluctuations of internal zero-lattice are
\textrm{U(1)} gauge field.

\subsubsection{Phase changing from writhe-twist locking condition}

In this section, we study the knot dynamics and entanglement evolution of
internal twisting after considering the twist-writhe locking condition.

There exist two intrinsic relationships between the number of level-1 T-type
zeroes and level-2 W-type zeroes. One is twist-writhe locking relation for
\textrm{A}-knot-crystal -- the number of level-1 T-type zeroes and level-2
W-type zeroes are equal to be the writhe number $W_{\left(  \mathrm{A}%
_{1}\mathrm{,A}_{2}\right)  ,1D}^{I}$ and the twist number $T_{\left(
\mathrm{A}_{1}\mathrm{,A}_{2}\right)  ,1D}^{I}$, respectively. The other is
twist-writhe locking relation for \textrm{B}-knot-crystal -- the number of
level-1 T-type zeroes and level-2 W-type zeroes are equal to be the writhe
number $W_{\left(  \mathrm{B}_{1}\mathrm{,B}_{2}\right)  ,1D}^{I}$ and the
twist number $T_{\left(  \mathrm{B}_{1}\mathrm{,B}_{2}\right)  ,1D}^{I}$, respectively.

Now, the knot with half-linking number on a composite knot-crystal cannot be a
free "particle". Instead, it becomes a composite object by trapping half of
writhe number (a level-2 W-type zero) and half of internal twist number (a
level-1 T-type zero), i.e.,%
\begin{equation}
\delta \zeta_{\left(  \mathrm{A,B}\right)  ,1D}^{I}=\delta W_{\left(
\mathrm{A}_{1}\mathrm{,A}_{2}\right)  ,1D}^{I}=\delta T_{\left(
\mathrm{A}_{1}\mathrm{,A}_{2}\right)  ,1D}^{I}=\pm \frac{1}{2}%
\end{equation}
or
\begin{equation}
\delta \zeta_{\left(  \mathrm{A,B}\right)  ,1D}^{I}=\delta W_{\left(
\mathrm{B}_{1}\mathrm{,B}_{2}\right)  ,1D}^{I}=\delta T_{\left(
\mathrm{B}_{1}\mathrm{,B}_{2}\right)  ,1D}^{I}=\pm \frac{1}{2}.
\end{equation}
That means a knot traps half of global winding of the center-membranes and
half of internal twisting between two vortex-membranes $\mathrm{A}_{1},$
$\mathrm{A}_{2}$ or $\mathrm{B}_{1},$ $\mathrm{B}_{2}$.

According to twist-writhe locking conditions, $\delta W_{\left(
\mathrm{A}_{1}\mathrm{A}_{2}\right)  ,1D}^{I}=-\delta T_{\left(
\mathrm{A}_{1}\mathrm{A}_{2}\right)  ,1D}^{I}=\delta W_{1D,\mathrm{A}}^{I}$
and $\delta W_{\left(  \mathrm{B}_{1}\mathrm{B}_{2}\right)  ,1D}^{I}=\delta
W_{1D,\mathrm{B}}^{I}=-\delta T_{\left(  \mathrm{B}_{1}\mathrm{B}_{2}\right)
,1D}^{I}$, a global winding of \textrm{A/B}-knot-crystal leads to additional
internal twisting between vortex-membranes $\mathrm{A}_{1}$, $\mathrm{A}_{2}$
or vortex-membranes $\mathrm{B}_{1},$ $\mathrm{B}_{2}$. The quantum phase
angle of knot state is the phase angle of $\psi(\vec{x},t)$ (or the phase
angle of $\mathrm{z}_{\mathrm{A/B},2}(\vec{x},t)$) are $\phi(\vec{x}%
,t)$\ (that is $\phi_{\mathrm{A/B},2}(\vec{x},t)$); The phase angle of the
internal twisting is $\phi_{\mathrm{A/B},1}(\vec{x},t)$. Under the phase
transformation of knots, according to twist-writhe locking conditions, the
phase of internal twisting changes
\begin{equation}
\Delta \phi_{\mathrm{A/B},2}(\vec{x},t)+\Delta \phi_{\mathrm{A/B},1}(\vec
{x},t)=0
\end{equation}
or
\begin{equation}
\Delta \phi(\vec{x},t)+\Delta \phi_{\mathrm{A/B},1}(\vec{x},t)=0.
\end{equation}

\subsubsection{Emergent U(1) gauge symmetry}

The \textrm{U(1)} gauge symmetry comes from indistinguishable phase of
internal twistings inside the composite knot. Based on the hidden information
of internal twistings (or internal knots), we show the physics picture of
gauge symmetry. It is known that gauge symmetry appears as the redundancy to
define the particles. There exists redundancy to define composite knots: the
exact initial phase of an internal zero inside a composite knot is not a
physical observable value. The rule to settle down the redundancy is the rule
to fix the gauge for gauge field. The \textrm{U(1)} gauge symmetry indicates
that we may locally reorganize the knots by different ways and get the same
result. In Fig.11, we show the theoretical structure of emergent U(1) Abelian
gauge symmetry and gauge fields. \begin{figure}[ptb]
\includegraphics[clip,width=0.53\textwidth]{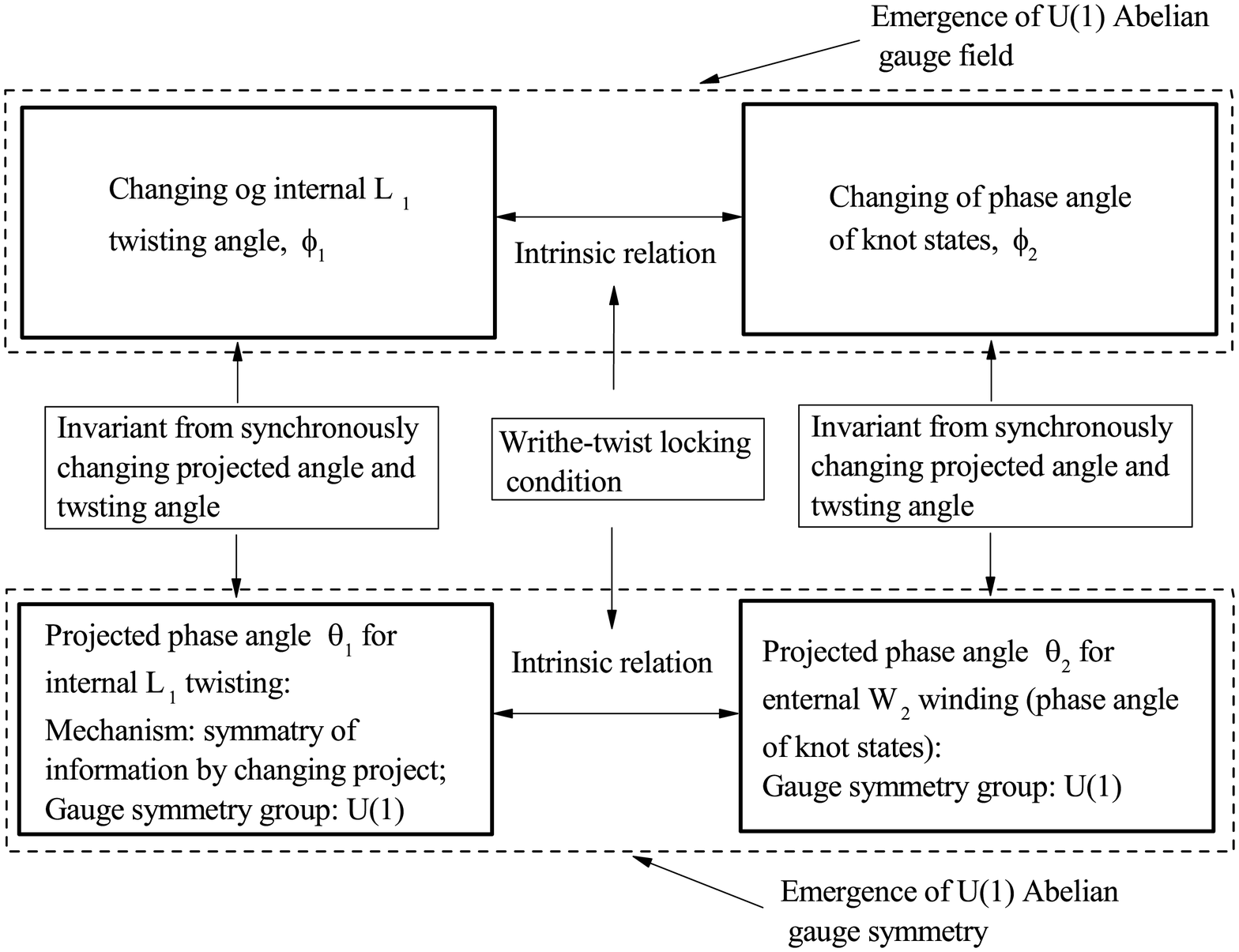}\caption{An illustration
of the theoretical structure of emergent U(1) Abelian gauge symmetry and gauge
fields}%
\end{figure}

Let us show the details. To well define a composite knot with an addition
T-type zero, we must choose a given initial phase angle of internal twistings
$(\phi_{\mathrm{A/B},1}(\vec{x},t))_{0}$ on a given point ($\vec{x},t$). Owing
to twist-writhe locking conditions, the initial phase angle of knot state
$(\phi(\vec{x},t))_{0}=\left(  \phi_{\mathrm{A/B},2}(\vec{x},t)\right)  _{0}$
on a given point ($\vec{x},t$) is chosen as $(\phi(\vec{x},t))_{0}=\left(
\phi_{\mathrm{A/B},2}(\vec{x},t)\right)  _{0}=(\phi_{\mathrm{A/B},1}(\vec
{x},t))_{0}+\phi_{0}$ where $\phi_{0}$ is constant. However, the absolute
phase angle of internal twistings is independent on the quantum phase angle of
knots. For example, we can set the initial phase angle of internal twistings
to $(\phi_{\mathrm{A/B},1}(\vec{x},t))_{0}^{\prime}\neq(\phi_{\mathrm{A/B}%
,1}(\vec{x},t))_{0}$. Different choices of initial phase angle of internal
twistings lead to same physics result. This mechanism leads to the existence
of local $\mathrm{U(1)}$ gauge symmetry.

Fig.12(a) is illustration of entanglement pattern of a 2D SOC knot-crystal
with ($\mathcal{N}=4$, $\mathcal{M}=2$) and $\Delta_{(1,2)}=1$. Each circle
denotes a level-2 W-type zero. The solid lines denote the entanglement pattern
for level-2 T-type zero-lattice. The dotted lines denote the entanglement
pattern for level-1 T-type zero-lattice. The purple dots denote internal
zeroes. Fig.12(d) is an illustration of a knot.

\begin{figure}[ptb]
\includegraphics[clip,width=0.53\textwidth]{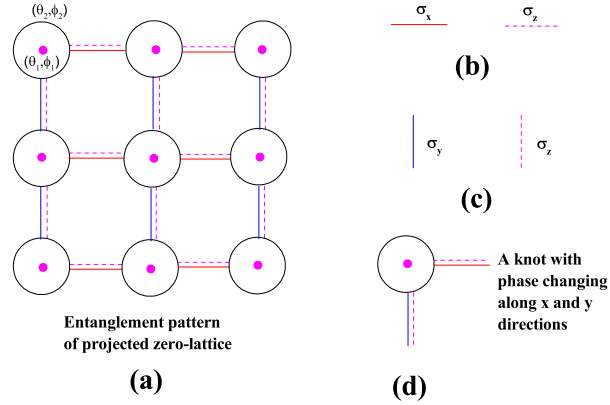}\caption{(a) An
illustration of entanglement pattern of a 2D SOC knot-crystal with
($\mathcal{N}=4$, $\mathcal{M}=2$) and $\Delta_{(1,2)}=1$. Each circle denotes
a level-2 W-type zero. The solid lines denote the entanglement pattern for
level-2 T-type zero-lattice. The dotted lines denote the entanglement pattern
for level-1 T-type zero-lattice. The purple dots denote internal zeroes; (b)
An illustration of entanglement matrices of level-1 and level-2 along
x-direction; (c) An illustration of entanglement matrices of level-1 and
level-2 along y-direction; (d) An illustration of a knot.}%
\end{figure}

According to the local \textrm{U(1)} gauge symmetry, the internal states of
knots change as following equation,
\begin{equation}
\left \vert \psi(\vec{x},t)\right \rangle \rightarrow \left \vert \psi^{\prime
}(\vec{x},t)\right \rangle =U_{\mathrm{U(1)}}\left(  \vec{x},t\right)
\left \vert \psi(\vec{x},t)\right \rangle .
\end{equation}
Due to the local \textrm{U(1)} gauge symmetry, one cannot distinguish the
state $\left \vert \psi(\vec{x},t)\right \rangle $ with $\left \vert \psi
^{\prime}(\vec{x},t)\right \rangle =U_{\mathrm{U(1)}}\left(  \vec{x},t\right)
\left \vert \psi(\vec{x},t)\right \rangle $. And the two knot states $\left \vert
\psi(\vec{x},t)\right \rangle $ and $\left \vert \psi^{\prime}(\vec
{x},t)\right \rangle =U_{\mathrm{U(1)}}\left(  \vec{x},t\right)  \left \vert
\psi(\vec{x},t)\right \rangle $ can be same by changing the initial phase angle
of internal twistings.

\subsubsection{Emergent $\mathrm{U(1)}$ gauge field}

Due to the existence of the local \textrm{U(1)} gauge symmetry, there exists
\textrm{U(1)} gauge field.

On a composite knot-crystal, the phase of knots are locked to the phase of the
internal twisting. Under a local twisting of level-1 T-type zero-lattice
$\Delta \phi_{\mathrm{A/B},1}(\vec{x},t)$, there exists corresponding global
winding of the level-2 W-type zero-lattice%
\begin{equation}
\Delta \phi(\vec{x},t)=\Delta \phi_{\mathrm{A/B},2}(\vec{x},t)=-\Delta
\phi_{\mathrm{A/B},1}(\vec{x},t).
\end{equation}
As a result, the phase of the knot at site $\vec{j}$ changes as%
\begin{equation}
\psi_{\vec{j}}\rightarrow e^{i\Delta \phi_{\vec{j}}n_{k}}\psi_{\vec{j}%
}=e^{-i\Delta \phi_{\vec{j},\mathrm{A/B},1}n_{k}}\psi_{\vec{j}}.
\end{equation}
Here, the phase changing $\Delta \phi_{\vec{j}}$ is relative to phase angle of
internal twistings $\Delta \phi_{\vec{j},\mathrm{A/B},1}$.

After considering the local changing of phase angle induced by additional
internal twisting $\Delta \phi_{\vec{j},\mathrm{A/B},1}$, the local coupling
between two knot states changes, i.e.,
\begin{align}
J\psi_{\vec{j}}^{\dagger}T_{\vec{j},\vec{j}^{\prime}}\psi_{\vec{j}^{\prime}}
&  \rightarrow J\psi_{\vec{j}}^{\dagger}e^{-in_{k}\Delta \phi_{\vec{j}}}\cdot
T_{\vec{j},\vec{j}^{\prime}}\cdot e^{i\Delta \phi_{\vec{j}^{\prime}}n_{k}}%
\psi_{\vec{j}^{\prime}}\\
&  =J\psi_{\vec{j}}^{\dagger}e^{in_{k}\Delta \phi_{\vec{j},\mathrm{A/B},1}%
}\cdot T_{\vec{j},\vec{j}^{\prime}}\cdot e^{-i\Delta \phi_{\vec{j}^{\prime
},\mathrm{A/B},1}n_{k}}\psi_{\vec{j}^{\prime}}\nonumber
\end{align}
where $T_{\vec{j},\vec{j}^{\prime}}$ is translation operator from $\vec{j}%
$-site to $\vec{j}^{\prime}$-site. We define a vector field to characterize
the local additional internal twistings $\Delta \phi_{\vec{j},1,\mathrm{A/B}},$%
\begin{align}
A_{\vec{j},\vec{j}^{\prime}}  &  =\Delta \phi_{\vec{j},\mathrm{A/B},1}%
-\Delta \phi_{\vec{j}^{\prime},\mathrm{A/B},1}\nonumber \\
&  =-\Delta \phi_{\vec{j}}+\Delta \phi_{\vec{j}^{\prime}}.
\end{align}
The local coupling between two knot states becomes
\begin{equation}
J\psi_{\vec{j}}^{\dagger}T_{\vec{j},\vec{j}^{\prime}}e^{ie_{0}n_{k}A_{\vec
{j},\vec{j}^{\prime}}}\psi_{\vec{j}^{\prime}}.
\end{equation}
The total kinetic energy for knots becomes
\begin{equation}
\mathcal{\hat{H}}_{\mathrm{coupling}}=J%
%TCIMACRO{\dsum \nolimits_{\left\langle \vec{j},\vec{j}^{\prime}\right\rangle
%}}%
%BeginExpansion
{\displaystyle \sum \nolimits_{\left \langle \vec{j},\vec{j}^{\prime
}\right \rangle }}
%EndExpansion
\psi_{\vec{j}}^{\dagger}T_{\vec{j},\vec{j}^{\prime}}e^{ie_{0}n_{k}A_{\vec
{j},\vec{j}^{\prime}}}\psi_{\vec{j}^{\prime}}+h.c..
\end{equation}

It is obvious that the vector field $A_{\vec{j},\vec{j}^{\prime}}$ that
characterizes the local position perturbation of internal zero-lattice plays
the role of \textrm{U(1)} gauge field. To illustrate the local \textrm{U(1)}
gauge symmetry, we do a local \textrm{U(1)} gauge transformation $U_{\vec
{j},\mathrm{U(1)}}=e^{i\Delta \phi_{0,\vec{j}}n_{k}}$ via changing the initial
phase angle of internal twistings
\begin{equation}
\phi_{0,\vec{j},\mathrm{A/B},1}\rightarrow \Delta \phi_{0,\vec{j},\mathrm{A/B}%
,1}^{\prime}=\phi_{0,\vec{j},\mathrm{A/B},1}+\Delta \phi_{0,\vec{j}%
,\mathrm{A/B},1}.
\end{equation}
Under above local \textrm{U(1)} gauge transformation, we have
\begin{equation}
\psi_{\vec{j}}\rightarrow \psi_{\vec{j}}^{\prime}=U_{\vec{j},\mathrm{U(1)}}%
\psi_{\vec{j}}=e^{-i\Delta \phi_{0,\vec{j}}n_{k}}\psi_{\vec{j}},
\end{equation}
and
\begin{align}
e_{0}n_{k}A_{\vec{j},\vec{j}^{\prime}}  &  \rightarrow e_{0}n_{k}A_{\vec
{j},\vec{j}^{\prime}}^{\prime}\nonumber \\
&  =e_{0}n_{k}A_{\vec{j},\vec{j}^{\prime}}-n_{k}(\Delta \phi_{0,\vec
{j},\mathrm{A/B},1}-\Delta \phi_{0,\vec{j}^{\prime},\mathrm{A/B},1}).
\end{align}
The total kinetic energy for knots turns into
\begin{equation}
\mathcal{\hat{H}}_{\mathrm{coupling}}\rightarrow \mathcal{\hat{H}%
}_{\mathrm{coupling}}^{\prime}=J%
%TCIMACRO{\dsum \nolimits_{\left\langle \vec{j},\vec{j}^{\prime}\right\rangle
%}}%
%BeginExpansion
{\displaystyle \sum \nolimits_{\left \langle \vec{j},\vec{j}^{\prime
}\right \rangle }}
%EndExpansion
\psi_{\vec{j}}^{\prime \dagger}T_{\vec{j},\vec{j}^{\prime}}e^{ie_{0}%
n_{k}A_{\vec{j},\vec{j}^{\prime}}^{\prime}}\psi_{\vec{j}^{\prime}}^{\prime
}+h.c..
\end{equation}
According to the twist-writhe locking condition $\phi_{\vec{j},\mathrm{A/B}%
,1}=-\Delta \phi_{\vec{j}}$, the Hamiltonian doesn't change,%
\begin{equation}
\mathcal{\hat{H}}_{\mathrm{coupling}}=\mathcal{\hat{H}}_{\mathrm{coupling}%
}^{\prime}.
\end{equation}

On the other hand, the situation for tempo phase changing is similar to that
for spatial phase changing. To characterize the tempo twist-writhe locking
condition, we introduce a Lagrangian variable $A_{0,\vec{j}}$ to path-integral
formulation as
\begin{equation}
A_{0,\vec{j}}\psi_{\vec{j}}^{\dagger}\psi_{\vec{j}}.
\end{equation}

In continuum limit, we have $U_{\vec{j},\mathrm{U(1)}}(t)\rightarrow
U_{\mathrm{U(1)}}(\vec{x},t)$, $A_{\vec{j},\vec{j}^{\prime}}\rightarrow \vec
{A}(x)\ $and $A_{0,\vec{j}}\rightarrow A_{0}(x).$ The Abelian gauge symmetry
is represented by%
\begin{equation}
\psi^{\prime}\rightarrow U_{\mathrm{SU(n)}}(\vec{x},t)\psi
\end{equation}
and
\begin{align}
A_{\mu}(\vec{x},t)  &  \rightarrow A_{\mu}(\vec{x},t)+i\left(  \partial_{\mu
}U_{\mathrm{U(1)}}(\vec{x},t)\right)  \left(  U_{\mathrm{U(1)}}(\vec
{x},t)\right)  ^{-1}\nonumber \\
&  =A_{\mu}(\vec{x},t)+\frac{1}{e_{0}n_{k}}\partial_{\mu}\phi(\vec{x},t).
\end{align}

Finally, we derive the path-integral formulation to characterize the effective
Hamiltonian in continuum limit for gauge field
\begin{equation}
\int \mathcal{D}A_{0}\mathcal{D}\vec{A}e^{i\mathcal{S}_{\mathrm{EM}}/\hbar}%
\end{equation}
where $\mathcal{S}_{\mathrm{EM}}=\int \mathcal{L}_{\mathrm{EM}}dtd^{3}x$ and
\begin{equation}
\mathcal{L}_{\mathrm{EM}}=\mathrm{e}A_{\mu}{j}_{(em)}^{\mu}-\frac{1}{4}%
F_{\mu \nu}F^{\mu \nu}. \label{e}%
\end{equation}
The gauge field strength $F_{\mu \nu}$ is defined by $F_{\mu \nu}=\partial_{\mu
}A_{\nu}-\partial_{\nu}A_{\mu}.$ The electric charge for a internal twisting
zero is $e_{0}.$ As a result, the total electric charge of a composite knot
with $n$-internal zeroes is just $\mathrm{e}=ne_{0}$. It is obvious that
$\mathcal{L}_{\mathrm{EM}}$ has local $\mathrm{U(1)}$ gauge symmetry. On the
contrary, people always obtain above formula ($\mathcal{L}_{\mathrm{EM}%
}=\mathrm{e}A_{\mu}{j}_{(em)}^{\mu}-\frac{1}{4}F_{\mu \nu}F^{\mu \nu}$) from
point view of local gauge symmetry.

From Eq.\ref{e}, we can derive the Maxwell equations
\begin{equation}
\partial_{\mu}F^{\mu \nu}={j}_{(em)}^{\mu}%
\end{equation}
where ${j}_{(em)}^{\mu}$ is the electric current. In addition, to give a
correct definition of knots in a composite knot-crystal, we must set down the
"gauge" $f(A_{\mu})=0$ that leads to a fixed $\phi_{0,\vec{j},\mathrm{A/B}%
,1}(\vec{x},t)$ at a give position $(\vec{x},t)$ for internal twisting.

For the case of $n=1$, we derive the $\mathrm{U(1)}$ gauge theory for
composite knot-crystal. Now, the charge of an electron is $e_{0}$. As a
result, the effective Lagrangian for knots turns into%
\begin{equation}
\mathcal{L}_{\mathrm{fermion}}(x)={\bar{e}}(x)i\gamma^{\mu}\partial_{\mu
}e(x)+m_{e}{\bar{e}}(x)e(x)+e_{0}{A}_{\mu}(x){j}_{(em)}^{\mu}(x)
\end{equation}
where
\begin{equation}
{j}_{(em)}^{\mu}(x)=i{\bar{e}}(x)\gamma^{\mu}e(x).
\end{equation}

For the case of $n=2$, the charge of an electron is $2e_{0};$ the charge of a
quark is $n_{\mathrm{quark}}\cdot e_{0}=e_{0}$. As a result, the effective
Lagrangian for knots turns into%
\begin{align}
\mathcal{L}_{\mathrm{fermion}}(x)  &  ={\bar{e}}(x)i\gamma^{\mu}\partial_{\mu
}e(x)+{\bar{q}}(x)i\gamma^{\mu}\partial_{\mu}q(x)\nonumber \\
&  +m_{e}{\bar{e}}(x)e(x)+m_{q}{\bar{q}}(x)q(x)+\mathrm{e}{A}_{\mu}%
(x){j}_{(em)}^{\mu}(x)
\end{align}
where%
\begin{equation}
{j}_{(em)}^{\mu}(x)=i{\bar{e}}(x)\gamma^{\mu}e(x)+i\frac{1}{2}{\bar{q}%
}(x)\gamma^{\mu}q(x).
\end{equation}

For the case of $n=3$, the charge of an electron is $3e_{0};$ the charge of a
u-quark is $n_{\mathrm{u-quark}}\cdot e_{0}=2e_{0},$ the charge of a d-quark
is $n_{\mathrm{d-quark}}=e_{0}$. As a result, the effective Lagrangian for
knots turns into%
\begin{align}
\mathcal{L}_{\mathrm{fermion}}(x)  &  ={\bar{e}}(x)i\gamma^{\mu}\partial_{\mu
}e(x)+{\bar{d}}(x)i\gamma^{\mu}\partial_{\mu}d(x)+{\bar{u}}(x)i\gamma^{\mu
}\partial_{\mu}u(x)\nonumber \\
&  +m_{e}{\bar{e}}(x)e(x)+m_{d}{\bar{d}}(x)d(x)+m_{u}{\bar{u}}%
(x)u(x)+\mathrm{e}{A}_{\mu}(x){j}_{(em)}^{\mu}(x)
\end{align}
where%
\begin{align}
{j}_{(em)}^{\mu}(x)  &  =i{\bar{e}}(x)\gamma^{\mu}e(x)+i\frac{2}{3}{\bar{u}%
}(x)\gamma^{\mu}u(x)\\
&  +i\frac{1}{3}{\bar{d}}(x)\gamma^{\mu}d(x).\nonumber
\end{align}
Here, we have set the charge of an electron $3e_{0}$ to be $\mathrm{e}$.

Finally, we show the physical picture of the $\mathrm{U(1)}$ gauge field.
$\mathrm{U(1)}$ gauge field comes from phase fluctuations of internal
zero-lattice. Two composite knots interact by exchanging phase fluctuations of
internal zero-lattice. An extra internal zeroes plays the role of source of
electric field. When there exists an extra internal-winding, the composite
knots will be attractive or repulsive to the source. When there exists finite
magnetic field, the path of moving composite knots will be curved. A photon is
a local internal twisting changing. Because the longitudinal fluctuation of
internal twisting is "eaten" knots due to writhe-twist locking condition, the
fluctuations of internal twisting have only transverse modes. This is why the
gauge fields are transverse waves with spin-1.

In addition, we compare the physics picture of $\mathrm{U(1)}$ gauge symmetry
in composite knot-crystal and that in Kaluza--Klein theory\cite{kk}. In knot
physics, each internal zero corresponds to a $\pi$-flux in extra space
($x_{d+1}$-$x_{d+2}$ space) and the deformation of internal zero-lattice
become\textrm{ }electromagnetic field, the fluctuations of internal
zero-lattice become Maxwell waves. For composite knot with a internal zero,
the total flux in extra space is $\pi,$ of which the total electric charge is
$e_{0}$. The situation is very similar to the emergent gauge symmetry in
Kaluza--Klein theory, in which the particles have finite angular momentum $m$
in extra space that corresponds to a half internal twisting (a internal zero)
in knot physics.

\subsection{Emergent $\mathrm{SU(n)}$ gauge field}

\subsubsection{Emergent SU(n) gauge symmetry}

In this part we discuss the emergent quantum field theory for composite knots
with taking into account the dynamics of internal zeroes/knots. For a 2-level
composite knot-crystal with ($\mathcal{N}=4$, $\mathcal{M}=2$) and an integer
hierarchy number $\Delta_{\{1,2\}}=n$, there are $n$ different types of knots
with $[1,$ $0],$ $[1,$ $1],$ ... $[1,$ $n-1].$ The composite knots with $[1,$
$n]$ correspond to electrons in particle physics and the composite knots with
$[1,$ $n_{k}]$ ($n_{k}$ is an integer number, $0<n_{k}<n$)\ correspond to
quarks. We point out that an emergent $\mathrm{SU(n)}$ non-Abelian gauge field
describes the knot dynamics on a composite knot-crystal. The \textrm{SU(n)}
gauge symmetry comes from indistinguishable states of internal-knots inside
the composite knot. In Fig.13, we show the theoretical structure of emergent
\textrm{SU(n)} non-Abelian gauge symmetry and non-Abelian gauge fields.

\begin{figure}[ptb]
\includegraphics[clip,width=0.53\textwidth]{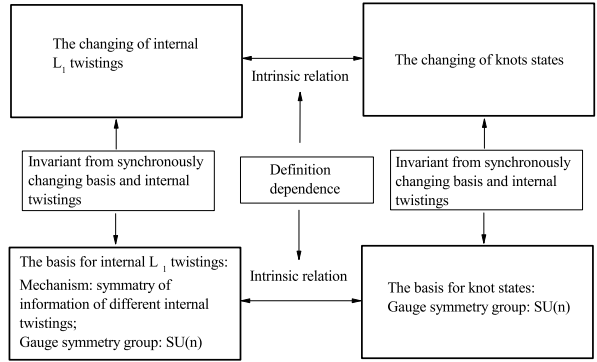}\caption{An illustration
of the theoretical structure of emergent \textrm{SU(n)} non-Abelian gauge
symmetry and non-Abelian gauge fields}%
\end{figure}

For the knot-crystal, each level-2 type-T zero (for example, a knot at
$\left(  \vec{x},t\right)  $) corresponds to $n$ level-1 internal type-T
zeroes. Because each internal zero corresponds to an internal knot, we label
the $n$ internal zeroes by $1$, $2$, ..., $n$ and the corresponding states of
internal knots by $\left \vert \psi_{\mathrm{inter},1}\right \rangle ,$
$\left \vert \psi_{\mathrm{inter},2}\right \rangle ,$ ..., $\left \vert
\psi_{\mathrm{inter},n}\right \rangle ,$ respectively. $\left \vert
\psi_{\mathrm{inter},1}\right \rangle ,$ $\left \vert \psi_{\mathrm{inter}%
,2}\right \rangle ,$ ..., $\left \vert \psi_{\mathrm{inter},n}\right \rangle $
have the same properties and $\left \vert \psi_{\mathrm{inter},1}\right \rangle
,$ $\left \vert \psi_{\mathrm{inter},2}\right \rangle ,$ ..., $\left \vert
\psi_{\mathrm{inter},n}\right \rangle $ make up a complete basis. In general,
due to symmetry of $\left \vert \psi_{\mathrm{inter},1}\right \rangle ,$
$\left \vert \psi_{\mathrm{inter},2}\right \rangle ,$ ..., $\left \vert
\psi_{\mathrm{inter},n}\right \rangle $, we can re-label the corresponding
states of internal knots by $\left \vert \psi_{\mathrm{inter},1}^{\prime
}\right \rangle ,$ $\left \vert \psi_{\mathrm{inter},2}^{\prime}\right \rangle ,$
..., $\left \vert \psi_{\mathrm{inter},n}^{\prime}\right \rangle $. The
relationship between the two basis is
\begin{equation}
\left(
\begin{array}
[c]{c}%
\left \vert \psi_{\mathrm{inter},1}^{\prime}\left(  \vec{x},t\right)
\right \rangle \\
\left \vert \psi_{\mathrm{inter},2}^{\prime}\left(  \vec{x},t\right)
\right \rangle \\
...\\
\left \vert \psi_{\mathrm{inter},n}^{\prime}\left(  \vec{x},t\right)
\right \rangle
\end{array}
\right)  =U_{\mathrm{SU(n)}}\left(  \vec{x},t\right)  \left(
\begin{array}
[c]{c}%
\left \vert \psi_{\mathrm{inter},1}\right \rangle \\
\left \vert \psi_{\mathrm{inter},2}\right \rangle \\
...\\
\left \vert \psi_{\mathrm{inter},n}\right \rangle
\end{array}
\right)
\end{equation}
where $U_{\mathrm{SU(n)}}\left(  \vec{x},t\right)  =e^{i\Theta \left(  \vec
{x},t\right)  }$ is the matrix of the representation of $\mathrm{SU(n)}$
group. $\Theta \left(  \vec{x},t\right)  =\sum_{a=1}^{n^{2}-1}\theta^{a}\left(
\vec{x},t\right)  \tau^{a}$ and $\theta^{a}$ are a set of $n^{2}-1$ constant
parameters, and $\tau^{a}$ are $n^{2}-1$ $n\times n$ matrices representing the
$n^{2}-1$ generators of the Lie algebra of $\mathrm{SU(n)}$\cite{yang}. This
result leads to the concept of non-Abelian gauge symmetry.

Thus, we have a local $\mathrm{SU(n)}$ symmetry that denotes the
indistinguishable states of internal-knots inside the composite knot.

For the case of $n=2$, there are two internal zeroes inside a knot that is
labeled by $1$, $2$. The corresponding basis of states of internal knots is
$\left(
\begin{array}
[c]{c}%
\left \vert \psi_{\mathrm{inter},1}(\vec{x},t)\right \rangle \\
\left \vert \psi_{\mathrm{inter},2}(\vec{x},t)\right \rangle
\end{array}
\right)  .$ Due to the local \textrm{SU(2)} gauge symmetry, the basis can be
arbitrary re-defined, i.e.,%
\begin{align}
\left(
\begin{array}
[c]{c}%
\left \vert \psi_{\mathrm{inter},1}(\vec{x},t)\right \rangle \\
\left \vert \psi_{\mathrm{inter},2}(\vec{x},t)\right \rangle
\end{array}
\right)   &  \rightarrow \left(
\begin{array}
[c]{c}%
\left \vert \psi_{\mathrm{inter},1}^{\prime}(\vec{x},t)\right \rangle \\
\left \vert \psi_{\mathrm{inter},2}^{\prime}(\vec{x},t)\right \rangle
\end{array}
\right) \\
&  =U_{\mathrm{SU(2)}}\left(  \vec{x},t\right)  \left(
\begin{array}
[c]{c}%
\left \vert \psi_{\mathrm{inter},1}(\vec{x},t)\right \rangle \\
\left \vert \psi_{\mathrm{inter},2}(\vec{x},t)\right \rangle
\end{array}
\right) \nonumber
\end{align}
where $U_{\mathrm{SU(2)}}\left(  \vec{x},t\right)  =e^{i\Theta \left(  \vec
{x},t\right)  }$ is the matrix of the representation of $\mathrm{SU(2)}$
group. $\Theta \left(  \vec{x},t\right)  =\sum_{a=1}^{3}\theta^{a}\left(
\vec{x},t\right)  \tau^{a}$ and $\theta^{a}$ are a set of $3$ constant
parameters, and $\tau^{a}$ are $2\times2$ Pauli matrices representing the $3$
generators of the Lie algebra of $\mathrm{SU(2)}$\cite{yang}.

When there exists a knot with $[1,$ $1],$ the knot states are denoted by
\begin{align}
\left(
\begin{array}
[c]{c}%
\left \vert q_{1}(\vec{x},t)\right \rangle \\
\left \vert q_{2}(\vec{x},t)\right \rangle
\end{array}
\right)   &  =\left(
\begin{array}
[c]{c}%
q^{\dagger}(\vec{x},t)\left \vert \psi_{\mathrm{inter},1}(\vec{x}%
,t)\right \rangle \\
q^{\dagger}(\vec{x},t)\left \vert \psi_{\mathrm{inter},2}(\vec{x}%
,t)\right \rangle
\end{array}
\right) \\
&  =\left(
\begin{array}
[c]{c}%
q_{1}^{\dagger}(\vec{x},t)\left \vert \mathrm{vac}\right \rangle \\
q_{2}^{\dagger}(\vec{x},t)\left \vert \mathrm{vac}\right \rangle
\end{array}
\right) \nonumber
\end{align}
where $\left \vert q_{1}(\vec{x},t)\right \rangle $ denotes a knot state with an
internal knot of 1-th internal-zero and $\left \vert q_{2}(\vec{x}%
,t)\right \rangle $ denotes a knot state with an internal knot of 2-th
internal-zero. $q^{\dagger}(\vec{x},t)$ is the creation operator of level-2
knot with two internal knot states.\ Because both the internal knot state
$\left \vert q_{1}(\vec{x},t)\right \rangle $ and the internal knot state
$\left \vert q_{2}(\vec{x},t)\right \rangle $ change level-1 half linking
number, we cannot distinguish the knots with the two different internal states
by detecting phase changing on level-2. As a result, the knot states $\left(
\begin{array}
[c]{c}%
\left \vert q_{1}(\vec{x},t)\right \rangle \\
\left \vert q_{2}(\vec{x},t)\right \rangle
\end{array}
\right)  $ has local \textrm{SU(2)} gauge symmetry, i.e.,
\begin{align}
\left(
\begin{array}
[c]{c}%
\left \vert q_{1}(\vec{x},t)\right \rangle \\
\left \vert q_{2}(\vec{x},t)\right \rangle
\end{array}
\right)   &  \rightarrow \left(
\begin{array}
[c]{c}%
\left \vert q_{1}^{\prime}(\vec{x},t)\right \rangle \\
\left \vert q_{2}^{\prime}(\vec{x},t)\right \rangle
\end{array}
\right)  =\left(
\begin{array}
[c]{c}%
q^{\dagger}(\vec{x},t)\left \vert \psi_{\mathrm{inter},1}^{\prime}(\vec
{x},t)\right \rangle \\
q^{\dagger}(\vec{x},t)\left \vert \psi_{\mathrm{inter},2}^{\prime}(\vec
{x},t)\right \rangle
\end{array}
\right) \\
&  =\left(
\begin{array}
[c]{c}%
q^{\dagger}(\vec{x},t)U_{\mathrm{SU(2)}}\left(  \vec{x},t\right)  \left \vert
\psi_{\mathrm{inter},1}(\vec{x},t)\right \rangle \\
q^{\dagger}(\vec{x},t)U_{\mathrm{SU(2)}}\left(  \vec{x},t\right)  \left \vert
\psi_{\mathrm{inter},2}(\vec{x},t)\right \rangle
\end{array}
\right) \nonumber \\
&  =U_{\mathrm{SU(2)}}\left(  \vec{x},t\right)  \left(
\begin{array}
[c]{c}%
q^{\dagger}(\vec{x},t)\left \vert \psi_{\mathrm{inter},1}(\vec{x}%
,t)\right \rangle \\
q^{\dagger}(\vec{x},t)\left \vert \psi_{\mathrm{inter},2}(\vec{x}%
,t)\right \rangle
\end{array}
\right) \nonumber \\
&  =U_{\mathrm{SU(2)}}\left(  \vec{x},t\right)  \left(
\begin{array}
[c]{c}%
\left \vert q_{1}(\vec{x},t)\right \rangle \\
\left \vert q_{2}(\vec{x},t)\right \rangle
\end{array}
\right)  .\nonumber
\end{align}
Therefore, the quarks also have local \textrm{SU(2)} gauge symmetry. This
leads to the existence of \textrm{SU(2) }non-Abelian gauge fields.

Fig.14(a) is an illustration of entanglement pattern of a 2D SOC knot-crystal
with ($\mathcal{N}=4$, $\mathcal{M}=2$) and $\Delta_{(1,2)}=2$. Each circle
denotes a level-2 W-type zero. The solid lines denote the entanglement pattern
for level-2 T-type zero-lattice. The dotted lines denote the entanglement
pattern for level-1 T-type zero-lattice. The dots inside the circles denote
internal zeroes. Fig.14(b) and Fig.14(c) illustrate $[1,$ $1]$ knot and
$\left[  1,\text{ }2\right]  $ knot, respectively.

\begin{figure}[ptb]
\includegraphics[clip,width=0.53\textwidth]{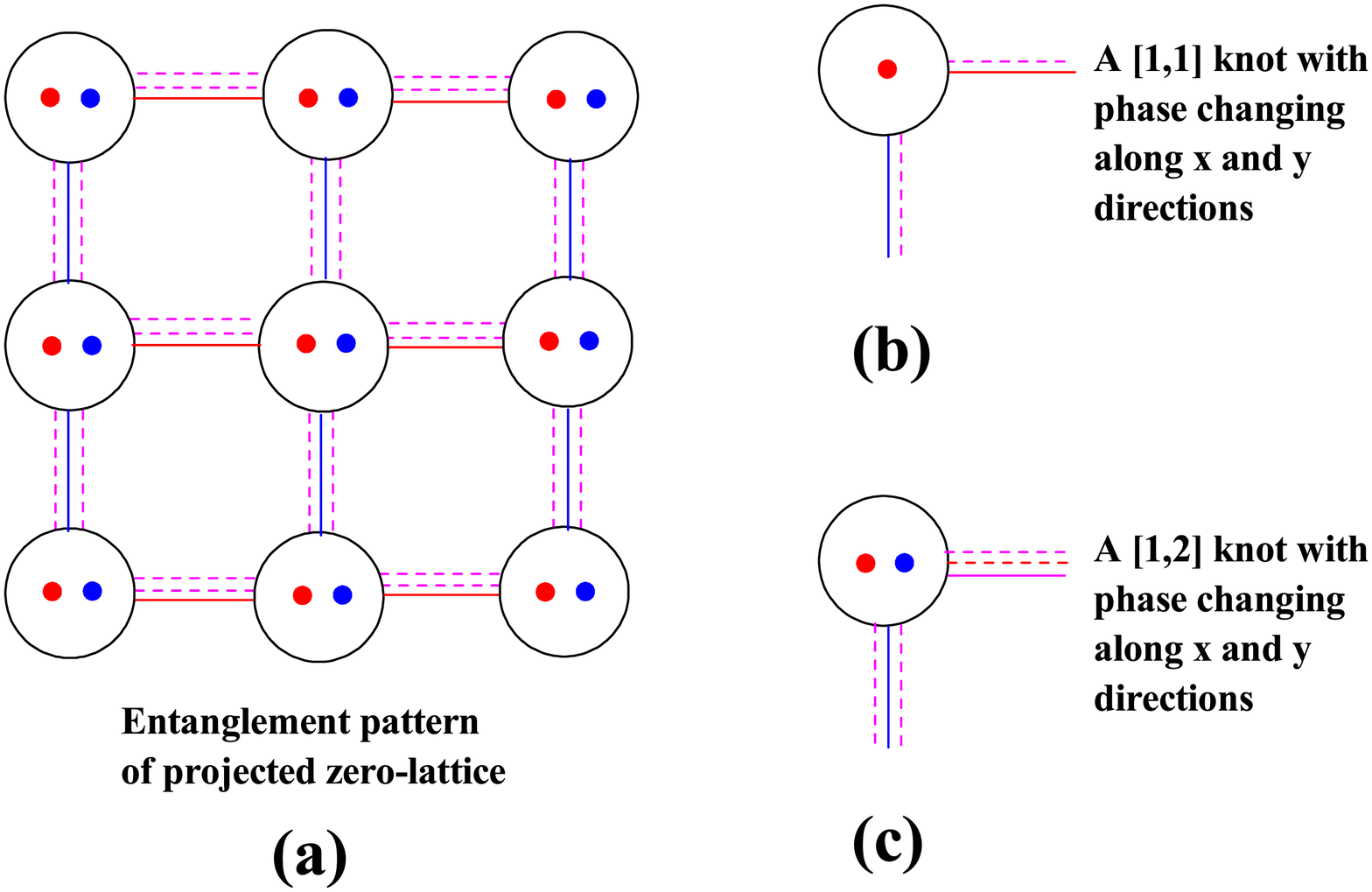}\caption{(a) An
illustration of entanglement pattern of a 2D SOC knot-crystal with
($\mathcal{N}=4$, $\mathcal{M}=2$) and $\Delta_{(1,2)}=2$. Each circle denotes
a level-2 W-type zero. The solid lines denote the entanglement pattern for
level-2 T-type zero-lattice. The dotted lines denote the entanglement pattern
for level-1 T-type zero-lattice. The dots inside the circles denote internal
zeroes; (b) An illustration of a $\left[  1,1\right]  $ knot; (c) An
illustration of a $\left[  1,2\right]  $ knot.}%
\end{figure}

According to the local \textrm{SU(2)} gauge symmetry, the internal states of
quarks change as following equation,
\begin{equation}
\left(
\begin{array}
[c]{c}%
\left \vert q_{1}(\vec{x},t)\right \rangle \\
\left \vert q_{2}(\vec{x},t)\right \rangle
\end{array}
\right)  \rightarrow \left(
\begin{array}
[c]{c}%
\left \vert q_{1}^{\prime}(\vec{x},t)\right \rangle \\
\left \vert q_{2}^{\prime}(\vec{x},t)\right \rangle
\end{array}
\right)  =U_{\mathrm{SU(2)}}\left(  \vec{x},t\right)  \left(
\begin{array}
[c]{c}%
\left \vert q_{1}(\vec{x},t)\right \rangle \\
\left \vert q_{2}(\vec{x},t)\right \rangle
\end{array}
\right)  .
\end{equation}
Due to the local \textrm{SU(2)} gauge symmetry, one cannot distinguish the
state $\left(
\begin{array}
[c]{c}%
\left \vert q_{1}(\vec{x},t)\right \rangle \\
\left \vert q_{2}(\vec{x},t)\right \rangle
\end{array}
\right)  $ with $\left(
\begin{array}
[c]{c}%
\left \vert q_{1}^{\prime}(\vec{x},t)\right \rangle \\
\left \vert q_{2}^{\prime}(\vec{x},t)\right \rangle
\end{array}
\right)  =U_{\mathrm{SU(2)}}\left(  \vec{x},t\right)  \left(
\begin{array}
[c]{c}%
\left \vert q_{1}(\vec{x},t)\right \rangle \\
\left \vert q_{2}(\vec{x},t)\right \rangle
\end{array}
\right)  $. The knot states $\left(
\begin{array}
[c]{c}%
\left \vert q_{1}(\vec{x},t)\right \rangle \\
\left \vert q_{2}(\vec{x},t)\right \rangle
\end{array}
\right)  $ and the knot states $\left(
\begin{array}
[c]{c}%
\left \vert q_{1}^{\prime}(\vec{x},t)\right \rangle \\
\left \vert q_{2}^{\prime}(\vec{x},t)\right \rangle
\end{array}
\right)  =U_{\mathrm{SU(2)}}\left(  \vec{x},t\right)  \left(
\begin{array}
[c]{c}%
\left \vert q_{1}(\vec{x},t)\right \rangle \\
\left \vert q_{2}(\vec{x},t)\right \rangle
\end{array}
\right)  $ can be same by changing the basis of states of internal knots.

\begin{figure}[ptb]
\includegraphics[clip,width=0.53\textwidth]{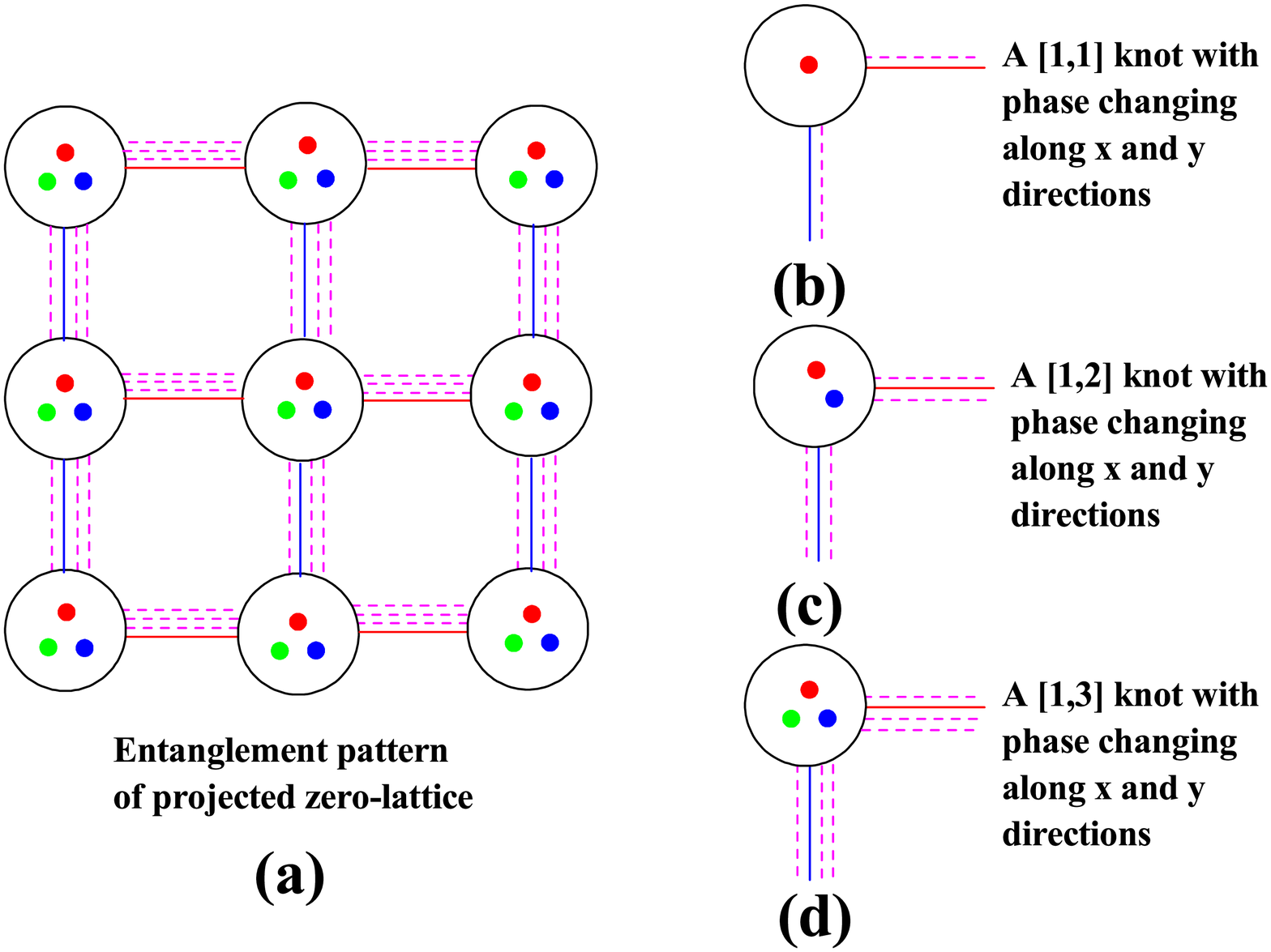}\caption{(a) An
illustration of entanglement pattern of a 2D SOC knot-crystal with
($\mathcal{N}=4$, $\mathcal{M}=2$) and $\Delta_{(1,2)}=3$. Each circle denotes
a level-2 W-type zero. The solid lines denote the entanglement pattern for
level-2 T-type zero-lattice. The dotted lines denote the entanglement pattern
for level-1 T-type zero-lattice. The dots inside the circles denote internal
zeroes; (b) An illustration of a $\left[  1,1\right]  $ knot; (c) An
illustration of a $\left[  1,2\right]  $ knot; (c) An illustration of a
$\left[  1,3\right]  $ knot.}%
\end{figure}

For the case of $n=3,$ there are three internal zeroes inside a composite knot
that is labeled by $1$, $2$, $3.$ The basis of states of internal knots has
local \textrm{SU(3)} gauge symmetry $\left(
\begin{array}
[c]{c}%
\left \vert \psi_{\mathrm{inter},1}(\vec{x},t)\right \rangle \\
\left \vert \psi_{\mathrm{inter},2}(\vec{x},t)\right \rangle \\
\left \vert \psi_{\mathrm{inter},3}(\vec{x},t)\right \rangle
\end{array}
\right)  .$ After re-ordering the internal zeroes, the states of d-quarks
$\left(
\begin{array}
[c]{c}%
d_{1,\tau,\sigma}\left \vert \mathrm{vac}\right \rangle \\
d_{2,\tau,\sigma}\left \vert \mathrm{vac}\right \rangle \\
d_{3,\tau,\sigma}\left \vert \mathrm{vac}\right \rangle
\end{array}
\right)  $ and those of u-quarks $\left(
\begin{array}
[c]{c}%
u_{1,\tau,\sigma}\left \vert \mathrm{vac}\right \rangle \\
u_{2,\tau,\sigma}\left \vert \mathrm{vac}\right \rangle \\
u_{3,\tau,\sigma}\left \vert \mathrm{vac}\right \rangle
\end{array}
\right)  $ also have local \textrm{SU(3)} gauge symmetry, i.e.,
\begin{align}
\left(
\begin{array}
[c]{c}%
d_{1,\tau,\sigma}\left \vert \mathrm{vac}\right \rangle \\
d_{2,\tau,\sigma}\left \vert \mathrm{vac}\right \rangle \\
d_{3,\tau,\sigma}\left \vert \mathrm{vac}\right \rangle
\end{array}
\right)   &  \rightarrow \left(
\begin{array}
[c]{c}%
d_{1,\tau,\sigma}^{\prime}\left \vert \mathrm{vac}\right \rangle \\
d_{2,\tau,\sigma}^{\prime}\left \vert \mathrm{vac}\right \rangle \\
d_{3,\tau,\sigma}^{\prime}\left \vert \mathrm{vac}\right \rangle
\end{array}
\right)  =U_{\mathrm{SU(3)}}\left(  \vec{x},t\right)  \left(
\begin{array}
[c]{c}%
d_{1,\tau,\sigma}\left \vert \mathrm{vac}\right \rangle \\
d_{2,\tau,\sigma}\left \vert \mathrm{vac}\right \rangle \\
d_{3,\tau,\sigma}\left \vert \mathrm{vac}\right \rangle
\end{array}
\right) \\
\left(
\begin{array}
[c]{c}%
u_{1,\tau,\sigma}\left \vert \mathrm{vac}\right \rangle \\
u_{2,\tau,\sigma}\left \vert \mathrm{vac}\right \rangle \\
u_{3,\tau,\sigma}\left \vert \mathrm{vac}\right \rangle
\end{array}
\right)   &  \rightarrow \left(
\begin{array}
[c]{c}%
u_{1,\tau,\sigma}^{\prime}\left \vert \mathrm{vac}\right \rangle \\
u_{2,\tau,\sigma}^{\prime}\left \vert \mathrm{vac}\right \rangle \\
u_{3,\tau,\sigma}^{\prime}\left \vert \mathrm{vac}\right \rangle
\end{array}
\right)  =U_{\mathrm{SU(3)}}\left(  \vec{x},t\right)  \left(
\begin{array}
[c]{c}%
u_{1,\tau,\sigma}\left \vert \mathrm{vac}\right \rangle \\
u_{2,\tau,\sigma}\left \vert \mathrm{vac}\right \rangle \\
u_{3,\tau,\sigma}\left \vert \mathrm{vac}\right \rangle
\end{array}
\right)  .\nonumber
\end{align}
Here $U_{\mathrm{SU(3)}}\left(  \vec{x},t\right)  =e^{i\Theta \left(  \vec
{x},t\right)  }$ is the matrix of the representation of $\mathrm{SU(3)}$
group. $\Theta \left(  \vec{x},t\right)  =\sum_{a=1}^{8}\theta^{a}\left(
\vec{x},t\right)  \tau^{a}$ and $\theta^{a}$ are a set of $8$ constant
parameters, and $\tau^{a}$ are $8$ $3\times3$ matrices representing the $8$
generators of the Lie algebra of $\mathrm{SU(3)}$\cite{yang}.

Fig.15 is an illustration of entanglement pattern of a 2D SOC knot-crystal
with ($\mathcal{N}=4$, $\mathcal{M}=2$) and $\Delta_{(1,2)}=3$. Each circle
denotes a level-2 W-type zero. The solid lines denote the entanglement pattern
for level-2 T-type zero-lattice. The dotted lines denote the entanglement
pattern for level-1 T-type zero-lattice. The dots inside the circles denote
internal zeroes; Fig.15(b), Fig.15(c), and Fig.15(d) illustrate $\left[
1,1\right]  $ knot, $\left[  1,2\right]  $ knot, $\left[  1,3\right]  $ knot, respectively.

We then derive the formulation of the gauge theory for the quarks with local
\textrm{SU(n) }symmetry. For simplicity, we denote the basis of quarks with
$n$ internal knot states by $\Psi_{\mathrm{quark}}=\left(
\begin{array}
[c]{c}%
\psi_{1,\mathrm{quark}}\\
\psi_{2,\mathrm{quark}}\\
...\\
\psi_{n,\mathrm{quark}}%
\end{array}
\right)  $ where $\psi_{i,\mathrm{quark}}$ describes the quantum state of the
composite knots. Due to local \textrm{SU(n) }symmetry of internal states for
internal-knots, $\psi_{i,\mathrm{quark}}$ and $\psi_{i^{\prime},\mathrm{quark}%
}$ can be changed into each other by choosing different local gauges. Thus,
$\Psi_{\mathrm{quark}}$ can be transformed locally according to an
$n$-dimensional representation,
\begin{equation}
\Psi_{\mathrm{quark}}\rightarrow \Psi_{\mathrm{quark}}^{\prime}\rightarrow
U_{\mathrm{SU(n)}}(x)\Psi_{\mathrm{quark}},
\end{equation}
where $U_{\mathrm{SU(n)}}(x)=e^{i\Theta(x)}$ is the matrix of the
representation of $\mathrm{SU(n)}$ group.

\subsubsection{Emergent \textrm{SU(n)} gauge theory for composite
knot-crystal}

The local symmetry of internal states ($U_{\mathrm{SU(n)}}(\vec{x}%
,t)\left \vert \psi \right \rangle =\left \vert \psi \right \rangle ^{\prime}$) for
internal-knots leads to an \textrm{SU(n)} gauge symmetry for composite knots
(elementary fermions), The \textrm{SU(n)} gauge symmetry for composite knots
(elementary fermions) leads emergent \textrm{SU(n)} gauge theory for composite knot-crystal.

To well define a composite knot with $n$ internal zeroes, we must choose
arbitrary $n$ nearest-neighbor zeroes. Gauge symmetries appear as the
redundancies to choose which $n$ internal zeroes to make up this composite
knot. The redundancy protected by internal symmetry for $n$ internal zeroes
leads to a non-Abelian gauge symmetry that is really the $\mathrm{SU(n)}$
gauge symmetry for strong interaction. As a result, we have an $\mathrm{SU(n)}%
$ gauge symmetry that will never be broken. The rule to settle down the
redundancy is the rule to fix the gauge for gauge field.

Due to the existence of the local $\mathrm{SU(n)}$ gauge symmetry, there
exists $\mathrm{SU(n)}$ gauge field. Under changing internal knots inside a
composite knot, the quark state at site $\vec{j}$ changes as%
\begin{equation}
q_{\vec{j}}\rightarrow q_{\vec{j}}^{\prime}=\tilde{U}_{\vec{j},\mathrm{SU(n)}%
}q_{\vec{j}}.
\end{equation}
Here $\tilde{U}_{\mathrm{SU(n)}}$ is a changing of internal knot\ states based
on certain basis $\left(
\begin{array}
[c]{c}%
\psi_{1,\mathrm{quark}}\\
\psi_{2,\mathrm{quark}}\\
...\\
\psi_{n,\mathrm{quark}}%
\end{array}
\right)  $. After considering the local changing of basis induced by
$U_{\vec{j},\mathrm{SU(n)}}$, the local coupling between two quark states
changes, i.e.,
\begin{equation}
Jq_{\vec{j}}^{\dagger}T_{\vec{j},\vec{j}^{\prime}}q_{\vec{j}^{\prime}%
}\rightarrow J(q_{\vec{j}}\tilde{U}_{\vec{j},\mathrm{SU(n)}})^{\dagger}\cdot
T_{\vec{j},\vec{j}^{\prime}}\cdot(\tilde{U}_{\vec{j}^{\prime},\mathrm{SU(n)}%
}q_{\vec{j}^{\prime}})
\end{equation}
where $T_{\vec{j},\vec{j}^{\prime}}$ is translation operator from $\vec{j}%
$-sie to $\vec{j}^{\prime}$-site. We define a vector field $\mathcal{A}%
_{\vec{j},\vec{j}^{\prime}}$ to characterize the local changing of basis%
\begin{equation}
e^{ig\mathcal{A}_{\vec{j},\vec{j}^{\prime}}}=(\tilde{U}_{\vec{j}%
,\mathrm{SU(n)}})^{-1}\tilde{U}_{\vec{j}^{\prime},\mathrm{SU(n)}}%
\end{equation}
where $g$ is coupling constant of $\mathrm{SU(n)}$ non-Abelian gauge field.
So, for perturbation case $\mathcal{A}_{\vec{j},\vec{j}^{\prime}}\sim0$, we
have
\begin{equation}
ig\mathcal{A}_{\vec{j},\vec{j}^{\prime}}\simeq(\delta \tilde{U}_{\vec
{j},\mathrm{SU(n)}})^{-1}\tilde{U}_{\vec{j}^{\prime},\mathrm{SU(n)}}%
\end{equation}
where $\delta \tilde{U}_{\vec{j},\mathrm{SU(n)}}=\tilde{U}_{\vec{j}%
,\mathrm{SU(n)}}-\tilde{U}_{\vec{j}^{\prime},\mathrm{SU(n)}}.$ The local
coupling between two knot states becomes
\begin{equation}
Jq_{\vec{j}}^{\dagger}e^{ig\mathcal{A}_{\vec{j},\vec{j}^{\prime}}}T_{\vec
{j},\vec{j}^{\prime}}q_{\vec{j}^{\prime}}.
\end{equation}
The total kinetic energy for knots becomes
\begin{equation}
\mathcal{\hat{H}}_{\mathrm{coupling}}=J%
%TCIMACRO{\dsum \nolimits_{\left\langle \vec{j},\vec{j}^{\prime}\right\rangle
%}}%
%BeginExpansion
{\displaystyle \sum \nolimits_{\left \langle \vec{j},\vec{j}^{\prime
}\right \rangle }}
%EndExpansion
Jq_{\vec{j}}^{\dagger}e^{ig\mathcal{A}_{\vec{j},\vec{j}^{\prime}}}T_{\vec
{j},\vec{j}^{\prime}}q_{\vec{j}^{\prime}}+h.c..
\end{equation}

It is obvious that the vector field $\mathcal{A}_{\vec{j},\vec{j}^{\prime}}$
that characterizes the local position perturbation of internal zero-lattice
plays the role of $\mathrm{SU(n}$\textrm{)} gauge field. To illustrate the
local $\mathrm{SU(n}$\textrm{)} gauge symmetry, we do a local $\mathrm{SU(n}%
$\textrm{)} gauge transformation $U_{\vec{j},\mathrm{SU(n)}}$ that is the
transformation of basis of internal knot states, i.e.,
\begin{equation}
\left(
\begin{array}
[c]{c}%
\psi_{1,\mathrm{quark}}^{\prime}\\
\psi_{2,\mathrm{quark}}^{\prime}\\
...\\
\psi_{n,\mathrm{quark}}^{\prime}%
\end{array}
\right)  =U_{\vec{j},\mathrm{SU(n)}}\left(
\begin{array}
[c]{c}%
\psi_{1,\mathrm{quark}}\\
\psi_{2,\mathrm{quark}}\\
...\\
\psi_{n,\mathrm{quark}}%
\end{array}
\right)  .
\end{equation}
Under a local $\mathrm{SU(n}$\textrm{)} gauge transformation, we have
\begin{equation}
q_{\vec{j}}\rightarrow q_{\vec{j}}^{\prime}=U_{\vec{j},\mathrm{SU(n)}}%
q_{\vec{j}},
\end{equation}
and
\begin{align}
g\mathcal{A}_{\vec{j},\vec{j}^{\prime}}  &  \rightarrow g\mathcal{A}_{\vec
{j},\vec{j}^{\prime}}^{\prime}=gU_{\vec{j},\mathrm{SU(n)}}\mathcal{A}_{\vec
{j},\vec{j}^{\prime}}(U_{\vec{j}^{\prime},\mathrm{SU(n)}})^{-1}\\
&  +i(\delta U_{\vec{j},\mathrm{SU(n)}})(U_{\vec{j},\mathrm{SU(n)}}%
)^{-1}\nonumber
\end{align}
where $\delta U_{\vec{j},\mathrm{SU(n)}}=(U_{\vec{j},\mathrm{SU(n)}}%
-U_{\vec{j}^{\prime},\mathrm{SU(n)}}).$ Here, we have used the following
result,
\begin{align}
e^{ig\mathcal{A}_{\vec{j},\vec{j}^{\prime}}}  &  =(\tilde{U}_{\vec
{j},\mathrm{SU(n)}})^{-1}\tilde{U}_{\vec{j}^{\prime},\mathrm{SU(n)}%
}\nonumber \\
&  \rightarrow e^{ig\mathcal{A}_{\vec{j},\vec{j}^{\prime}}^{\prime}}%
=U_{\vec{j},\mathrm{SU(n)}}e^{ig\mathcal{A}_{\vec{j},\vec{j}^{\prime}}%
}(U_{\vec{j}^{\prime},\mathrm{SU(n)}})^{-1}\nonumber \\
&  =U_{\vec{j},\mathrm{SU(n)}}(1+ig\mathcal{A}_{\vec{j},\vec{j}^{\prime}%
})(U_{\vec{j}^{\prime},\mathrm{SU(n)}})^{-1}\nonumber \\
&  =U_{\vec{j},\mathrm{SU(n)}}(U_{\vec{j}^{\prime},\mathrm{SU(n)}}%
)^{-1}+igU_{\vec{j},\mathrm{SU(n)}}\mathcal{A}_{\vec{j},\vec{j}^{\prime}%
}(U_{\vec{j}^{\prime},\mathrm{SU(n)}})^{-1}\nonumber \\
&  =1+ig\mathcal{A}_{\vec{j},\vec{j}^{\prime}}^{\prime}\simeq e^{ig\mathcal{A}%
_{\vec{j},\vec{j}^{\prime}}^{\prime}}.
\end{align}
The total kinetic energy for knots turns into
\begin{equation}
\mathcal{\hat{H}}_{\mathrm{coupling}}\rightarrow \mathcal{\hat{H}%
}_{\mathrm{coupling}}^{\prime}=J%
%TCIMACRO{\dsum \nolimits_{\left\langle \vec{j},\vec{j}^{\prime}\right\rangle
%}}%
%BeginExpansion
{\displaystyle \sum \nolimits_{\left \langle \vec{j},\vec{j}^{\prime
}\right \rangle }}
%EndExpansion
(q_{\vec{j}}^{\prime})^{\dagger}e^{ig\mathcal{A}_{\vec{j},\vec{j}^{\prime}%
}^{\prime}}T_{\vec{j},\vec{j}^{\prime}}q_{\vec{j}^{\prime}}^{\prime}+h.c..
\end{equation}
The Hamiltonian doesn't change,%
\begin{equation}
\mathcal{\hat{H}}_{\mathrm{coupling}}=\mathcal{\hat{H}}_{\mathrm{coupling}%
}^{\prime}.
\end{equation}

On the other hand, the situation for tempo phase changing is similar to that
for spatial phase changing. To characterize the tempo twist-writhe locking
condition, we introduce a Lagrangian variable $\mathcal{A}_{0,\vec{j}}$ to
path-integral formulation as
\begin{equation}
\mathcal{A}_{0,\vec{j}}q_{\vec{j}}^{\dagger}q_{\vec{j}}.
\end{equation}

In continuum limit, we have, $U_{\vec{j},\mathrm{SU(n)}}(t)\rightarrow
U_{\mathrm{SU(n)}}(\vec{x},t),$ $\mathcal{A}_{\vec{j},\vec{j}^{\prime}%
}\rightarrow \mathcal{\vec{A}}(x)\ $and $\mathcal{A}_{0,\vec{j}}\rightarrow
\mathcal{A}_{0}(x).$ The non-Abelian gauge symmetry is represented by%
\begin{equation}
q_{\mathrm{quark}}^{\prime}\rightarrow U_{\mathrm{SU(n)}}(\vec{x}%
,t)q_{\mathrm{quark}}%
\end{equation}
and%
\begin{align}
\mathcal{A}_{\mu}(\vec{x},t)  &  \rightarrow U_{\mathrm{SU(n)}}(\vec
{x},t)\mathcal{A}_{\mu}(\vec{x},t)\left(  U_{\mathrm{SU(n)}}(\vec
{x},t)\right)  ^{-1}\nonumber \\
&  +\frac{i}{g}\left(  \partial_{\mu}U_{\mathrm{SU(n)}}(\vec{x},t)\right)
\left(  U_{\mathrm{SU(n)}}(\vec{x},t)\right)  ^{-1}.
\end{align}
The gauge strength is defined by $\mathcal{G}_{\mu \nu}$ as%
\begin{equation}
\mathcal{G}_{\mu \nu}=\partial_{\mu}\mathcal{A}_{\nu}-\partial_{\nu}%
\mathcal{A}_{\mu}-ig\left[  \mathcal{A}_{\mu},\mathcal{A}_{\nu}\right]
\end{equation}
or
\begin{equation}
\mathcal{G}_{\mu \nu}=G_{\mu \nu}^{a}t^{a}\text{, }G_{\mu \nu}^{a}=\partial_{\mu
}A_{\nu}^{a}-\partial_{\nu}A_{\mu}^{a}+gf^{abc}A_{\mu}^{b}A_{\nu}^{c}.
\end{equation}
The Lagrangian of Yang-Mills field can only be written as:
\begin{equation}
\mathcal{L}_{\mathrm{YM}}(\mathrm{SU(n)})=-\frac{1}{2}\mathrm{Tr}%
\mathcal{G}_{\mu \nu}\mathcal{G}^{\mu \nu}+\mathrm{Tr}J_{\mathrm{YM}}^{\mu
}\mathcal{A}_{\mu}%
\end{equation}
where $J_{\mathrm{YM}}^{\mu}=i\bar{q}_{\mathrm{quark}}\gamma^{\mu
}q_{\mathrm{quark}}.$ The Lagrangian density $\mathcal{L}_{\mathrm{YM}%
}(\mathrm{SU(n)})$ is invariant under the gauge transformations with an
$x$-dependent $U_{\mathrm{SU(n)}}(\vec{x},t)$.

Finally, we show the physical picture of the $\mathrm{SU(n)}$ gauge field. An
extra internal zero also plays the role of source of $\mathrm{SU(n)}$ gauge
field and carries color degree of freedom. There are $n$ different colors. For
example, for the case of $\mathrm{SU(3)}$, there are three colors called red,
blue and green, respectively. The collective modes of $\mathrm{SU(3)}$ gauge
field are always called gluons that are the fluctuations of internal zeroes.
Two colored composite knots interact by exchanging gluons. However, a
composite object with $n_{k}=N\cdot n$ internal zeroes ($N$ is an integer
number) are colorless. Because electron is a composite knot without additional
internal zeroes ($n_{e}\equiv3$), the fluctuations reorganizing internal
zeroes will never affect electrons. So, electrons don't couple the gluons.

\subsection{Examples}

A composite knot-crystal that is projected to coupled zero-lattices naturally
gives rise to gauge bosons (such as photons) and fermions (such as electrons).
Gauge bosons are vibrations of zero-lattice, while fermions are composite
knots of zero-lattice. We get the universal effective Lagrangian for a
composite knot-crystal as
\begin{equation}
\mathcal{L}=\mathcal{L}_{\mathrm{fermion}}+\mathcal{L}_{\mathrm{EM}%
}(\mathrm{U(1)})+\mathcal{L}_{\mathrm{YM}}(\mathrm{SU(n)}).
\end{equation}
The \textrm{U(1)} gauge field characterizes the interaction from the phase
fluctuations of internal zero-lattice. The $\mathrm{SU(n)}$ gauge field
characterizes the interaction from the fluctuations of twisting number of
internal zero-lattice.

\subsubsection{Physics of simple knot-crystal with $n=0$}

For a simple knot-crystal, the knot becomes chargeless electron and the
effective model is a free Dirac model, of which the effective Lagrangian is
\begin{equation}
\mathcal{L}={\bar{e}}(x)i\gamma^{\mu}\partial_{\mu}e(x)+m_{e}{\bar{e}}(x)e(x),
\end{equation}
where $m_{e}=2\omega^{\ast}$.

\subsubsection{Physics of composite knot-crystal with $n=1$}

For the composite knot-crystal without internal winding ($n=0$), the composite
knot becomes electron with $n_{e}=1$. The effective model is \textrm{U(1)}
gauge theory, of which the effective Lagrangian is given by
\begin{align}
\mathcal{L}  &  ={\bar{e}}(x)i\gamma^{\mu}\partial_{\mu}e(x)+m_{e}{\bar{e}%
}(x)e(x)\\
&  -\frac{1}{4}F_{\mu \nu}F^{\mu \nu}+\mathrm{e}{A}_{\mu}(x){j}_{(em)}^{\mu
}(x)\nonumber
\end{align}
where ${j}_{(em)}^{\mu}(x)=i{\bar{e}}(x)\gamma^{\mu}e(x)$. $m_{e}%
=2\omega^{\ast}$ are masses for electron. The phonons ${A}_{\mu}(x)$
characterize the fluctuations of internal zero-lattice. This model gives QED.

\subsubsection{Physics of composite knot-crystal with $n=2$}

For the composite knot-crystal with $n=2$, the (composite) knots are electron
with $n_{e}=2$, quarks with $n_{\mathrm{quark}}=1$. The effective model is
$\mathrm{SU(2)}\otimes$\textrm{U(1)} gauge theory, of which the effective
Lagrangian is given by
\begin{align}
\mathcal{L}  &  ={\bar{e}}(x)i\gamma^{\mu}\partial_{\mu}e(x)+{\bar{\psi}%
}_{\mathrm{quark}}(x)i\gamma^{\mu}\partial_{\mu}\psi_{\mathrm{quark}}(x)\\
&  +m_{e}{\bar{e}}(x)e(x)+m_{\mathrm{quark}}{\bar{\psi}}_{\mathrm{quark}%
}(x)\psi_{\mathrm{quark}}(x)\nonumber \\
&  -\frac{1}{4}F_{\mu \nu}F^{\mu \nu}+\mathrm{e}{A}_{\mu}(x){j}_{(em)}^{\mu
}(x)\nonumber \\
&  -\frac{1}{2}\mathrm{Tr}\mathcal{G}_{\mu \nu}\mathcal{G}^{\mu \nu}%
+\mathrm{Tr}J_{\mathrm{YM}}^{\mu}\mathcal{A}_{\mu}\nonumber
\end{align}
where $\mathrm{e}=2e_{0}$. $m_{\mathrm{quark}}=2\omega^{\ast}$ are masses for
quark and electron, respectively. The electric current is
\begin{equation}
{j}_{(em)}^{\mu}(x)=i{\bar{e}}(x)\gamma^{\mu}e(x)+\frac{1}{2}i{\bar{\psi}%
}_{\mathrm{quark}}(x)\gamma^{\mu}\psi_{\mathrm{quark}}(x).
\end{equation}
The $\mathrm{SU(2)}$ color current is $J_{\mathrm{YM}}^{a,\mu}=i\bar{\Psi
}_{\mathrm{quark}}\gamma^{\mu}T^{a}\Psi_{\mathrm{quark}}.$

The \textrm{U(1)} gauge field characterizes the interaction from the phase
fluctuations of internal zero-lattice. The $\mathrm{SU(2)}$ gauge field
characterizes the interaction from the number fluctuations of internal zero-lattice.

\section{Emergent quantum field theory for 3-level composite knot-crystal with
($\mathcal{N}=4$, $\mathcal{M}=3$)}

In this section we study the effective theory for knot on a 3-level composite
knot-crystal with ($\mathcal{N}=4$, $\mathcal{M}=3$). To generate a 3-level
composite knot-crystal with ($\mathcal{N}=4$, $\mathcal{M}=3$), we
symmetrically wind a 2-level double-helix knot-crystal with ($\mathcal{N}=4$,
$\mathcal{M}=2$) along different spatial directions. As a result, there exists
an additional W-type zero-lattice by W-type projection on the center membrane
of \textrm{A}-knot-crystal and \textrm{B}-knot-crystal -- the level-3 W-type
zero-lattice. The effective theory of a 3-level composite knot-crystal with
($\mathcal{N}=4$, $\mathcal{M}=3$) reproduces Standard model, an
$\mathrm{SU}_{\mathrm{Strong}}\mathrm{(3)}\otimes \mathrm{SU}_{\mathrm{weak}%
}\mathrm{(2)}\otimes$\textrm{U}$_{\mathrm{Y}}$\textrm{(1)} gauge theory with
Higgs mechanism due to spontaneous symmetry breaking. $\Delta_{\{2,3\}}$ is a
very large number, for example, $\Delta_{\{2,3\}}\sim10^{10}$. We call this
composite knot-crystal with ($\mathcal{N}=4$, $\mathcal{M}=3$) to be the
\emph{standard knot-crystal}. Fig.5(a) is an illustration of the standard knot-crystal.

\subsection{Classification of composite knots}

In particle physics, there exist different types fermionic elementary
particles: neutrino, electron and quarks. To explain the existence of
different types of elementary particles, people try to go beyond SM. Pati and
Salam \cite{Pati} had proposed preons to be the fundamental constituent
particles. Then, other types of models were developed, for example, the Rishon
Model proposed simultaneously by Harari and Shupe \cite{Harari,Shupe}, the
helon model by S. O. Bilson-Thompson\cite{helon}, the tangles by C.
Schiller\cite{Strand}. In knot physics, we show a correspondence between
different types of elementary particles and different types of composite knots.

By trapping different types of zeroes, there exist different types of knots.
We use the following number series to label different types of knots,
\begin{equation}
\lbrack n_{\mathrm{L}_{2}},\text{ }n_{\mathrm{L}_{1}}]
\end{equation}
where $n_{\mathrm{L}_{2}}$ is the half linking number of level-2 entangled
knot-crystals that is equal to the sum of the number of level-2 T-type zeroes
$n_{\mathrm{L}_{2}\mathrm{T}}$ and the number of level-3 W-type zeroes
$n_{\mathrm{L}_{3}\mathrm{W}}$ as
\begin{equation}
n_{\mathrm{L}_{2}}=n_{\mathrm{L}_{2}\mathrm{T}}+n_{\mathrm{L}_{3}\mathrm{W}},
\end{equation}
$n_{\mathrm{L}_{1}}$ is the half linking number of level-1 entangled
vortex-membranes that is equal to the sum of the number of level-1 T-type
zeroes $n_{\mathrm{L}_{1}\mathrm{T}}$ and the number of level-2 W-type zeroes
$n_{\mathrm{L}_{2}\mathrm{W}},$
\begin{equation}
n_{\mathrm{L}_{1}}=n_{\mathrm{L}_{1}\mathrm{T}}+n_{\mathrm{L}_{2}\mathrm{W}}.
\end{equation}

For different types of knots, due to trapping half linking number of two
entangled knot-crystals, we must have $n_{\mathrm{L}_{2}}=1$. So the
classification of the knot type is based on the half internal linking number
$n_{\mathrm{L}_{1}}$ (the half linking number of level-1 entangled
vortex-membranes). For a 2-level double-helix knot-crystal with ($\mathcal{N}%
=4$, $\mathcal{M}=2$) with an integer hierarchy number $\Delta_{\{1,2\}}=n$,
There are $n$ different types of knots with $[1,$ $0],$ $[1,$ $1],$ ... $[1,$
$n-1].$ The composite knots with $[1,$ $0]$ correspond to electrons in
particle physics and the composite knots with $[1,$ $n_{k}]$ ($n_{k}$ is an
integer number, $0<n_{k}<n$)\ correspond to quarks. However, there are two
types of knots with $[1,$ $0]$, one is $[1_{\mathrm{L}_{2}\mathrm{T}},$ $3]$
that is knot with a level-2 T-type zero, the other is $[1_{\mathrm{L}%
_{3}\mathrm{W}},$ $0]$ that is knot with a level-3 W-type zero. The knot with
$[1_{\mathrm{L}_{3}\mathrm{W}},$ $0]$ corresponds to neutrino and the knot
with $[1_{\mathrm{L}_{2}\mathrm{T}},$ $3]$ corresponds to electron.

For example, for the case of $n=3$, the correspondence between four types of
knots and elementary particles is given by
\begin{align}
\text{A knot with }[1_{\mathrm{L}_{2}\mathrm{T}},3]  &  \longleftrightarrow
\text{An electron,}\\
\text{A knot with }[1_{\mathrm{L}_{3}\mathrm{W}},0]  &  \longleftrightarrow
\text{An neutrino,}\nonumber \\
\text{A knot with }[1_{\mathrm{L}_{2}\mathrm{T}},1]  &  \longleftrightarrow
\text{An u-qurak,}\nonumber \\
\text{A knot with }[1_{\mathrm{L}_{2}\mathrm{T}},2]  &  \longleftrightarrow
\text{An d-qurak}\nonumber
\end{align}
or
\begin{align}
\psi_{\lbrack1_{\mathrm{L}_{2}\mathrm{T}},3]}  &  \longleftrightarrow
e\text{,}\\
\psi_{\lbrack1_{\mathrm{L}_{3}\mathrm{W}},0]}  &  \longleftrightarrow
\nu \text{,}\nonumber \\
\psi_{\lbrack1_{\mathrm{L}_{2}\mathrm{T}},1]}  &  \longleftrightarrow
u\text{,}\nonumber \\
\psi_{\lbrack1_{\mathrm{L}_{2}\mathrm{T}},2]}  &  \longleftrightarrow
d\text{.}\nonumber
\end{align}
As a result, in low energy limit, the effective Lagrangian of knots on a
3-level composite knot-crystal with ($\mathcal{N}=4$, $\mathcal{M}=3$) and
$\Delta_{\{1,2\}}=n=3$ becomes%
\begin{align}
\mathcal{L}_{\mathrm{fermion}}(x)  &  ={\bar{\nu}}i\gamma^{\mu}\partial_{\mu
}\nu+{\bar{e}}i\gamma^{\mu}\partial_{\mu}e+{\bar{d}}i\gamma^{\mu}\partial
_{\mu}d\nonumber \\
&  +{\bar{u}}i\gamma^{\mu}\partial_{\mu}u+m_{e}{\bar{e}}e+m_{d}{\bar{d}%
}d+m_{u}{\bar{u}}u.
\end{align}
There is no mass term for the left-hand neutrinos.

\subsection{Emergent $\mathrm{SU_{\mathrm{Strong}}(3)\otimes U_{\mathrm{EM}%
}(1)}$ gauge theory for the standard knot-crystal}

It was known that a 2-level composite knot-crystal with ($\mathcal{N}=4$,
$\mathcal{M}=2$) is described by $\mathrm{SU_{\mathrm{Strong}}(n)\otimes
U_{\mathrm{EM}}(1)}$ gauge theory. The \textrm{U(1)} gauge field characterizes
the interaction from the phase fluctuations of internal zero-lattice and the
$\mathrm{SU(n)}$ gauge field characterizes the interaction from the
fluctuations of twisting number of internal zero-lattice.

For the standard knot-crystal (a composite knot-crystal 3-level composite
knot-crystal with ($\mathcal{N}=4$, $\mathcal{M}=3$) and $n=3$), we have four
types of composite knots: electron with $n_{e}=3$, d-quarks with
$n_{\mathrm{d-quark}}=1,$ u-quarks with $n_{\mathrm{u-quark}}=2.$ From point
view of fluctuations of level-1 internal twistings, without considering the
fluctuations of level-3 windings, the effective model is an
$\mathrm{SU_{\mathrm{Strong}}(3)}\otimes$\textrm{U}$_{\mathrm{EM}}%
$\textrm{(1)} gauge theory, of which the Lagrangian is
\begin{align}
\mathcal{L}  &  ={\bar{\nu}}i\gamma^{\mu}\partial_{\mu}\nu+{\bar{e}}%
i\gamma^{\mu}\partial_{\mu}e+{\bar{d}}i\gamma^{\mu}\partial_{\mu}d\\
&  +{\bar{u}}i\gamma^{\mu}\partial_{\mu}u+m_{e}{\bar{e}}e+m_{d}{\bar{d}%
}d+m_{u}{\bar{u}}u\nonumber \\
&  -\frac{1}{4}F_{\mu \nu}F^{\mu \nu}+\mathrm{e}{A}_{\mu}(x){j}_{(em)}^{\mu
}(x)\nonumber \\
&  -\frac{1}{2}\mathrm{Tr}\mathcal{G}_{\mu \nu}\mathcal{G}^{\mu \nu}%
+\mathrm{Tr}J_{\mathrm{YM}}^{\mu}\mathcal{A}_{\mu}\nonumber
\end{align}
where $\mathrm{e}=3e_{0}.$ $m_{\nu}=0,$ $m_{u}=2\omega^{\ast},$ $m_{d}%
=2\omega^{\ast},$ and $m_{e}=2\omega^{\ast}$ are masses for neutrino, u-quark,
d-quark and electron, respectively. The electric current is
\begin{align}
{j}_{(em)}^{\mu}(x)  &  =i{\bar{e}}(x)\gamma^{\mu}e(x)+i\frac{2}{3}{\bar{u}%
}(x)\gamma^{\mu}u(x)\\
&  +i\frac{1}{3}{\bar{d}}(x)\gamma^{\mu}d(x).\nonumber
\end{align}
The $\mathrm{SU}_{\mathrm{Strong}}\mathrm{(3)}$ color current for strong
interaction is
\begin{equation}
J_{\mathrm{YM}}^{a,\mu}={\bar{u}}(x)i\gamma^{\mu}T^{a}u(x)+{\bar{d}}%
(x)i\gamma^{\mu}T^{a}d(x).
\end{equation}
This model gives QED and QCD. In particular, neutrino doesn't couple to
$\mathrm{SU_{\mathrm{Strong}}(3)}\otimes$\textrm{U}$_{\mathrm{EM}}%
$\textrm{(1)} gauge fields.

\subsection{Emergent electro-weak $\mathrm{SU}_{\mathrm{weak}}%
\mathrm{(2)\otimes U}_{\mathrm{Y}}\mathrm{(1)}$ gauge theory for the standard
knot-crystal}

In above section, from point view of fluctuations of level-1 internal
twistings, without considering the fluctuations of level-3 windings, the
effective model is an $\mathrm{SU_{\mathrm{Strong}}(3)}\otimes$\textrm{U}%
$_{\mathrm{EM}}$\textrm{(1)} gauge theory. After considering the fluctuations
of level-3 windings, the situation changes. In this section, we focus on the
effect from point view of the fluctuations of level-3 windings and get an
electro-weak $\mathrm{SU}_{\mathrm{weak}}\mathrm{(2)\otimes U}_{\mathrm{Y}%
}\mathrm{(1)}$ gauge field theory. An $\mathrm{SU}_{\mathrm{weak}%
}\mathrm{(2)\otimes U}_{\mathrm{Y}}\mathrm{(1)}$ gauge field that couples
neutrino (a knot with $[1_{\mathrm{L}_{3}\mathrm{W}},0]$) and electron (a knot
with $[1_{\mathrm{L}_{2}\mathrm{T}},3]$) characterizes the interaction by
exchanging the fluctuations of linking number in the unit cell of level-3
W-type zero-lattice. The fluctuations of $\mathrm{SU}_{\mathrm{weak}%
}\mathrm{(2)}$ gauge theory come from the fluctuations of level-2 linking
number on a unit cell of level-3 W-type zero-lattice and the fluctuations of
$\mathrm{U}_{\mathrm{Y}}\mathrm{(1)}$ gauge theory come from the fluctuations
of level-1 linking number (the internal zeroes) on a unit cell of level-3
W-type zero-lattice.

\subsubsection{Electro-weak $\mathrm{SU}_{\mathrm{weak}}\mathrm{(2)\otimes
U}_{\mathrm{Y}}\mathrm{(1)}$ gauge symmetry for leptons}

A neutrino is really half-winding with level-3 W-type zero and an electron is
a half-twisting with internal twisting $n_{e}=\Delta_{(1,2)}=3$. Without
considering the charge degree of freedom, we cannot distinguish a uniform
distributed left-hand electron from a uniform distributed left-hand neutrino
on a unit cell of level-3 zero-lattice. Therefore, we may regard a uniform
distributed left-hand neutrino $\nu$ and a uniform distributed left-hand
electron $e$ to be same object of an \textrm{SU}$_{\mathrm{weak}}$\textrm{(2)}
gauge symmetry on a unit cell of level-3 W-type zero-lattice. The fluctuations
of internal zeroes on the unit cell of level-3 W-type zero-lattice play the
role of the Abelian gauge fields for the \textrm{U}$_{\mathrm{Y}}$\textrm{(1)}
gauge symmetry.

We then consider a left-hand electron $e$ to be a composite knot with
\begin{equation}
n_{e}=\left(  n_{e}\right)  _{\mathrm{SU}_{\mathrm{weak}}\mathrm{(2)}}+\left(
n_{e}\right)  _{\mathrm{U}_{\mathrm{Y}}\mathrm{(1)}}=3
\end{equation}
where $\left(  n_{e}\right)  _{\mathrm{SU}_{\mathrm{weak}}\mathrm{(2)}}%
=\frac{3}{2}$ and $\left(  n_{e}\right)  _{\mathrm{U}_{\mathrm{Y}}%
\mathrm{(1)}}=\frac{3}{2}$ and a left-hand neutrino $\nu$ to be a composite
knot with
\begin{equation}
n_{\nu}=\left(  n_{\nu}\right)  _{\mathrm{SU}_{\mathrm{weak}}\mathrm{(2)}%
}+\left(  n_{\nu}\right)  _{\mathrm{U}_{\mathrm{Y}}\mathrm{(1)}}=0
\end{equation}
where $\left(  n_{\nu}\right)  _{\mathrm{SU}_{\mathrm{weak}}\mathrm{(2)}%
}=-\frac{3}{2}$ and $\left(  n_{\nu}\right)  _{\mathrm{U}_{\mathrm{Y}%
}\mathrm{(1)}}=\frac{3}{2}$.

To characterize the property of opposite-component of internal zeroes for
leptons
\begin{equation}
\left(  n_{e}\right)  _{\mathrm{SU}_{\mathrm{weak}}\mathrm{(2)}}=-\left(
n_{\nu}\right)  _{\mathrm{SU}_{\mathrm{weak}}\mathrm{(2)}}=\frac{3}{2},
\end{equation}
a left-hand neutrino $\nu$ and a left-hand electron $e$ make up a lepton
$\mathrm{SU}_{\mathrm{weak}}\mathrm{(2)}$ spinor
\begin{equation}
\psi_{\mathrm{Lepton}}=\left(
\begin{array}
[c]{c}%
\nu \\
e
\end{array}
\right)  .
\end{equation}
The left-handed components of the lepton fields are assigned to doublets of
$\mathrm{SU}_{\mathrm{weak}}\mathrm{(2)}$
\begin{equation}
\psi_{\mathrm{Lepton,}L}=\frac{1}{2}(1+\gamma_{5})\left(
\begin{array}
[c]{c}%
\nu \\
e
\end{array}
\right)  \label{leptrepr2}%
\end{equation}

The right-handed components are assigned to singlets of $\mathrm{SU}%
_{\mathrm{weak}}\mathrm{(2)}$ that has no neutrino and we have
\begin{equation}
\psi_{\mathrm{Lepton,}R}={e}_{R}=\frac{1}{2}(1-\gamma_{5}){e}.
\end{equation}
As a result, we have
\begin{align}
\psi_{\mathrm{Lepton,}L}  &  \rightarrow e^{i\vec{\tau}\vec{\theta}(\vec{X}%
)}\psi_{\mathrm{Lepton,}L},\text{ }\nonumber \\
\psi_{\mathrm{Lepton,}R}  &  \rightarrow \psi_{\mathrm{Lepton,}R}%
\end{align}
with $\vec{\tau}$ the three Pauli matrices.

On the other hand, to characterize the property of identical-component of
internal zeroes for leptons$\ $%
\begin{equation}
\left(  n_{e}\right)  _{\mathrm{U}_{\mathrm{Y}}\mathrm{(1)}}=\left(  n_{\nu
}\right)  _{\mathrm{U}_{\mathrm{Y}}\mathrm{(1)}}=\frac{3}{2},
\end{equation}
a left-hand neutrino $\nu$ and a left-hand electron $e$ have an additional
charge degree of freedom from $\mathrm{U}_{\mathrm{Y}}\mathrm{(1)}$ gauge
symmetry. Such charge degree of freedom is called the supercharge degrees of
freedom. The supercharge $\mathcal{Y}$ of left-hand electron and left-hand
neutrino with internal zeroes$\  \left(  n_{e}\right)  _{\mathrm{U}%
_{\mathrm{Y}}\mathrm{(1)}}=\left(  n_{\nu}\right)  _{\mathrm{U}_{\mathrm{Y}%
}\mathrm{(1)}}=\frac{3}{2}$ is
\begin{equation}
\mathcal{Y}(\psi_{\mathrm{Lepton,}L})=-1.
\end{equation}
An important fact is that \emph{unit internal zero has supercharge }$Y$\emph{
to be }$-\frac{2}{3}$, i.e.,
\begin{equation}
\mathcal{Y}(\emph{internal\ zero})=-\frac{2}{3}.
\end{equation}

For a right-hand electron, there are $3$ internal zeroes. The $3$ internal
zeroes have supercharge $\mathcal{Y}$ to be $3\times \left(  -\frac{2}%
{3}\right)  =-2$. Therefore, right-hand electrons ${e}_{R}$ have $-2$
supercharge as
\begin{equation}
\mathcal{Y}(\psi_{\mathrm{Lepton,}R})=-2.
\end{equation}

\subsubsection{Electro-weak $\mathrm{SU}_{\mathrm{weak}}\mathrm{(2)\otimes
U}_{\mathrm{Y}}\mathrm{(1)}$ gauge symmetry for quarks}

In addition, we consider the electro-weak interaction for quarks. In unit cell
of level-3 zero-lattice, a d-quark has 1 internal zero $n_{\mathrm{d-quark}%
}=1$ and an anti-u-quark has 2 internal zero $n_{\mathrm{u}^{\ast
}\mathrm{-quark}}=-2$. We may also regard a uniform distributed left-hand
d-quark and a uniform distributed left-hand anti-u-quark to be same object of
an \textrm{SU}$_{\mathrm{weak}}$\textrm{(2)} gauge symmetry on the unit cell
of level-3 W-type zero-lattice.

We consider a left-hand d-quark to be a composite knot as
\begin{equation}
n_{\mathrm{d-quark}}=\left(  n_{\mathrm{d-quark}}\right)  _{\mathrm{SU}%
_{\mathrm{weak}}\mathrm{(2)}}+\left(  n_{\mathrm{d-quark}}\right)
_{\mathrm{U}_{\mathrm{Y}}\mathrm{(1)}}%
\end{equation}
where $\left(  n_{\mathrm{d-quark}}\right)  _{\mathrm{SU}_{\mathrm{weak}%
}\mathrm{(2)}}=\frac{3}{2}$ and $\left(  n_{\mathrm{d-quark}}\right)
_{\mathrm{U}_{\mathrm{Y}}\mathrm{(1)}}=-\frac{1}{2}$ and a left-hand u-quark
to be a knot with 1.5 internal zero,
\begin{equation}
n_{\mathrm{u}^{\ast}\mathrm{-quark}}=\left(  n_{\mathrm{u}^{\ast
}\mathrm{-quark}}\right)  _{\mathrm{SU}_{\mathrm{weak}}\mathrm{(2)}}+\left(
n_{\mathrm{u}^{\ast}\mathrm{-quark}}\right)  _{\mathrm{U}_{\mathrm{Y}%
}\mathrm{(1)}}%
\end{equation}
where $\left(  n_{\mathrm{u-quark}}\right)  _{\mathrm{SU}_{\mathrm{weak}%
}\mathrm{(2)}}=-\frac{3}{2}$ and $\left(  n_{\mathrm{u-quark}}\right)
_{\mathrm{U}_{\mathrm{Y}}\mathrm{(1)}}=-\frac{1}{2}$.

To characterize the property of opposite-component of internal zeroes for
$\bar{u}$ and $d$ as
\begin{equation}
\left(  n_{\mathrm{d-quark}}\right)  _{\mathrm{SU}_{\mathrm{weak}}%
\mathrm{(2)}}=-\left(  n_{\mathrm{u-quark}}\right)  _{\mathrm{SU}%
_{\mathrm{weak}}\mathrm{(2)}}=\frac{3}{2},
\end{equation}
$\bar{u}$ and $d$ make up a quark $\mathrm{SU}_{\mathrm{weak}}\mathrm{(2)}$
spinor\cite{gla}%
\begin{equation}
\psi_{\mathrm{quark}}=\left(
\begin{array}
[c]{c}%
\bar{u}\\
d
\end{array}
\right)  .
\end{equation}
They are fractionally charged and come each in three \textquotedblleft
colours\textquotedblright,%
\begin{equation}
\psi_{\mathrm{quark},L}=\frac{1}{2}(1+\gamma_{5})\left(
\begin{array}
[c]{c}%
\bar{u}\\
d
\end{array}
\right)  .
\end{equation}

On the other hand, to characterize the property of identical-component of
internal zeroes for $\bar{u}$ and $d$ as$\ $%
\begin{equation}
\left(  n_{\mathrm{d-quark}}\right)  _{\mathrm{SU}_{\mathrm{weak}}%
\mathrm{(2)}}=\left(  n_{\mathrm{u-quark}}\right)  _{\mathrm{SU}%
_{\mathrm{weak}}\mathrm{(2)}}=-\frac{1}{2},
\end{equation}
a left-hand d-quark and a left-hand anti-u-quark have the supercharge
$\mathcal{Y}$ to be
\begin{equation}
\mathcal{Y}(\psi_{\mathrm{quark,}L})=-\frac{1}{2}\times \frac{2}{3}=\frac{1}%
{3}.
\end{equation}

Since a unit internal zero has supercharge $\mathcal{Y}$ to be $-\frac{2}{3}$,
we find that the super charge for quarks $u_{R},$ $d_{R}$ to be
\begin{equation}
\mathcal{Y}(u_{R})=-\frac{4}{3},\text{ }\mathcal{Y}(d_{R})=-\frac{2}{3}.
\end{equation}

\subsubsection{Effective Lagrangian of electro-weak $\mathrm{SU}%
_{\mathrm{weak}}\mathrm{(2)\otimes U}_{\mathrm{Y}}\mathrm{(1)}$ gauge theory}

Finally, we write down the effective Lagrangian of electro-weak $\mathrm{SU}%
_{\mathrm{weak}}\mathrm{(2)\otimes U}_{\mathrm{Y}}\mathrm{(1)}$ gauge theory%
\begin{align}
\mathcal{L}  &  =\mathcal{L}_{\mathrm{fermion}}+\mathcal{L}_{\mathrm{Y}%
}(\mathrm{U}_{\mathrm{Y}}\mathrm{(1)})+\mathcal{L}_{\mathrm{weak}}%
(\mathrm{SU}_{\mathrm{weak}}\mathrm{(2)})\\
&  =\mathrm{Tr}\bar{\psi}_{L}i\gamma^{\mu}(\partial_{\mu}-igW_{\mu}%
+i\frac{g^{\prime}}{2}B_{\mu})\psi_{L}\nonumber \\
&  +\bar{\psi}_{R}i\gamma^{\mu}(\partial_{\mu}+ig^{\prime}B_{\mu})\psi
_{R}\nonumber \\
&  -\frac{1}{4}B_{\mu \nu}B^{\mu \nu}-\mathrm{Tr}\frac{1}{2}W_{\mu \nu}W^{\mu \nu
}\nonumber
\end{align}
where $W_{\mu}$ and $B_{\mu}$ denote the gauge fields associated to weak
$\mathrm{SU}_{\mathrm{weak}}\mathrm{(2)}$ and super-charge $\mathrm{U}%
_{\mathrm{Y}}\mathrm{(1)}$ respectively, of which the corresponding field
strengths are $W_{\mu \nu}$ and $B_{\mu \nu}$. The two coupling constants $g$
and $g^{\prime}$ correspond to the groups $\mathrm{SU}_{\mathrm{weak}%
}\mathrm{(2)}$ and $\mathrm{U}_{\mathrm{Y}}\mathrm{(1)}$ respectively. Because
neutrino has only left-hand degrees of freedom, the charged $W$'s couple only
to the left-handed components of the lepton fields.

Electro-weak $\mathrm{SU}_{\mathrm{weak}}\mathrm{(2)\otimes U}_{\mathrm{Y}%
}\mathrm{(1)}$ gauge symmetry comes from the redundancies of 2-level composite
knots and those of 1-level internal knots in unit cell of level-3
zero-lattices. As a result, the fluctuations of $\mathrm{SU}_{\mathrm{weak}%
}\mathrm{(2)}$ gauge theory comes from the fluctuations of linking number on a
unit cell of level-3 W-type zero-lattice and the fluctuations of
$\mathrm{U}_{\mathrm{Y}}\mathrm{(1)}$ gauge theory come from the residue phase
fluctuations of internal-twisting on a unit cell of level-3 zero-lattice.

\subsubsection{Higgs mechanism and spontaneous symmetry breaking for standard
knot-crystal}

We then study the properties of angular velocity of leapfrogging motion
$\omega(\vec{X},t)$ (that is really the Higgs field $\Phi(\vec{X},t)/2$ in
Standard model), that is
\begin{equation}
\psi^{\prime}(\vec{X},t)=e^{2i\tau_{x}\omega(\vec{X},t)\cdot t}\cdot \psi
(\vec{X},t).
\end{equation}
Consequently, the effect of leapfrogging motion is to change $\psi_{L}(\vec
{X})$ to $\psi_{R}(\vec{X})$ and there appears an extra term in Hamiltonian as
$\psi^{\dagger}\omega \tau_{x}\psi=\psi_{R}^{\dagger}\omega \psi_{L}.$ Because
of
\begin{equation}
\mathcal{Y}(\psi_{\mathrm{Lepton,}L})=-1,\text{ }\mathcal{Y}(\psi
_{\mathrm{Lepton,}R})=0,
\end{equation}
the $\mathrm{U}_{\mathrm{Y}}\mathrm{(1)}$ super-charge of $\omega(\vec{X},t)$
must be
\begin{align}
\mathcal{Y}(\omega(\vec{X},t))  &  =\mathcal{Y}(\psi_{\mathrm{Lepton,}%
R})-\mathcal{Y}(\psi_{\mathrm{Lepton,}L})\\
&  =1.\nonumber
\end{align}
On the other hand, due to
\begin{align}
\psi_{L}(\vec{X})  &  \rightarrow e_{L}^{i\vec{\tau}\vec{\theta}(\vec{X})}%
\psi(\vec{X}),\text{ }\nonumber \\
\psi_{R}(\vec{X})  &  \rightarrow \psi_{R}(\vec{X}),
\end{align}
$\omega(\vec{X},t)$ must be an $\mathrm{SU}_{\mathrm{weak}}\mathrm{(2)}$
complex doublet as
\begin{align}
\omega(\vec{X},t)  &  =\left(
\begin{array}
[c]{c}%
\phi^{+}\\
\phi^{0}%
\end{array}
\right)  ,\text{ }\nonumber \\
\omega(\vec{X},t)  &  \rightarrow e^{i\vec{\tau}\vec{\theta}(X)}\omega(\vec
{X},t).
\end{align}
As a result, the fluctuating leapfrogging angular velocity of a standard
knot-crystal $\omega_{0}\rightarrow \omega(\vec{X},t)$ plays the role of Higgs
field $\Phi(\vec{X},t)$ in Standard model\cite{higgs}.

Next, we write down the Lagrangian of the leapfrogging field $\omega(\vec
{X},t).$ Because the leapfrogging field $\omega(\vec{X},t)$ is an
$\mathrm{SU}_{\mathrm{weak}}\mathrm{(2)}$ complex doublet and has supercharge
$\mathcal{Y}=1$, we get the kinetic term of leapfrogging field $\omega(\vec
{X},t)$ as
\begin{equation}
|(\partial_{\mu}-ig\frac{\vec{\tau}}{2}\cdot \vec{W}_{\mu}-i\frac{g^{\prime}%
}{2}B_{\mu})\omega(\vec{X},t)|^{2}.
\end{equation}
To obtain the finite leapfrogging velocity, we also add a phenomenological
term $V(\omega(\vec{X},t)).$

Finally, by adding Yukawa coupling between the Higgs field and fermions, the
full Lagrangian of $\omega(\vec{X},t)$ is given by%
\begin{align}
\mathcal{L}_{\mathrm{Higgs}}  &  =|(\partial_{\mu}-ig\frac{\vec{\tau}}{2}%
\cdot \vec{W}_{\mu}-i\frac{g^{\prime}}{2}B_{\mu})\omega(\vec{X},t)|^{2}%
\nonumber \\
&  -V(\omega(\vec{X},t))+\bar{\psi}_{\mathrm{Lepton,}L}G_{\mathrm{Lepton}%
}\omega(\vec{X},t)\bar{\psi}_{\mathrm{Lepton,}R}\nonumber \\
&  +\bar{\psi}_{\mathrm{quark},L}G_{\mathrm{quark}}\omega(\vec{X}%
,t)\psi_{\mathrm{quark},R}+h.c.
\end{align}
where $G_{\mathrm{Lepton}}=G_{\mathrm{quark}}=2$.

A finite leapfrogging angular velocity is given by minimizing $\omega(\vec
{X},t),$ of which the expected value is $\omega^{\ast}$. Then, the weak gauge
symmetry is spontaneously broken, we get a finite angular velocity of
leapfrogging motion as
\begin{equation}
\left \langle \omega(\vec{X},t)\right \rangle =\frac{1}{\sqrt{2}}\left(
\begin{array}
[c]{c}%
0\\
\omega^{\ast}%
\end{array}
\right)  +\delta \omega(\vec{X},t).
\end{equation}
A finite angular velocity of leapfrogging motion creates a mass term for the
electrons $m_{e}=\frac{1}{\sqrt{2}}G_{e}\omega^{\ast}.$For the system with
finite angular velocity of leapfrogging motion, we produce masses for the
quarks given by $m_{u}=G_{u}\omega^{\ast}$, $m_{d}=G_{d}\omega^{\ast}.$
Because there is no right-hand neutrino, the mass of neutrino is zero,
$m_{\nu}\equiv0$. In addition, the Higgs field also has mass that is
$m_{\mathrm{Higgs}}\neq0.$

The finite angular velocity of leapfrogging motion plays the role of Higgs
condensation and the Higgs mechanism of 3-level composite knot-crystal with
($\mathcal{N}=4$, $\mathcal{M}=3$) breaks the original gauge symmetry
according to $\mathrm{SU}_{\mathrm{weak}}\mathrm{(2)\otimes U}_{\mathrm{Y}%
}\mathrm{(1)}\rightarrow \mathrm{U}_{\mathrm{EM}}\mathrm{(1)}$.

As a result, the $\mathrm{SU}_{\mathrm{weak}}\mathrm{(2)}$ gauge fields obtain
masses from the following terms\cite{wein}
\begin{equation}
\frac{1}{8}(\omega^{\ast})^{2}[g^{2}(W_{\mu}^{1}W^{1\mu}+W_{\mu}^{2}W^{2\mu
})+(g^{\prime}B_{\mu}-gW_{\mu}^{3})^{2}].
\end{equation}
The mass for the charged vector bosons $W_{\mu}^{\pm}=(W_{\mu}^{1}\mp iW_{\mu
}^{2})/\sqrt{2}\ $is
\begin{equation}
m_{W}=\frac{\omega^{\ast}g}{2}.
\end{equation}
After diagonalization, the gauge fields $B_{\mu}$ and $W_{\mu}^{3}$ are
transformed into gauge fields $Z_{\mu}$ and $A_{\mu}$ from the following
relations%
\begin{align}
Z_{\mu}  &  =\cos \theta_{W}B_{\mu}-\sin \theta_{W}W_{\mu}^{3},\\
A_{\mu}  &  =\cos \theta_{W}B_{\mu}+\sin \theta_{W}W_{\mu}^{3},\nonumber
\end{align}
with $\tan \theta_{W}=g^{\prime}/g$, of which the masses are
\begin{align}
m_{Z}  &  =\frac{\omega^{\ast}(g^{2}+{g^{\prime}}^{2})^{1}/2}{2}=\frac{m_{W}%
}{\cos \theta_{W}},\\
m_{A}  &  =0.\nonumber
\end{align}
The \textquotedblleft Weinberg angle\textquotedblright \ $\theta_{W}$ becomes
the angle between the original $\mathrm{U(1)}$ and the one left unbroken. \ 

The neutral gauge bosons $A_{\mu}$ are massless and will be identified with
the photons. Now, the gauge symmetry $\mathrm{U}_{\mathrm{EM}}\mathrm{(1)}$
accompanying $A_{\mu}$ is to change the position of the internal zeroes of the
composite knots that will never be broken. That is
\begin{align}
\psi(\vec{X})  &  =\left(
\begin{array}
[c]{c}%
\psi_{L}(\vec{X})\\
\psi_{R}(\vec{X})
\end{array}
\right)  \rightarrow \left(
\begin{array}
[c]{c}%
\psi_{L}(x)e^{-i\mathrm{e}\phi(\vec{X})}\\
\psi_{R}(x)e^{-i\mathrm{e}\phi(\vec{X})}%
\end{array}
\right) \\
&  \rightarrow e^{-i\mathrm{e}\phi(\vec{X})}\psi(\vec{X}).\nonumber
\end{align}
Then, the electric charge operator $\mathrm{e}$ will be a linear combination
of $T_{3}$ and $\mathcal{Y}$ as
\begin{equation}
\mathrm{e}=T_{3}+\frac{\mathcal{Y}}{2}. \label{charge}%
\end{equation}

\subsection{Unified theory of standard knot-crystal}

Finally, we derive the unified theory of standard knot-crystal by considering
all points of view. The low energy effective theory is just the Standard model
-- an $\mathrm{SU}_{\mathrm{Strong}}\mathrm{(3)}\otimes \mathrm{SU}%
_{\mathrm{weak}}\mathrm{(2)}\otimes$\textrm{U}$_{\mathrm{Y}}$\textrm{(1)}
gauge theory with Higgs mechanism due to spontaneous symmetry breaking.

Before considering Higgs condensation or $\omega_{0}=0,$ the low energy
effective Lagrangian density is
\begin{align}
\mathcal{L}_{\mathrm{SM}}  &  =\mathcal{L}_{\mathrm{fermion}}+\mathcal{L}%
_{\mathrm{Y}}(\mathrm{U}_{\mathrm{Y}}\mathrm{(1)})+\mathcal{L}%
_{\mathrm{strong}}(\mathrm{SU}_{\mathrm{Strong}}\mathrm{(3)})\\
&  +\mathcal{L}_{\mathrm{weak}}(\mathrm{SU}_{\mathrm{weak}}\mathrm{(2)}%
)+\mathcal{L}_{\mathrm{Higgs}}\nonumber
\end{align}
where%
\begin{align}
\mathcal{L}_{\mathrm{fermion}}  &  =\mathrm{Tr}\bar{\psi}_{L}i\gamma^{\mu
}(\partial_{\mu}-ig\frac{\vec{\tau}}{2}\cdot \vec{W}_{\mu}+i\frac{g^{\prime}%
}{2}B_{\mu})\psi_{L}\\
&  +\bar{\psi}_{R}i\gamma^{\mu}(\partial_{\mu}+ig^{\prime}B_{\mu})\psi
_{R}\nonumber \\
&  +\mathrm{Tr}\bar{\psi}_{\mathrm{quark},L}i(\partial_{\mu}-ig\frac{\vec
{\tau}}{2}\cdot \vec{W}_{\mu}-i\frac{g^{\prime}}{6}B_{\mu})\psi_{\mathrm{quark}%
,L}\nonumber \\
&  +\bar{u}_{R}i(\partial_{\mu}+i\frac{2g^{\prime}}{3}B_{\mu})u_{R}+\bar
{d}_{R}i(\partial_{\mu}+i\frac{g^{\prime}}{3}B_{\mu})d_{R},\nonumber
\end{align}%
\begin{equation}
\mathcal{L}_{\mathrm{Y}}(\mathrm{U}_{\mathrm{Y}}\mathrm{(1)})=-\frac{1}%
{4}B_{\mu \nu}B^{\mu \nu},
\end{equation}%
\begin{equation}
\mathcal{L}_{\mathrm{strong}}(\mathrm{SU}_{\mathrm{Strong}}\mathrm{(3)}%
)=-\frac{1}{2}\mathrm{Tr}\mathcal{G}_{\mu \nu}\mathcal{G}^{\mu \nu},
\end{equation}%
\begin{equation}
\mathcal{L}_{\mathrm{weak}}(\mathrm{SU}_{\mathrm{weak}}\mathrm{(2)}%
)=-\mathrm{Tr}\frac{1}{2}W_{\mu \nu}W^{\mu \nu},
\end{equation}%
\begin{align}
\mathcal{L}_{\mathrm{Higgs}}  &  =|(\partial_{\mu}-ig\frac{\vec{\tau}}{2}%
\cdot \vec{W}_{\mu}-i\frac{g^{\prime}}{2}B_{\mu})\omega|^{2}\\
&  -V(\omega)+\bar{\psi}_{\mathrm{quark},L}G_{\mathrm{quark}}\omega
\psi_{\mathrm{quark},R}\nonumber \\
&  +\bar{\psi}_{\mathrm{Lepton,}L}G_{\mathrm{Lepton}}\omega \psi
_{\mathrm{Lepton,}R}+h.c..\nonumber
\end{align}

After considering the Higgs condensation $\omega^{\ast}\neq0,$ we have the low
energy effective Lagrangian as
\begin{align}
\mathcal{L}_{\mathrm{SM}}  &  =\mathcal{L}_{\mathrm{fermion}}+\mathcal{L}%
(\mathrm{U}_{\mathrm{EM}}\mathrm{(1)})+\mathcal{L}_{\mathrm{weak}}%
(\mathrm{SU}_{\mathrm{weak}}\mathrm{(2)})\\
&  +\mathcal{L}_{\mathrm{strong}}(\mathrm{SU}_{\mathrm{Strong}}\mathrm{(3)}%
)+\mathcal{L}_{\mathrm{Higgs}}\nonumber \\
&  ={\bar{\nu}}i\gamma^{\mu}\partial_{\mu}\nu+\bar{e}_{R}i\gamma^{\mu}%
\partial_{\mu}e\nonumber \\
&  +\bar{u}i\gamma^{\mu}\partial_{\mu}u+\bar{d}i\gamma^{\mu}\partial_{\mu
}d+m_{e}{\bar{e}}e\nonumber \\
&  +m_{d}{\bar{d}}d+m_{u}{\bar{u}}u.\nonumber \\
&  -\frac{1}{4}F_{\mu \nu}F^{\mu \nu}-\mathrm{Tr}\frac{1}{2}W_{\mu \nu}W^{\mu \nu
}-\frac{1}{2}\mathrm{Tr}\mathcal{G}_{\mu \nu}\mathcal{G}^{\mu \nu}\nonumber \\
&  +\mathrm{e}{A}_{\mu}(x){j}_{(em)}^{\mu}+\mathrm{Tr}J_{\mathrm{YM}}^{\mu
}\mathcal{A}_{\mu}\nonumber \\
&  +\frac{1}{8}(\omega^{\ast})^{2}[g^{2}(W_{\mu}^{1}W^{1\mu}+W_{\mu}%
^{2}W^{2\mu})+(g^{\prime}B_{\mu}-gW_{\mu}^{3})^{2}]\nonumber \\
&  +{g}_{w}(W_{\mu}^{+}{j}_{w\,-}^{\mu}+W_{\mu}^{-}{j}_{w\,+}^{\mu})+{g}%
_{z}W_{\mu}^{z}{j}_{w\,z}^{\mu}\nonumber \\
&  +|\partial_{\mu}\omega|^{2}+m_{\mathrm{Higgs}}\left \vert \omega \right \vert
^{2}+...\nonumber
\end{align}
where the electric current is%
\begin{align}
{j}_{(em)}^{\mu}  &  =i{\bar{e}}\gamma^{\mu}e+i\frac{2}{3}{\bar{u}}\gamma
^{\mu}u\\
&  +i\frac{1}{3}{\bar{d}}\gamma^{\mu}d,\nonumber
\end{align}
the weak current is%
\begin{align}
{j}_{w\,-}^{\mu}  &  =i\bar{e}\gamma_{\mu}\nu+iu^{T}\gamma_{\mu}d,\\
{j}_{w\,+}^{\mu}  &  =i\bar{\nu}\gamma_{\mu}e+i{d}^{T}\gamma_{\mu}u,\nonumber
\end{align}
and the color current is
\begin{equation}
J_{\mathrm{YM}}^{a,\mu}={\bar{u}}i\gamma^{\mu}T^{a}u+{\bar{d}}i\gamma^{\mu
}T^{a}d.
\end{equation}
This is exact one-flavor Standard model, an $\mathrm{SU}_{\mathrm{Strong}%
}\mathrm{(3)}\otimes \mathrm{(SU(2))}_{\mathrm{weak}}\otimes$\textrm{U}%
$_{\mathrm{Y}}$\textrm{(1)} gauge theory with Higgs mechanism due to
spontaneous symmetry breaking.

\section{Summary and discussion}

In the end, we give a summary. In this paper, knot dynamics on composite
knot-crystal is studied. From knot physics, the knot-crystal becomes
fundamental physical object, of which elementary excitations are gauge fields
and fermionic particles (knots). In particular, for a special composite
knot-crystal -- standard knot-crystal, the low energy effective theory is just
the Standard model -- an $\mathrm{SU}_{\mathrm{Strong}}\mathrm{(3)}%
\otimes \mathrm{SU}_{\mathrm{weak}}\mathrm{(2)}\otimes$\textrm{U}$_{\mathrm{Y}%
}$\textrm{(1)} gauge theory with Higgs mechanism due to spontaneous symmetry
breaking. The knot physics provides a way to unify all gauge fields and
elementary fermionic excitations. In table.1, we give the correspondence
between different quantum field theories in modern physics and the different
composite knot-crystals in\ knot physics.

\begin{widetext}
\begin{table*}[t]%
\begin{tabular}
[c]{|c|cccc|}\hline Composite knot-crystal& Quantum field theory & & &\\
\hline   $\mathcal{N}=1$ and $\mathcal{M}=1$& Weyl Fermion model && &
\\ \hline $\mathcal{N}=2$ and $\mathcal{M}=1$& Dirac Fermion model & & &
\\ \hline $\mathcal{N}=2$ and $\mathcal{M}=2$& Weak SU(2) gauge field theory & & &
\\ \hline $\mathcal{N}=4$ and $\mathcal{M}=2$& SU(n)*U(1) gauge field theory & & &
\\ \hline $\mathcal{N}=4$ and $\mathcal{M}=3$& One flavor Standard model& &
&\\ \hline
\end{tabular}
\caption{The correspondence between different quantum field theories in modern physics and the different composite knot-crystals in knot physics}
\end{table*}
\end{widetext}

Grand Unified Theory (GUT) is a dream of physicists to unify all
non-gravitational interactions. There are several approaches towards GUT.
String theory is a possible theory of GUT\cite{ss}. According to string
theory, matter consists of vibrating strings (or strands) and different
oscillatory patterns of strings become different particles with different
masses. In condensed matter physics, the idea of our universe as an
\textquotedblleft \emph{emergent}\textquotedblright \ phenomenon has become
increasingly popular. In emergence approach, a deeper and unified
understanding of the universe is developed based on a complicated many-body
system. Different quantum fields correspond to different many-body systems:
the vacuum corresponds to the ground state and the elementary particles
correspond to the excitations of the systems. According to string-net picture
proposed by Wen, our universe (such as gauge interaction, Fermi statistics,
...) emerges from a frustrated quantum spin model\cite{wen}. In addition,
there exist many other proposals of GUT from different points of view, such as
loop quantum gravity theory\cite{loop,s}, G. Lisi's \textrm{E8}
theory\cite{e8}, C. Schiller's Strand Model\cite{Strand}, ...

Based on knot physics, we may guess that our universe becomes a composite
knot-crystal -- standard knot-crystal, of which the hierarchy series is $\{3,$
$N\}$ where $N$ is a very large number, $N\gg1$ (for example, $N\sim10^{15}$).

In this paper there are two important issues that we don't discuss: 1) the
flavor physics, including the origin of three-flavor, the value of each
elements of Cabibbo-Kobayashi-Maskawa mass matrix\cite{ckn} and the mechanism
of weak charge-parity (CP) violence; 2) quark confinement that may be relevant
to the dynamics of internal knots inside the quarks. In the future, we will
study these issues and develops a complete knot theory for particle physics
and quantum field theory.

\acknowledgments This work is supported by NSFC Grant No. 11674026.


\begin{thebibliography}{99}                                                                                               %


\bibitem {Thomson1880a}W. Thomson, Phil. Mag. \textbf{10}, 155 (1880).

\bibitem {Donnelly1991a}R.J. Donnelly, \textit{Quantized Vortices in Helium
II} (Cambridge University Press, Cambridge 1991).

\bibitem {S95}B. V. Svistunov, Phys. Rev. B\textbf{52}, 3647 (1995).

\bibitem {V00}W. F. Vinen, Phys. Rev. B\textbf{61}, 1410 (2000).

\bibitem {dys93}F.~W. Dyson, Part II Phil. Trans. R. Soc. Lond. A
\textbf{184}, 1041 (1893).

\bibitem {hic22}W.~M. Hicks, Proc. R. Soc. Lond. A \textbf{102}, 111 (1922).

\bibitem {bor13}A.~V. Borisov, A.~A. Kilin and I.~S. Mamaev, Regul. Chaotic
Dyn. \textbf{18}, 33 (2013).

\bibitem {wac14}D.~H. Wacks, A.~W. Baggaley and C.~F. Barenghi, Phys. Fluids
\textbf{26}, 027102 (2014).

\bibitem {cap14}R.~M. Caplan, J.~D. Talley, R. Carretero-Gonz\'{a}lez and
P.~G. Kevrekidis, Phys. Fluids \textbf{26}, 097101 (2014).

\bibitem {1}N. Hietala, R. H\"{a}nninen, H. Salman, C. F. Barenghi, arXiv:1603.06403.

\bibitem {Kleckner2013}D. Kleckner, and W. T. M. Irvine, Nat. Phys.
\textbf{9}, 253(2013).

\bibitem {Hall2016}D. S. Hall, M. W. Ray, K. Tiurev, E. Ruokokoski, A. H.
Gheorghe, and M. M\"{o}tt\"{o}nen, Nat. Phys. \textbf{12}, 478(2016).

\bibitem {kou}S. P. Kou, Int. J Mod. Phys. \textbf{B 31}, 1750241 (2017).

\bibitem {leap}Boris Khesin, arXiv:1201.5914.

\bibitem {2}H. K. Moffatt, and R. L. Ricca, Proc. R. Soc. Lond. A 439, 411 (1992).

\bibitem {3}K. Klenin and J. Langowski, Biopolymers, V 54, 307 (2000).

\bibitem {kou1}S. P. Kou, arXiv:1706.06879.

\bibitem {we}H. Weyl, I. Z. Phys. \textbf{56}: 330 (1929).

\bibitem {yang}C. N. Yang and R. L. Mills, Phys. Rev. 96, 191 (1954).

\bibitem {ly}T. D. Lee and C. N. Yang, Phys. Rev. 104, 254 (1956).

\bibitem {higgs}P. W. Higgs, Phys. Lett. 12, 132 (1964).

\bibitem {wein}S. Weinberg, Phys. Rev. Lett. 19, 1264 (1967).

\bibitem {kk}T. Kaluza, Sitzungsber. Preuss. Akad. Wiss. Berlin. (Math.
Phys.): 966 (1921). O. Klein, . Zeitschrift f\"{u}r Physik A 37. 895 (1926).
O. Klein, Nature 118. 516 (1926).

\bibitem {Pati}J. C. Pati, A. Salam, Phys. Rev. D10, 275 (1974).

\bibitem {Harari}H. Harari, Phys. Lett. B86, 83 (1979).

\bibitem {Shupe}M. Shupe, Phys. Lett. B86, 87 (1979).

\bibitem {helon}S. Bilson-Thompson,. arXiv:hep-ph/0503213 (2005).

\bibitem {Strand}C. Schiller, \textit{A fascinating speculation: The strand
model}.

\bibitem {gla}S. L. Glashow, J. Iliopoulos, and L. Maiani, Phys. Rev. D 2,
1285 (1970).

\bibitem {stand}C. Quigg, \textit{Gauge Theories of the Strong, Weak, and
Electromagnetc Interactions,} Addison--Wesley Pub. Co., Menlo Park, (1983).

\bibitem {so}H. Georgi and S. Glashow, Phys. Rev. Letts. 32. 438, (1974).

\bibitem {ss}M. Kaku, \textit{Introduction to Superstring and M-Theory 2nd
edition}. New York, Springer-Verlag (1999).

\bibitem {wen}{X.-G. Wen}, \textit{Quantum Field Theory of Many-Body Systems},
(Oxford Univ. Press, Oxford, 2004).

\bibitem {loop}A. Ashtekar, Phys. Rev. Lett. 57, 2244 (1986); C. Rovelli,
arXiv:1102.3660; C. Rovelli, Cambridge University Press (2004).

\bibitem {s}S. O. Bilson-Thompson, F. Markopoulou, L. Smolin, Class. Quant.
Grav. 24, 3975 (2007).

\bibitem {e8}A. G. Lisi, arXiv:0711.0770 [hep-th] (2007).

\bibitem {ckn}N. Cabibbo, Phys. Rev. Letts. 10. 531 (1963). M. Kobayashi, T.
Maskawa, Prog. Theo. Phys. 49. 652 (1973).
\end{thebibliography}
\end{document}